\documentclass[pre,singlecolumn,noshowpacs,preprintnumbers,amsmath,amssymb,nofootinbib]{revtex4}

\usepackage{mathrsfs}
\usepackage[dvipdfmx]{graphicx}
\usepackage{dcolumn}
\usepackage{bm}
\usepackage{color}
\usepackage{comment}
\usepackage{ascmac}
\usepackage{diagbox}
\usepackage{multirow}
\usepackage{rotating}
\usepackage{diagbox}
\usepackage{ulem}

\newcommand{\mcL}{\mathcal{L}}
\newcommand{\mcD}{\mathcal{D}}
\newcommand{\tl}{\tilde}
\newcommand{\ra}{\rangle}
\newcommand{\la}{\langle}
\newcommand{\tb}{\textbf}

\newcommand{\eps}{\epsilon}

\newcommand{\hxi}{\hat{\xi}}

\newcommand{\hz}{\hat{z}}

\newcommand{\ve}{\varepsilon}

\newcommand{\mrG}{\mathrm{G}}
\newcommand{\mrP}{\mathrm{P}}

\newcommand{\hbmz}{\hat{\bm{z}}}
\newcommand{\pd}{\partial}

\newcommand{\tilh}{\tilde{h}}
\newcommand{\tilbh}{\tilde{\bm{h}}}

\newcommand{\htt}{\hat{t}}

\newcommand{\cut}{\mathrm{cut}}

\newcommand{\hnu}{\hat{\nu}}
\newcommand{\hN}{\hat{N}}
\newcommand{\hxiCP}{\hat{\xi}^{\mathrm{CP}}}

\newcommand{\mrss}{\mathrm{ss}}
\newcommand{\mcS}{\mathcal{S}_F}
\newcommand{\mcR}{\mathcal{R}}
\newcommand{\mcH}{\mathcal{H}}
\newcommand{\eu}{u}

\begin{document}
\title{
				Exact asymptotic solutions to nonlinear Hawkes processes: \\
				a systematic classification of the steady-state solutions
			}

\author{Kiyoshi Kanazawa$^{1,2}$ and Didier Sornette$^{3,4}$}

\affiliation{
	$^1$ Faculty of Engineering, Information and Systems, The University of Tsukuba, Tennodai, Tsukuba, Ibaraki 305-8573, Japan\\
	$^2$ JST, PRESTO, 4-1-8 Honcho, Kawaguchi, Saitama 332-0012, Japan\\
	$^3$ ETH Zurich, Department of Management, Technology and Economics, Zurich, Switzerland\\
	$^4$ Institute of Risk Analysis, Prediction and Management, Academy for Advanced Interdisciplinary Studies, Southern University of Science and Technology, Shenzhen, China
}
\date{\today}

\begin{abstract}
	Hawkes point processes are first-order non-Markovian stochastic models of intermittent bursty dynamics with applications to physical, seismic, epidemic, biological, financial, and social systems. While accounting for positive feedback loops that may lead to critical phenomena in complex systems, the standard linear Hawkes process only describes excitative phenomena. To describe the co-existence of excitatory and inhibitory effects (or negative feedbacks) as occurs for instance in seismic (so-called stress shadows) and neural systems (via glycine and gamma-aminobutyric acid (GABA) transmitters), extensions involving nonlinear dependences of the intensity as a function of past activity are needed. However, such nonlinear Hawkes processes have been found hitherto to be analytically intractable due to the interplay between their non-Markovian and nonlinear characteristics, with no analytical solutions available. Here we present various exact and robust asymptotic solutions to nonlinear Hawkes processes using the field Master equation (ME) approach introduced previously by the authors. We report explicit power law formulas for the steady-state intensity distributions $P_{\mrss}(\lambda)\propto \lambda^{-1-a}$, where the tail exponent $a$ is expressed analytically as a function of parameters of the nonlinear Hawkes models. 
	We present three robust interesting characteristics of the nonlinear Hawkes process:
	(i)~for one-sided positive marks (i.e., in the absence of inhibitory effects), the nonlinear Hawkes process can exhibit any power law relation either as intermediate asymptotics ($a\leq 0$) or as true asymptotics ($a>0$) by appropriate model selection; 
	(ii)~for distribution of marks with zero mean (i.e., for balanced excitatory and inhibitory effects), the Zipf law ($a\approx 1$) is universally observed for a wide class of nonlinear Hawkes processes with fast-accelerating intensity map;
	(iii)~for marks with a negative mean, the asymptotic power law tail becomes lighter as the mean mark becomes more negative. 
	We introduce the basic analytical tools for advanced Hawkes modeling, particularly for model calibration to real time-series data in various complex systems. 
\end{abstract}
\pacs{}

\maketitle

\section{Introduction}

	Intermittent bursts are ubiquitously observed with temporal and spatial clustering characters in physical~\cite{ScherMontroll75,Scheretal2002}, seismic~\cite{Ogata1988,Ogata1999,HelmsSor02,Shyametal2019}, epidemic~\cite{Feng-epidemic2019}, financial~\cite{Errais-Giesecke2010,Chakraetal11,Jiangealmultifract19}, and social systems~\cite{SorDeschatres04,CraneSor08}. Such bursty dynamics can be well described by the Hawkes process~\cite{Hawkes1,Hawkes2,Hawkes3}, a non-Markovian self-excited point process capturing both long memory effects and critical bursts, such that past events keep their potential influence to trigger future bursty events for a long time, potentially leading to critical bursts. However, the essential non-Markovian nature of this model has been an obstacle preventing the development of a unified analytical theory because the established framework of Markovian stochastic processes is not applicable. 
	
	Recently, however, a new theoretical scheme was developed to address such non-Markovian stochastic processes directly, in particular for the Hawkes process~\cite{KzDidier2019PRL,KzDidier2019PRR}. This scheme is based on a mapping from the non-Markovian Hawkes model to an equivalent stochastic partial differential equation (SPDE). The SPDE is then mapped to an equivalent field master equation (field ME), i.e., a functional-differential equation for the probability density functional (PDF) of the intensity. The solutions of this equation can be obtained analytically in their asymptotic form, in particular near criticality. This theoretical framework predicted a novel non-universal power law relation for the intensity as an intermediate asymptotics~\cite{Barenblatt}. It has the potential for further explorations of the theoretical properties of more general Hawkes processes.
	
	Since the basic linear Hawkes (LHawkes) process is analytically solved in this framework, it is natural to seek further generalisation of the framework, such as for nonlinear Hawkes (NLHawkes) processes~\cite{Bremaud1996,BouchaudBook}. NLHawkes processes are particularly important to account for the presence of inhibitory effects: in addition to positive feedbacks, many systems are characterised by co-existing negative feedbacks. In the context of point processes, while the standard Hawkes process describes only excitatory processes, many systems are kept in balance by the additional occurrence of inhibitory processes. For instance, inhibitory effects naturally appear in seismicity \cite{MSA_PRL,MSA_Geophys} as any earthquake creates a tensorial stress perturbation within the visco-plasto-elastic Earth crust with the presence of ``stress shadows'' in certain regions around the ruptured fault where future earthquakes are less likely \cite{Nandan-stress-shadow_16}, while other regions are brought closer to rupture by an increase in the local relevant stress component. Similarly, neurobiological brains are kept in balance by the interplay between excitatory and inhibitory neurotransmitters, with the resulting cascades of excitations exhibiting power law statistics \cite{Plenz-Niebur_critical-brain_2014,Osorio1,SorOso_2010}. 
	 
	The long-standing problem of combining inhibitory and excitatory effects in point processes requires considering nonlinear extensions of the Hawkes processes in order to fulfill the condition that the intensity (a probability per unit time) remains nonnegative. Our recent Letter~\cite{KzDidier2021} has presented a step toward a general theory of NLHawkes processes by applying the framework of the field ME~\cite{KzDidier2019PRL,KzDidier2019PRR}. In this Letter, we discovered the existence of an asymptotic ubiquitous power law distribution of the intensity for NLHawkes processes in the case of mark distributions with non-positive mean. Since NLHawkes processes may have a huge variety of forms, and thus of control parameters, for instance in the tension-intensity map defined below and in the mark distribution, it would be useful to further study various NLHawkes processes by systematically classifying their solutions according to the asymptotic analyses of the field MEs.  

	The present article supplements our short Letter~\cite{KzDidier2021} by providing a systematic classification of various NLHawkes processes, together with various explicit exact and asymptotic solutions. In this paper, we present a general formulation for the NLHawkes processes and provide their explicit solutions for various cases. In particular, we report three interesting asymptotic features which are valid for a wide class of memory kernels. (i)~In the absence of inhibitory effects (i.e., when events all have positive marks), we find a non-universal power law relation for the intensity distribution at criticality, with an exponent $a$ that can take any value, i.e., corresponding to a genuine power law $(a>0)$ or to an intermediate power law asymptotics $(a\leq 0)$. This is in contrast to the LHawkes model, where only a negative exponent $a<0$ exists ~\cite{KzDidier2019PRL,KzDidier2019PRR}. (ii)~In the presence of inhibitory effects (i.e., both positive and negative marks coexist), in the case where the mark distribution has zero mean corresponding to a balance between inhibitory and excitatory effects, a wide class of NLHawkes processes exhibit Zipf's law ($a\approx 1$) for their intensity distributions. (iii)~For negative mean marks, we derive the asymptotic formula for the intensity PDF, whose tail becomes thinner than in the zero mean mark case. This provides a new mechanism for the ubiquity of power laws, including Zipf's law, in the form of a universal property of the NLHawkes family composed of intensity maps growing sufficiently fast as a function of the tension (to be defined below) and with balanced marks.
	
	This article is organised as follows. We present the detailed mathematical formulation of the NLHawkes processes in Sec.~\ref{sec:setup}. In Sec.~\ref{sec:field_master_Eqs}, the NLHawkes processes are mapped onto Markovian SPDEs, whose time evolution are described by MEs. We also develop a mathematical scheme to analyse the MEs, such as the functional Kramers-Moyal (KM) expansion and system size expansion (SSE) for the diffusive limit. In Sec.~\ref{sec:sol1_expon_memory_wo_inhibitory}, we study the exact solutions to NLHawkes processes with an exponential memory kernel without inhibitory effect (i.e., only the positive feedback effects are taken into account). In Sec.~\ref{sec:inhibitory_effects} and~\ref{sec:inhibitory_effects_AS}, we study the exact solutions of NLHawkes processes with an exponential memory kernel and in the presence of inhibitory effects (i.e., when both positive and negative feedback are considered). In Sec.~\ref{sec:sol3_arbitrary_memory} and~\ref{sec:sol4_arbitrary_memory}, we present the asymptotic solutions of the NLHawkes models with an arbitrary memory in the absence and presence of inhibitory effects, respectively. Sec.~\ref{sec:discussion} discusses future possible extensions and progress that can derive from our present work. Sec.~\ref{sec:conclusion} concludes and is followed by nine appendices presenting detailed derivations omitted from the main text for the sake of conciseness. 

	For readers interested only in the overview of our results, go to Sec.~\ref{sec:setup} and Table~\ref{table:summary_classification_discussion}. Indeed, all our results are summarised in Table~\ref{table:summary_classification_discussion}, which maps the inputs of the model (i.e., setups) to the outputs (i.e., the resultant asymptotic PDFs).

\section{Setup}\label{sec:setup}
	We first introduce the mathematical notations used to define the NLHawkes model. We then review the NLHawkes processes and their applications for real data analysis of complex systems, to highlight their utility and importance in various contexts. 
	
	\subsection{Mathematical notation}
		We denote any stochastic variable $\hat{A}$ with a hat to distinguish it from a non-stochastic real number $A$. The ensemble average of any stochastic variable $\hat{A}$ is written as $\la \hat{A}\ra$. The probability density function (PDF) is denoted by $P_t(A):= \la \delta(A-\hat{A}(t))\ra$, which characterizes the probability that $\hat{A}(t) \in [A,A+dA)$ as $P_t(A)dA$. Using the PDF, the ensemble average can be rewritten as 
		\begin{equation}
			\la \hat{A}(t)\ra:= \int AP_t(A)dA.
		\end{equation}

		We define the real number space by $\mcR$. Its nonnegative part is denoted by $\mcR^+:=\{x \>|\> x\geq 0, \>x\in \mcR\}$. The $K$-dimensional real number space is denoted as $\mcR_K:=\{(x_1,...x_n) \>|\> x_k \in \mcR \mbox{ for } k=1,2,\dots,n\}$ and its nonnegative part is written as $\mcR_K^+:=\{(x_1,...x_K) \>|\> x_k \in \mcR, \> x_k\geq 0 \mbox{ for } k=1,2,\dots,K\}$. We also define the functional space by $\mcS$. For example, a function $f$ defined on $\mcR^+$ is in the function space $\mcS$, such that $\{f(x)\}_{x\in \mcR^+} \in \mcS$. 

		In this paper, functionals (i.e., maps from a function space $\mcS$ to a real number space $\mcR$) appear to characterise the ``path" probability density. For any $\{z(x)\}_{x\in \mcR_+} \in \mcS$, a functional $f$ is denoted as $f[z]:= f[\{z(x)\}_{x\in \mcR^+}]$. Here, the square bracket emphasizes that $f$ is a functional (i.e., its argument is a function), but not an ordinary function.
		
		For a stochastic variable $\{\hz(t,x)\}_{x\in \mcR^+}$ defined on a field $x\in \mcR^+$, the probability density functional (PDF) is written as $P_t[z]:= \la \delta [z-\hz] \ra = P_t[\{z(x)\}_{x\in \mcR^+}]$ with the $\delta$ functional $\delta [z-\hz]:= \prod_{x\in \mcR^+}\delta(z(x)-\hz(t,x))$. Here, the PDF is defined over paths so that probability weighted quantities involve path integrals. For instance, the ensemble average is defined by 
		\begin{equation}
			\la \hat A(t)\ra = \int  A(t)P_t[z]\mcD z, \>\>\> \mcD z:= \prod_{x\in \mcR^+}dz(x)
		\end{equation} 
		where $\mcD z$ is the path-integral volume element.

	\subsection{Model}
		\begin{figure*}
			\centering
			\includegraphics[width=130mm]{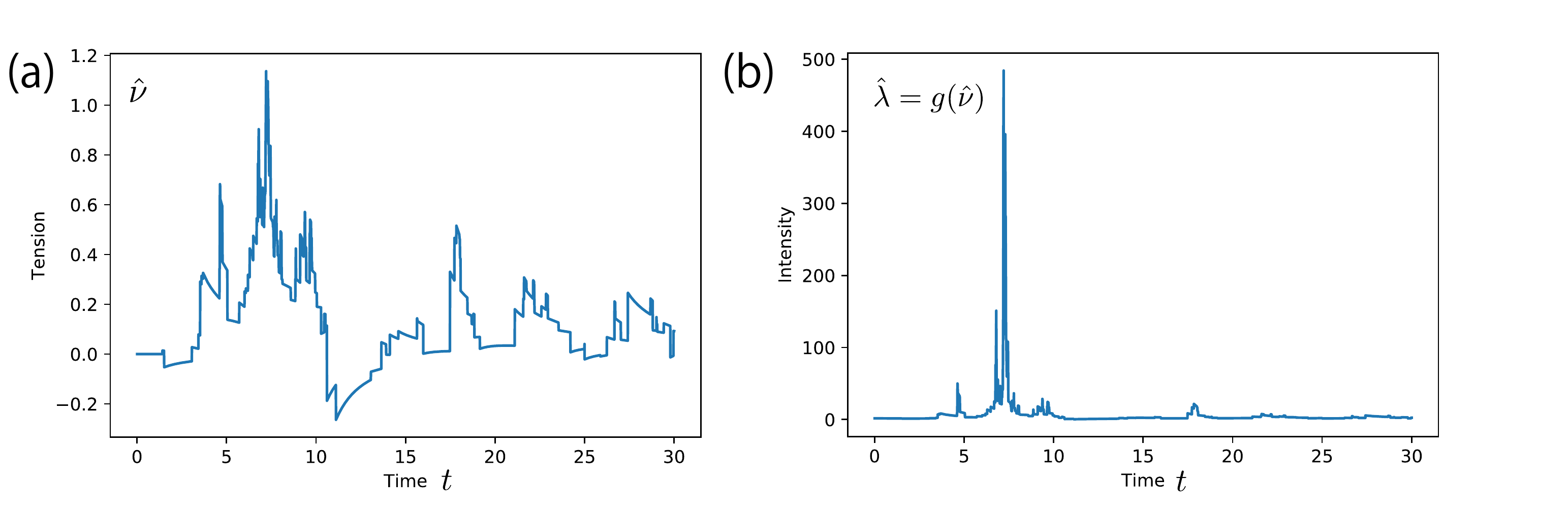}
			\caption	{
						A sample trajectory of the tension $\hnu$ (left panel) and the intensity $\hat{\lambda}=g(\hnu)$ (right panel) in the NLHawkes process~\eqref{def:NL_Hawkes_gen}. The functions and model parameters are: 
						$\rho(y)=e^{-x^2/2}/\sqrt{2\pi}$, $g(\nu)=e^{\beta \nu + \nu_0}$, $\beta=5$, $\nu_0 = 0.5$. The memory kernel $h(t)$ is defined by expression~\eqref{eq:MarkovEmbedding_Continuous} with $\tilh(x) = \{\eta_{\rm ini} + c_{\rm ini}(x-x_{\rm ini})\}/x$ for $x\in [x_{\rm ini}, x_{\rm fin}]$ and $\tilh(x)=0$ for $x\not \in [x_{\rm ini}, x_{\rm fin}]$ with $x_{\rm ini}=0.5$, $x_{\rm fin}=10$, $\eta_{\rm ini}=0.021$, $c_{\rm ini} = 0.0044$, and we use the discrete step size $dx=0.0475$. 
					}
			\label{fig:originai_trajectory}
		\end{figure*}
		
		Let us now formulate the marked NLHawkes process studied in this paper. Let us consider an internal variable $\hnu(t)$ that represents the total ``tension" of the system, and which obeys a non-Markovian stochastic differential equation (SDE),
		\begin{subequations}
			\label{def:NL_Hawkes_gen}
			\begin{equation}
				\hnu(t) = \sum_{i=1}^{\hN(t)}\hat{y}_ih(t-\htt_i), 			\label{def:NL_Hawkes_gen_a}
			\end{equation}
			where $\{\hat{y}_i\}_i$ is an independent and identically distributed (IID) random sequence of random numbers (``jumps'') obeying a distribution $\rho(y)$, $h(t)$ is a nonnegative memory kernel, $\{\htt_i\}_i$ is a Poisson process conditional on a time-dependent intensity $\hat{\lambda}(t)$, and $\hN(t)$ is the total number of events during $[0,t)$ (called ``counting process"). The jump size $\hat{y}_i$ is called a ``mark" in the point process literature. Here the intensity $\hat{\lambda}(t)$ is assumed to be stochastic and is a nonnegative nonlinear function of the total tension $\hnu(t)$, defined as
			\begin{equation}
				\hat{\lambda}(t) = g(\hnu(t)) > 0.
				\label{rth3yhb2gvq}
			\end{equation}
			In this paper, we call $g(\hnu)$ the tension-intensity map or intensity function. The intensity is the probability per unit time for an event to be triggered: assuming $\hN(t)=k$, $\hat{\lambda}(t)dt$ gives the probability that $\htt_{k+1} \in [t,t+dt)$ for an infinitely small time interval $dt\to 0$. We can rewrite Eqs.~\eqref{def:NL_Hawkes_gen_a} and \eqref{rth3yhb2gvq} as 
			\begin{equation}
				\hat{\lambda}(t) = g\left(\sum_{i=1}^{\hN(t)}\hat{y}_ih(t-\htt_i)\right).
			\end{equation}
		\end{subequations}
		This is the fundamental dynamical equation governing the NLHawkes processes.  See Fig.~\ref{fig:originai_trajectory} for a schematic trajectory. 
		
		In this article, we particularly focus on power law forms of the steady PDF of the intensity for large $\lambda$ as 
		\begin{equation}
			P_{\mrss}(\lambda) = \lim_{t\to \infty} \left< \delta(\lambda-\hat{\lambda}(t))\right> \propto \lambda^{-1-a}.
			\label{eq:power-law_PDF}
		\end{equation}
		where $a$ is the exponent of the complementary cumulative distribution function (CCDF)\footnote{
		We used the PDF exponent $a_{\mathrm{PDF}}=1+a_{\mathrm{CCDF}}$ for the description of the power law relations in Refs.~\cite{KzDidier2019PRL,KzDidier2019PRR}, where CCDF stands for the complementary cumulative distribution function.}.
		
		\paragraph*{Remark.}
		Model~\eqref{def:NL_Hawkes_gen} is a natural nonlinear generalisation of the conventional (linear) Hawkes process. Indeed, the LHawkes process is recovered by choosing a linear intensity function, 
		\begin{equation}
			\lambda = g(\nu) = \nu + \nu_0,
		\end{equation}
		assuming both $h(t)$ and $\nu$ are nonnegative. In contrast to the conventional Hawkes process, we do not assume nonnegativity of $\hnu$ and $\hat{y}_i$ for the case of general nonnegative nonlinear intensity function $g(\hnu)$.

		For the LHawkes process, the integral of the memory kernel
		\begin{equation}
			\label{eq:def:branching_ratio}
			\eta := \int_{0}^\infty h(t)dt
		\end{equation}
		is an important parameter (called the {\it branching ratio}) since it controls the fertility of events to trigger descendants (triggered events). Indeed, the LHawkes process is subcritical for $\eta<1$, critical at $\eta=1$, and supercritical for $\eta>1$.

	\subsection{Motivation and literature review}
		We now present a brief self-contained review of the existing literature on NLHawkes processes for statistical physics readers who may be unfamiliar with this topic. Readers interested only in our main results may skip this subsection. 
		
		NLHawkes processes were first introduced by Br\'emaud and Massouli\'e~\cite{Bremaud1996} in 1996, who were concerned with general conditions for the existence of the processes. Since then, there have been a few applications to seismic, financial, and neural modelling, in particular for empirical comparisons. However, beyond the derivation of general conditions for existence, obtaining analytical solutions of these models is very difficult due to the complex interplay between their nonlinear and non-Markovian structures. Only a few studies exist, such as the analysis of the stability of these processes (conditions for non-explosiveness)~\cite{Bremaud1996}, a special solution for the ZHawkes (Zumbach Hawkes) processes with an exponential memory in the diffusive limit~\cite{QHawkesBouchaud}, and an asymptotic analysis for high-baseline intensity using the functional central limit theorem~\cite{GaoZhu2018_NLHawkes}.
		
		There are several motivations for introducing NLHawkes processes. Here we focus on two following interesting properties: (i)~inhibitory effects and (ii)~physical underpinning of the nonlinear tension-intensity maps. Indeed, one of the motivations for introducing NLHawkes processes is to describe inhibitory effects~\cite{BouchaudBook}, such that previous events can produce negative feedback effects on the total tension $\hnu$. For simplicity, let us consider the case where the tension-intensity map $\lambda = g(\nu)$ is an increasing function. For this setup, an event with positive mark $\hat{y}_i > 0$ is likely to induce future events and, inversely, an event with negative mark $\hat{y}_i < 0$ is likely to inhibit future events. This means that negative marks $\hat{y}_i<0$ represent inhibitory effects, while positive marks $\hat{y}_i>0$ represent excitatory effects.
		
		To implement such inhibitory effects, nonlinearity in the tension-intensity map is essential because the LHawkes process cannot accommodate inhibitory effects. Indeed, if we assume an affine tension-intensity map $g(\nu)=\nu_0 + \nu$ with nonnegative constant $\nu_0$, $\nu$ must take value larger than $-\nu_0$ in order for the tension-intensity map to remain nonnegative. This condition requires that the mark distribution must be one-sided toward the positive direction (i.e., $\rho(y) = 0$ for $y<0$); otherwise, $\nu$ takes value smaller than $-\nu_0$ with non-zero probability and the model assumption is violated. In fact, the model cannot be defined as a negative intensity or probability density cannot be given mathematical sense.
		
		The second nice property of NLHawkes processes is that the nonlinearity of the tension-intensiy maps captures in a natural way the real mechanisms occurring in the modelled systems. Let us illustrate this point by reviewing several versions of the NLHawkes processes studied in the literature. 
						
		\subsubsection{Example 1: seismic modelling \label{rtwjhyrbgq}}
			One of the most illustrative cases is found in the modelling of statistical seismicity. Let us regard the tension $\hnu(t)$ as the total stress component along the fault best oriented for rupture at a given point $\vec r$ in the Earth crust. Let $\hat{t}_i$ be the time of occurrence of the $i$th earthquake. This earthquake creates a tensorial stress field that adds to the pre-existing stress field. Again, for our discussion, we simplify the picture by taking this stress perturbation as being a scalar, for instance the Mohr-Coulomb stress amplitude along the fault best oriented for rupture at point $\vec r$. Furthermore, we take into account the visco-elastic property of the crust, which means that a stress perturbation is progressively relaxed via a memory kernel $h(t-\hat{t}_i)$ that tends to $0$ at long times. Then, the total stress at $\vec r$ is obtained as the sum of the stress perturbations created by all past earthquakes 
			\begin{equation}
				\hnu(t) = \sum_{i=1}^{N(t)} \hat{y}_i h(t-\hat{t}_i).
				\label{yhybh3g2q}
			\end{equation}
			Note that the marks $ \hat{y}_i$ can be positive (resp. negative), corresponding to the $i$th earthquake bringing the point $\vec r$ closer to (resp. further away from) failure.  The former case is the most intuitive and represents the stress load on $\vec r$ due to the redistribution of forces by the earthquake fault slip in its neighborhood, especially close to its fault tips and in its stress lobes of positive influence. The later case is known as ``stress shadow'' \cite{Nandan-stress-shadow_16} and is associated with the tensorial nature of the stress disturbances induced by an earthquake. Given the stochasticity in the distribution of earthquake sizes, in their positions and orientations, the marks  $\hat{y}_i$ are stochastic variables. Given the total stress (tension) \eqref{yhybh3g2q}, the next ingredient is to recognise that mechanical rupture and earthquakes are thermally activated with an effective inverse temperature $\beta$ that is renormalised via the quenched heterogeneity of the medium \cite{Cili_1_01,Cili_1_02,Sai-Sor05}. Then, the probability for the next earthquake to occur is given by the Arrhenius formula, thus formulating the intensity $\lambda$ as a decreasing exponential function $e^{-\beta \Delta E(t)}$  of the energy barrier $\Delta E(t)$ for nucleation. The key point is to approximate the energy barrier as a decreasing affine function of the stress field: $\Delta E(t) = (E-\hnu(t))$, where $E$ is a constant. Putting all together, this yields 
			\begin{equation}
				\lambda (t) = \lambda_0 e^{\beta \hnu(t)}.
				\label{yj4u3hgq}
			\end{equation}
			We finally obtain the NLHawkes with an exponential intensity
			\begin{equation}
				\hat{\lambda}(t) = g\left(\sum_{i=1}^{N(t)} \hat{y}_i h(t-\hat{t}_i)\right), \>\>\> g(\hnu):= \lambda_0 e^{\beta \hnu}.
				\label{dqeth5y3jwn}
			\end{equation}
			In addition, given that the prediction of earthquake magnitudes is empirically very difficult (while the short-term prediction of their rates is rather possible \cite{HelmSor03}), it is a plausible assumption that the marks are drawn independently of the current tension $\hnu(t)$.
								
			In this simplified presentation, we have restricted our attention to the temporal version of the general formulation, which is known as the multifractal stress activation (MSA) model \cite{MSA_PRL,MSA_Geophys} and involves space in addition to time in the formulation of the tension and intensity. It is remarkable that both inhibitory effects and nonlinear intensity function appear naturally for this system, as the result of the random stress perturbations induced by earthquakes and from the Arrhenius law (renormalised by quenched disorder), respectively. 
			
			It should noted that Refs.~\cite{MSA_PRL,MSA_Geophys} offered only an approximate scaling theory to derive magnitude dependent Omori law exponents and that no analytical results exist for the MSA model or for its temporal-only version \eqref{dqeth5y3jwn}.

		\subsubsection{Example 2: financial modelling}\label{subsec:QHawkes_review}
			Ref.~\cite{Bowsher07} is one of the very first uses in finance of the LHawkes process (in its bivariate form) in order to model the joint dynamics of trades and mid-price changes of the NYSE. Ref.~\cite{FiliSor_12} provided the first quantitative framework using the LHawkes process to study and quantify the level of endogeneity (or ``reflexivity'') of market fluctuations. The basic idea is that trades and price changes are analysed by investors (humans or machines) as one of the useful information channels to improve trading decisions, on the basis (or belief) that past actions reveal intentions and that there is a persistence in price trends, volume, volatility and more generally of trading activity. In this sense, the self-exciting Hawkes process is a natural candidate to model the point processes of discrete trades and mid-price changes \cite{BacryMuzy2015}.
			 			
			As an improved model, a nonlinear version of the Hawkes process was introduced by Blanc, Donier, and Bouchaud \cite{QHawkesBouchaud}, where the intensity dynamics is given by a quadratic extension to the standard Hawkes process,
			\begin{equation}
				\hat{\lambda}(t) = \lambda_0 + \int_{-\infty}^t L(t-s)\hxi^{\mrP}_{\rho(y);\hat{\lambda}(s)}(s)ds + \int_{-\infty}^tds \int_{-\infty}^tdu K(t-s,t-u)\hxi^{\mrP}_{\rho(y);\hat{\lambda}(s)}(s)\hxi^{\mrP}_{\rho(y);\hat{\lambda}(u)}(u)
				\label{eq:QHawkes_review}
			\end{equation}
			with $\rho(y)=(1/2)[\delta(y-1)+\delta(y+1)]$ and the term $\hxi^{\mrP}_{\rho(y);\hat{\lambda}(s)}(s)$ is the compound Poisson process with intensity $\hat{\lambda}(t)$ and jump size distribution $\rho(y)$ as defined below by expression (\ref{rwhbgrvgqfvq}). This model is called the quadratic Hawkes (QHawkes) processes and has been theoretically analyzed in Ref.~\cite{QHawkesBouchaud}. 
			
			Since this model is nonlinear and non-Markovian, its systematic analysis is difficult and only limited results are available. However, by assuming $K(t,s) = h(s)h(s)$ and $L(t)=0$, this model reduces to a simpler NLHawkes process,
			\begin{equation}
				\hat{\lambda}(t) = g\left(\int_{-\infty}^t h(t-s)\hxi_{\rho(y); \hat{\lambda}(s)}(s)ds\right) = g\left(\sum_{i=1}^{\hat{N}(t)}\hat{y}_ih(t-\hat{t}_i) \right), \>\>\> g(\nu):= \lambda_0 + \nu^2,
			\end{equation}
			where we have used $\hxi_{\rho(y); \hat{\lambda}(s)}(s)=\sum_{i=1}^{\hat{N}(s)}\hat{y}_i\delta(s-\hat{t}_i)$. This NLHawkes process is a special case of the Zumbach Hawkes (ZHawkes) process, without the Hawkes feedback. While the ZHawkes process is simpler than the QHawkes, it is still difficult to solve analytically. Therefore, the analysis in Ref.~\cite{QHawkesBouchaud} focused on the special case of an exponential memory $h(t) = (\eta/\tau)e^{-t/\tau}$ and considered the diffusive limit\footnote{
				They call their analysis the low-frequency asymptotics, taking the long-time limit and a constant endogeneity rescaling. This asymptotic method is essentially equivalent to the diffusive limit in the framework of the system size expansion (SSE), a traditional asymptotic analyses developed for statistical physics, which is formulated in Sec.~\ref{sec:system_size_expansion}.  
				}. 
			For this special case, the steady-state PDF of the intensity obeys a power law with non-universal exponent
			\begin{equation}
				P_{\mrss}(\lambda) \propto \lambda^{-1-a}, \>\>\> a = \frac{1}{2} + \frac{1}{2n_Z}
				\label{eq:power-law_QHawkes_Bouchaud}
			\end{equation}
			with a constant $n_Z$ called Zumbach norm (see Ref.~\cite{QHawkesBouchaud} for details). It is remarkable that a power law relation~\eqref{eq:power-law_QHawkes_Bouchaud} appears even for short memory kernels without introducing any power law distributions. To the best of our knowledge, this special solution was the only available analytical solution for a NLHawkes process before our work~\cite{KzDidier2021}.
			
			One of the main claims in Ref.~\cite{QHawkesBouchaud} is that the power law relation~\eqref{eq:power-law_QHawkes_Bouchaud} provides a validation step supporting the relevance of the QHawkes process for financial data analyses, because it matches the empirical power law price-change distribution, which is well-known stylised fact in market microstructure. From this viewpoint, the authors of Ref.~\cite{QHawkesBouchaud} claim that the QHawkes process is a minimal generalisation beyond the LHawkes process that is essentially needed to account for empirical facts. 

		\subsubsection{Example 3: the self-excited multifractal model}
			It is also useful to mention the self-excited multifractal model \cite{FiliSorMulti_11}, which is not per se a point process but got its inspiration from self-excited point processes, the concept of reflexivity \cite{Soros88}, the multifractal random walk  model \cite{BacryMuzyMRW2001} and its generalisations \cite{SaichevSornettegen06,SaichevFilimonov2008}. Reminiscent of a NLHawkes model with a much stronger nonlinearity than quadratic, the self-excited multifractal model is defined such that the amplitudes of the increments of the process are expressed as exponentials of a long memory of past increments: 
			\begin{equation}\label{eq_dX}
				dX(t)=\sigma\exp\left\{
				-\frac1\sigma\int\limits_{-\infty}^t h(t-t')dX(t')
				\right\}dW(t),
			\end{equation}
			where $dW(t)$ is the increment of the regular Wiener noise process, $h(t)$ is a memory kernel function and $\sigma$ controls the amplitude of the noise as well as the dimension and scale of $X(t)$. Interpreting $dX(t)$ as a log-return of a financial price, the self-excited multifractal process recovers all the standard stylised facts documented in empirical financial time series. The exploration of the links between the self-excited multifractal model and the exponential NLHawkes process is left for the future.
	
		\subsubsection{Goal of this study: solutions for various nonlinear Hawkes processes}
			The above summaries highlight the fact that analytical solutions for NLHawkes processes have not been obtained yet, except for special cases (such as the ZHawkes case with exponential memory in the diffusive limit). In this context, our goal is to systematically classify NLHawkes processes and then provide analytical (both exact and asymptotic) solutions for various NLHawkes processes, in particular for the steady-state intensity PDF $P_{\mrss}(\lambda)$. All our results are summarised as Table~\ref{table:summary_classification_discussion}, with the mapping between the inputs of the model (i.e., setups) to the outputs (i.e., the resultant asymptotic PDFs). 

\begin{sidewaystable}[h]
	\centering
	\caption{
		Summary of the results obtained in the present work, for both one-sided and two-sided mark distributions. The obtained steady-state intensity distributions of intensities are systematically classified for various NLHawkes processes. FAI and MSA stand for fast-accelerating intensity ($g(\nu)\gg \nu^2$) and multifractal stress activation model ($g(\nu)\propto e^{\beta \nu}$, $\beta>0$). In this report, we assume that the moment-generating function $\Phi(x):=\int_{-\infty}^\infty \rho(y)(e^{xy}-1)dy$ exists and $c^*$ is the positive root of $\Phi(c^*)=0$ for $m<0$ or is equal to zero for $m=0$. In addition, we define $m:=\int_{-\infty}^\infty y\rho(y)dy$, $p_+:=\int_0^\infty \rho(y)dy$, and $p_-:=\int_{-\infty}^0 \rho(y)dy$. 
	}
	\label{table:summary_classification_discussion}
	\begin{tabular}{|c||c|c|c|c|c|c|c|} \hline
		Model & Mark PDF $\rho(y)$ & Tension-intensity map $g(\nu)$ & Critical? & PDF $P_{\mrss}(\lambda)$ & Exponent $a$ & \begin{tabular}{c} Section for \\ $h(t)=(\eta/\tau)e^{-t/\tau}$ \end{tabular} & \begin{tabular}{c} Section for  \\ general $h(t)$ \end{tabular} \\\hline\hline
		 Linear & \multicolumn{1}{c|}{\begin{tabular}{c} One sided: \\ $\rho(y)=0$ \\ for $y<0$ \end{tabular}} & $\nu_0+\nu_1$, $\nu_0>0$ & \multicolumn{1}{c|}{\begin{tabular}{c} Yes \\ $(\eta \uparrow 1)$ \end{tabular}} & \multicolumn{1}{c|}{\begin{tabular}{c} $\propto \lambda^{-1-a}e^{-\lambda/\lambda_{\mathrm{cut}}}$ \\ $(\lim_{\eta\uparrow 1}\nu_{\mathrm{cut}}=\infty)$ \end{tabular}} & \begin{tabular}{c} $a<0$ \\ (intermediate \\ asymptotics, \\ non-universal) \end{tabular} & \multicolumn{1}{c|}{Sec.~\ref{sec:sol1_expon_memory_wo_inhibitory}} & \multicolumn{1}{c|}{Sec.~\ref{sec:sol3_arbitrary_memory}} \\\cline{1-1}\cline{3-3}\cline{6-6}
		Ramp & \multicolumn{1}{c|}{} & $\max\{\nu_0,\nu - \nu_1\}$, $\nu_0>0$ & \multicolumn{1}{c|}{} & \multicolumn{1}{c|}{} & \begin{tabular}{c} any-real number $a$ \\ (non-universal) \end{tabular} & \multicolumn{1}{c|}{} & \multicolumn{1}{c|}{} \\\hline\hline
		Ramp & \multicolumn{1}{c|}{\begin{tabular}{c} Two-sided, \\ symmetric: \\ $\rho(y)=\rho(-y)$ \end{tabular}} & $\max\{\nu_0,\nu - \nu_1\}$, $\nu_0>0$ & \multicolumn{1}{c|}{No} & \begin{tabular}{c} $\propto e^{-\lambda/\lambda_{\mathrm{cut}}}$ \\ $(\lambda_{\mathrm{cut}}<\infty)$ \end{tabular} & Absent & \multicolumn{1}{c|}{Sec.~\ref{sec:inhibitory_effects}} & \multicolumn{1}{c|}{Missing} \\\cline{1-1}\cline{3-3}\cline{5-6}
		Quadratic & \multicolumn{1}{c|}{} & $\propto \nu^2$ & \multicolumn{1}{c|}{} & \multicolumn{1}{c|}{$\propto \lambda^{-1-a}$} & \begin{tabular}{c} $a>1/2$ \\ (non-universal) \end{tabular} & \multicolumn{1}{c|}{} & \multicolumn{1}{c|}{} \\\cline{1-1}\cline{3-3}\cline{6-6}\cline{8-8}
		\begin{tabular}{c} Polynomial \\ of order $n>2$ \\ (FAI) \end{tabular} & \multicolumn{1}{c|}{} & $\propto \nu^n, n>2$ & \multicolumn{1}{c|}{} & \multicolumn{1}{c|}{} & \begin{tabular}{c} $a=1-1/n$ \\ (universal) \end{tabular} & \multicolumn{1}{c|}{} & \multicolumn{1}{c|}{Sec.~\ref{sec:sol4_arbitrary_memory}} \\\cline{1-1}\cline{3-3}\cline{6-6}
		\begin{tabular}{c} Exponential \\ (FAI, MSA) \end{tabular} & \multicolumn{1}{c|}{} & $\propto e^{\beta \nu}$ & \multicolumn{1}{c|}{} & \multicolumn{1}{c|}{} & \multicolumn{1}{c|}{\begin{tabular}{c} $a=1$ \\ (universal, Zipf) \end{tabular}} & \multicolumn{1}{c|}{} & \multicolumn{1}{c|}{} \\\cline{1-1}\cline{3-3}
		\begin{tabular}{c} Gaussian \\ (FAI) \end{tabular} & \multicolumn{1}{c|}{} & $\propto e^{\beta \nu^2}$ & \multicolumn{1}{c|}{} & \multicolumn{1}{c|}{} & \multicolumn{1}{c|}{} & \multicolumn{1}{c|}{} & \multicolumn{1}{c|}{} \\\cline{1-1}\cline{3-3}\cline{5-6}
		FAI & \multicolumn{1}{c|}{} & $\gg \nu^2$ & \multicolumn{1}{c|}{} & \multicolumn{2}{c|}{$\propto \lambda^{-1}\left[\frac{dg(\nu)}{d\nu}\right]^{-1}_{\nu=g^{-1}(\lambda)}$} & \multicolumn{1}{c|}{} & \multicolumn{1}{c|}{} \\\hline\hline
		Ramp & \multicolumn{1}{c|}{\begin{tabular}{c} Two-sided, \\ non positive mean: \\ $p_+>0$, \\ $p_->0$, \\ $m \leq 0$ \end{tabular}} & $\max\{\nu_0,\nu - \nu_1\}$, $\nu_0>0$ & \multicolumn{1}{c|}{No} & \begin{tabular}{c} $\propto e^{-\lambda/\lambda_{\mathrm{cut}}}$ \\ $(\lambda_{\mathrm{cut}}<\infty)$ \end{tabular} & Absent & \multicolumn{1}{c|}{Sec.~\ref{sec:inhibitory_effects_AS}} & Missing \\\cline{1-1}\cline{3-3}\cline{5-6}\cline{8-8}
		\begin{tabular}{c} Exponential \\ (FAI, MSA) \end{tabular} & \multicolumn{1}{c|}{} & $\propto e^{\beta \nu}$ & \multicolumn{1}{c|}{} & $\propto \lambda^{-1-a}$ & \begin{tabular}{c}  $a\geq 1$ \\(non-universal) \end{tabular} & \multicolumn{1}{c|}{} & \multicolumn{1}{c|}{Sec.~\ref{sec:sol4_arbitrary_memory}} \\\cline{1-1}\cline{3-3}\cline{5-6}
		FAI & \multicolumn{1}{c|}{} & $\gg \nu^2$ & \multicolumn{1}{c|}{} & \multicolumn{2}{c|}{\begin{tabular}{c} $\propto \lambda^{-1}\left[e^{-u\nu}\left|\frac{dg(\nu)}{d\nu}\right|^{-1}\right]_{\nu=g^{-1}(\lambda)}$ \\ $u:=c^*/h(0)$ \end{tabular}} & \multicolumn{1}{c|}{} & \multicolumn{1}{c|}{} \\\hline
	\end{tabular}
\end{sidewaystable}

\section{Master equation formulation}\label{sec:field_master_Eqs}
	In this section, we introduce an analytical framework for the general NLHawkes process based on the field MEs. We first provide a Markovian mapping from the original non-Markovian NLHawkes process to a Markovian SPDE. We then derive the corresponding field ME for any memory kernel, which is shown to simplify for the special case of an exponential memory kernel. We next develop two useful tools that have a long tradition in the history of physical stochastic processes: the Kramers-Moyal (KM) expansion and the system size expansion (SSE) for the diffusive limit. The field ME is then shown to reduce to the functional Fokker-Planck equations (FPEs) for a special case. 
		
	\subsection{Mapping to Markovian SPDEs}
		\begin{figure*}
			\centering
			\includegraphics[width=130mm]{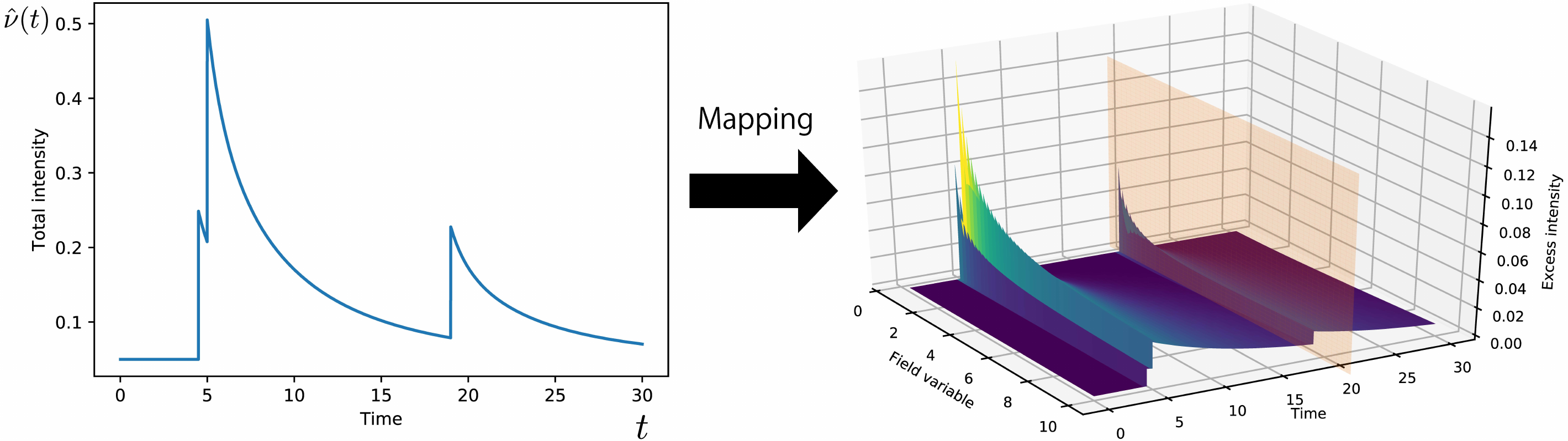}
			\caption	{
						Schematics of the Markovian embedding: the original non-Markovian one-dimensional dynamics (described by the SDE~\eqref{def:NL_Hawkes_gen}; left panel) is mapped onto the Markovian field dynamics (described by the SPDE~\eqref{eq:MarkovEmbedding_Continuous}; right panel). 
					}
			\label{fig:Mapping_MarkovEmbedding}
		\end{figure*}
		Following Ref.~\cite{KzDidier2019PRL}, let us present the mapping from the original non-Markovian stochastic process~\eqref{def:NL_Hawkes_gen} to Markovian SPDEs. Let us decompose the total tension $\hnu(t)$ and the memory kernel $h(t)$ as continuous sums
		\begin{subequations}
			\label{eq:MarkovEmbedding_Continuous}
			\begin{equation}
				\label{eq:decomposition}
				h(t) = \int_{0}^\infty dx \tilh(x)e^{-t/x}~,  \>\>\>
				\hnu(t) = \int_0^\infty dx \hz(t,x).
			\end{equation}
			The intuition behind this decomposition is that the memory kernel is decomposed into a continuous sum of exponential terms with amplitude $\tilh(x)$. This then suggests to use $x$ as an auxiliary field $x \in (0, \infty)$, and then to decompose the tension as a continuous sum over the ``excess tensions'' $\hz(t,x)$. The excess tensions are assumed to satisfy the following SPDEs:
			\begin{equation}\label{eq:SPDE_mapping}
				\frac{\partial \hz(t,x)}{\partial t} = -\frac{\hz(t,x)}{x} + \tilh(x)\hxiCP_{\rho(y);\hat{\lambda}(t)}, \>\>\> \hat{\lambda}(t) = G[\hz]:= g\left( \int_0^\infty dx \hz(t,x)\right)
			\end{equation}
			under the initial condition $\hz(t=0,x)=0$. The term $\hxiCP_{\rho(y);\hat{\lambda}(t)}$ is the compound Poisson process with intensity $\hat{\lambda}(t)$ and jump size distribution $\rho(y)$:
			\begin{equation}
				\hxiCP_{\rho(y);\hat{\lambda}(t)} = \sum_{i=1}^{\hN(t)}\hat{y}_i\delta(t-\htt_i),
				\label{rwhbgrvgqfvq}
			\end{equation}
		\end{subequations}
		which means that the random marks $y_i$ obeys the distribution $\rho(y)$. 

		This mapping can be schematically illustrated as shown in Fig.~\ref{fig:Mapping_MarkovEmbedding}: the original dynamics is one-dimensional, governed by the SDE~\eqref{def:NL_Hawkes_gen}. In this low-dimensional representation, the dynamics is non-Markovian. However, by applying the Markovian embedding, we can construct an infinite-dimensional Markovian dynamics governed by the SPDE~\eqref{eq:MarkovEmbedding_Continuous} by adding sufficiently many auxiliary variables $\hz(t,x)$. 
		
		\subsubsection*{Proof of equivalence.}
		The SPDE~\eqref{eq:SPDE_mapping} together with the decomposition formula~\eqref{eq:decomposition} is equivalent to the original marked NLHawkes process~\eqref{def:NL_Hawkes_gen}. Indeed, the formal solution of Eq.~\eqref{eq:SPDE_mapping} is given by
		\begin{equation}
			\hz(t,x) = \tilh(x)\int_0^t dt'e^{-(t-t')/x}\hxiCP_{\rho(y);\hat{\lambda}(t)} = \sum_{i=1}^{\hN(t)}\hat{y}_i\tilh(x)\int_0^t dt'e^{-(t-t')/x}\delta(t-\htt_i) = \sum_{i=1}^{\hN(t)}\hat{y}_i\tilh(x)e^{-(t-\htt_i)/x},
		\end{equation}
		leading to
		\begin{equation}
			\hnu(t) = \sum_{i=1}^{\hN(t)}\hat{y}_i\int_0^\infty dx \tilh(x)e^{-(t-\htt_i)/x} = \sum_{i=1}^{\hN(t)}\hat{y}_ih(t-\htt_i).
		\end{equation}
		It is noteworthy that this derivation does not make explicit reference to the definition of $\hat{\lambda}(t)=g(\hnu(t))$ and is independent of the specific function $g(\nu)$.

	\subsection{Field master equation}
		In this subsection, we study the functional ME corresponding to the SPDE~\eqref{eq:SPDE_mapping}. The field ME of the PDF $P_t[z]$ is given by
		\begin{subequations}
			\label{eq:functional_master_gen}
			\begin{equation}
				\frac{\partial P_t[z]}{\partial t} = \left(\mathcal{L}_{\rm adv} + \mathcal{L}_{\rm jump}\right)P_t[z]\label{eq:functional_master_gen2}
			\end{equation}
			with advective and jump Liouville operators
			\begin{align}
				\mathcal{L}_{\rm adv}P_t[z] &:=  \int_0^\infty dx \frac{\delta}{\delta z(x)}\left(\frac{z(x)}{x}P_t[z]\right), \\
				\mathcal{L}_{\rm jump}P_t[z] &:= \int_{-\infty}^{\infty}dy\rho(y)G\left[z-y\tilh\right]P_t\left[z-y\tilh\right] - G\left[z\right]P_t[z]. 
			\end{align}
		\end{subequations}
		In this paper, we provide various analytical exact or asymptotic solutions of \eqref{eq:functional_master_gen}.

		\subsubsection{Derivation}
			It is useful to provide a derivation of the field ME~\eqref{eq:functional_master_gen} via a discrete approach, which gives a sound mathematical interpretation and control of the functional derivatives~\cite{GardinerB}. Let us consider the case of the memory kernel composed of a discrete sum of $K$ exponentials (which we refer to as $K$-exponentials),
			\begin{subequations}
				\label{eq:der_K-expon_memory_Markov_embedding}
				\begin{equation}
					\label{eq:K-exponentials}
					h(t) = \sum_{k=1}^K \tilh_k e^{-t/\tau_k}. 
				\end{equation}
				The NLHawkes process~\eqref{def:NL_Hawkes_gen} together with the $K$-exponentials~\eqref{eq:K-exponentials} can be mapped onto a Markovian equation by introducing $\hbmz:= (\hz_1, \dots, \hz_K)$, 
				\begin{equation}
					\frac{d\hz_k(t)}{dt} = -\frac{\hz_k(t)}{\tau_k} + \tilh_k\hxiCP_{\rho(y); \hat{\lambda}(t)}, \>\>\>
					\hat{\lambda}(t)=g(\hnu(t)), \>\>\> 
					\hnu(t):= \sum_{k=1}^K \hz_k(t), 
				\end{equation}
				which is parallel to the Markovian embedding procedure for Eqs.~\eqref{eq:MarkovEmbedding_Continuous}. We introduce the following function $G$, which will be convenient for future developments,
			\begin{equation}
					G(\hbmz) := g\left(\hnu \right) = g\left( \sum_{k=1}^K \hz_k \right). 
				\end{equation}
			\end{subequations}
			
			The ME for the SDE~\eqref{eq:der_K-expon_memory_Markov_embedding} is derived as follows. Let us consider an arbitrary function $f(\hbmz)$ and its time-evolution $df(\hbmz(t)): = f(\hbmz(t+dt))-f(\hbmz(t))$ during $[t,t+dt)$:
			\begin{align}
				df(\hbmz) = 	\begin{cases}
								\displaystyle -\sum_{k=1}^K\frac{\hz_{k}}{\tau_k}\frac{\partial f(\hbmz)}{\partial \hz_k}dt & (\mbox{No jump during }[t,t+dt)\mbox{: probability = }1-\hat{\lambda}(t)dt) \cr
								f(\hbmz+\hat{y} \tilbh) - f(\hbmz) & (\mbox{Jump in }[t,t+dt)\mbox{ with }\hat{y}\in [y,y+dy)\mbox{: probability=}\hat{\lambda}(t)\rho(y)dtdy)
							\end{cases}
			\end{align}
			with $\tilbh:= (\tilh_1, \dots, \tilh_K)$. By taking the ensemble average of both sides over realisations of the excess tensions $\hbmz:= (\hz_1, \dots, \hz_K)$, we obtain
			\begin{align}
				\la df(\hbmz) \ra = \left< -\sum_{k=1}^K\frac{\hz_{k}}{\tau_k}\frac{\partial f(\hbmz)}{\partial \hz_k} dt\right> + \int_{-\infty}^\infty dy\rho(y) \left< \left( f(\hbmz + \hat{y} \tilbh) - f(\hbmz) \right)dt \right>,
			\end{align}
			which is equivalent to 
			\begin{align}
				\int_{-\infty}^\infty d\bm{z}  f(\bm{z})\frac{\partial P_t(\bm{z})}{\partial t} 
				&= \int_{-\infty}^\infty d\bm{z} P_t(\bm{z})\left[ -\sum_{k=1}^K\frac{z_{k}}{\tau_k}\frac{\partial f(\bm{z})}{\partial z_k} + \int_{-\infty}^\infty dy\rho(y) G(\bm{z})\left( f(\bm{z} + y\tilbh) - f(\bm{z}) \right) \right] \notag \\
				&= \int_{-\infty}^\infty d\bm{z} f(\bm{z})\left[ \sum_{k=1}^K \frac{\partial }{\partial z_k}\frac{z_{k}}{\tau_k} P_t(\bm{z}) + \int_{-\infty}^\infty dy\rho(y) \left[ G(\bm{z}-y \tilbh)P(\bm{z} - y \tilbh) - G(\bm{z})P(\bm{z}) \right] \right],
				\label{eq:master_trans_identity1}
			\end{align}
			by using the following relation
			\begin{equation}
				\la df(\hbmz) \ra = \la f(\hbmz(t+dt))- f(\hbmz(t))\ra = \int_{-\infty}^\infty d\bm{z}  f(\bm{z})P_{t+dt}(\bm{z}) - \int_{-\infty}^\infty d\bm{z}  f(\bm{z})P_{t}(\bm{z}) = dt \int_{-\infty}^\infty d\bm{z}  f(\bm{z})\frac{\partial P_t(\bm{z})}{\partial t} + O(dt^2).
			\end{equation}
			Since Eq.~\eqref{eq:master_trans_identity1} is an identity holding for any $f(\bm{z})$, we obtain the corresponding ME
			\begin{equation}
				\label{eq:master_eq_discrete_K}
				\frac{\partial P_t(\bm{z})}{\partial t} = \sum_{k=1}^K \frac{\partial }{\partial z_k}\frac{z_{k}}{\tau_k} P_t(\bm{z}) + \int_{-\infty}^\infty dy\rho(y) \left[ G(\bm{z}-y \tilbh)P(\bm{z} - y \tilbh) - G(\bm{z})P(\bm{z}) \right],
			\end{equation}
			where we have performed an integration par part and have used the variable transformation $\bm{z}+y\tilh \to \bm{z}$.
			
			We then proceed with the continuous limit for the memory kernel. We first rewrite
			\begin{equation}
				\label{eq:K-exponentials_limit}
				h(t) = \sum_{k=1}^K \tilh_k e^{-t/\tau_k}\to \sum_{k=1}^Kdx\tilh(x_k)e^{-t/x_k}, \>\>\> \hnu(t)=g\left(\sum_{k=1}^K \hz_k(t)\right) \to g\left(\sum_{k=1}^K dx\hz(t,x_k)\right)
			\end{equation}
			for the formal replacement
			\begin{equation}
				\tau_k \to x_k, \>\>\>   \tilh_k \to \tilh(x_k)dx, \>\>\> \hz_k \to \hz(t,x_k)dx
			\end{equation}
			obtained by introducing the lattice interval $dx$ and $x_k:= kdx$. By introducing the formal functional derivative and integration for the limit $K\to \infty$ and $dx\to 0$
			\begin{equation}
				\frac{\delta }{\delta z(x_k)}[\dots ] := \lim_{dx\to 0} \frac{1}{dx}\frac{\partial }{\partial z(x_k)}[\dots ], \>\>\> \int_0^\infty dx [\dots] := \lim_{dx\to 0} \sum_{k=1}^K dx [\dots],
			\end{equation}
			we obtain 
			\begin{equation}
				\frac{\partial P_t[z]}{\partial t} = \int_0^\infty dx\frac{\delta }{\delta z(x)}\frac{z(x)}{x} P_t[z] + \int_{-\infty}^\infty dy\rho(y) G[z-y\tilh]P_t[z-y\tilh] - G[z]P_t[z]
			\end{equation}
			and
			\begin{equation}
				h(t) = \int_0^\infty dx \tilh(x)e^{-t/x}, \>\>\> \hnu(t)=g\left(\int_0^\infty dx\hz(t,x)\right),
			\end{equation}
			which is equivalent to Eq.~\eqref{eq:functional_master_gen} (see Appendix~\ref{sec:app:dirac_delta} for the definition of the Dirac delta function and the functional derivative). See Appendix~\ref{sec:app:field_master} for another derivation based on direct manipulation of functional derivatives.

		\subsubsection{Mathematical remark}
			Master (or FP) equations based on functional derivatives often appear in the description of SPDEs, such as for stochastic chemical reactions~\cite{GardinerB}. While this continuous description is a useful tool for formal calculations, unfortunately, its mathematical foundation has not been established yet. Indeed, one can easily observe that there is the potential problem of encountering a divergence, such as $[\delta/\delta z(x)] z(x)P[z]=\delta(0)P[z]+z(x)\delta P[z]/\delta z(x)$. This problem might be serious for nonlinear SPDEs even for physical observables (see the divergence problem of nonlinear stochastic chemical reaction; Chapter 13.3.3 in~\cite{GardinerB}), while it might not be for linear SPDE. One can find that this divergence is not serious for the LHawkes processs and the generalised Langevin equation~\cite{KzDidier2019PRL} at least in understanding physical observables. Remarkably, for the generalised Langevin equation, this divergence problem is essentially the same as the one encountered in quantum field theory and can be renormalised in the same manner with which the divergence problem of the zero-point energy is solved in quantum electrodynamics. We note that, in the case of the NLHawkes process, the SPDE~\eqref{eq:MarkovEmbedding_Continuous} itself is fortunately linear, while the intensity function $g(\hnu)$ is nonlinear. 
			
			To avoid such mathematically delicate issues, our strategy is to follow a safer interpretation that follows Ref.~\cite{GardinerB}: we regard the field ME~\eqref{eq:functional_master_gen} (or the FPE~\eqref{eq:functional_FP_diffusive}) as a formal limit of the discrete ME~\eqref{eq:master_eq_discrete_K}. If we encounter a potential problem of divergence in Eq.~\eqref{eq:functional_master_gen}, we return to the discrete ME~\eqref{eq:master_eq_discrete_K} to proceed with the calculations, and then come back to its formal limit~\eqref{eq:functional_master_gen}. We confirm that our main results hold  for the general discrete cases~\eqref{eq:der_K-expon_memory_Markov_embedding} and we then generalise them to the continuous limit. 

	\subsection{Special case: exponential memory kernel}
		Let us here focus on the simplest case of the single exponential memory kernel:
		\begin{equation}\label{eq:exponential_memory}
			h(t) = \frac{\eta}{\tau} e^{-t/\tau},
		\end{equation}
		or equivalently 
		\begin{equation}
			\tilh(x) = \frac{\eta}{\tau}\delta(x-\tau)
		\end{equation}
		with positive real numbers $\eta$ and $\tau$. 
		Consistent with the definition \eqref{eq:def:branching_ratio}, parameter $\eta$ is the branching ratio.
		This special case is easier to analyse analytically, since the functional ME~\eqref{eq:functional_master_gen} reduces to the ME for a PDF of the total tension $\nu$,
		\begin{equation}\label{eq:master_expon_kernel}  
			\frac{\partial P_t(\nu)}{\partial t} = \frac{1}{\tau}\frac{\partial }{\partial \nu}[\nu P_t(\nu)] + \int dy\rho(y)g(\nu-\eta y/\tau)P_t(\nu-\eta y/\tau)-g(\nu)P_t(\nu). 
		\end{equation}
			
	\subsection{Functional Kramers-Moyal expansion}
		One of the standard analytical prescriptions to analyse MEs is the KM expansion. The KM expansion was historically introduced for a formal validation of the FP description from MEs. This formal expansion was criticised by van Kampen due to its ambiguous validity as an asymptotic series. Later, van Kampen developed a mathematically sophisticated formulation in the form of the SSE~\cite{VanKampen}. Let us present a sound formulation of the KM functional expansion for the field ME, which will be utilised for a further generalisation of the SSE in Sec~\ref{sec:system_size_expansion}. 
	
		\subsubsection{Exponential memory case}
			To first present the key idea, let us focus on the exponential-memory case~\eqref{eq:exponential_memory}. By considering the expansion
			\begin{equation}
				\int dy\rho(y)g(\nu-\eta y/\tau)P_t(\nu-\eta y/\tau) = \sum_{k=0}^\infty \frac{\alpha_k}{k!}\left(-\frac{\eta}{\tau}\right)^k \frac{\partial^k}{\partial \nu^k} g(\nu)P_t(\nu), \>\>\>
				\alpha_k := \int_{-\infty}^\infty dy\rho(y) y^k, 
			\end{equation}
			the ME~\eqref{eq:master_expon_kernel} can be rewritten as 
			\begin{equation}\label{eq:KM_expon}
				\frac{\partial P_t(\nu)}{\partial t} = \frac{1}{\tau}\frac{\partial }{\partial \nu}[\nu P_t(\nu)] + \sum_{k=1}^\infty \frac{\alpha_k}{k!}\left(-\frac{\eta}{\tau}\right)^k \frac{\partial^k}{\partial \nu^k} g(\nu)P_t(\nu).
			\end{equation}
			This is the Kramers-Moyal (KM) expansion for the ME~\eqref{eq:master_expon_kernel} for this special case. We have assumed that all the KM coefficients $\{\alpha_k\}_{k\geq 1}$ are finite, which excludes some singular classes of mark distributions (e.g., power law mark distributions).

		\subsubsection{General cases}
			The above formulation can be generalised by considering the functional Taylor expansion (see Appendix~\ref{sec:app:dirac_delta})
			\begin{align}
				\int_{-\infty}^{\infty}dy\rho(y)G\left[ z-y\tilh \right] P_t\left[z-y\tilh \right] 
				= \sum_{k=0}^\infty \frac{\alpha_{k}}{k!}\left(-\int_0^\infty dx y\tilh(x) \frac{\delta }{\delta z(x)}\right)^k G\left[ z \right]P_t[z]
			\end{align}
			with KM coefficients defined by
			\begin{equation}
				\alpha_k :=\int_{-\infty}^\infty dy\rho(y) y^k,
			\end{equation}
			and assuming that all the KM coefficients $\{\alpha_k\}_{k\geq 1}$ are finite. Using this relation, the field ME can be rewritten as 
			\begin{equation}
				\frac{\partial P_t[z]}{\partial t} = \int_0^\infty dx \frac{\delta}{\delta z(x)}\left(\frac{z(x)}{x} P_t[z]\right) + 
				\sum_{k=1}^\infty \frac{\alpha_{k}}{k!}\left(-\int_0^\infty dx \tilh (x)\frac{\delta }{\delta z(x)}\right)^k G\left[ z\right]P_t[z].
			\end{equation}

	\subsection{Diffusive limit: system size expansion}\label{sec:system_size_expansion}
		\begin{figure*}
			\centering
			\includegraphics[width=150mm]{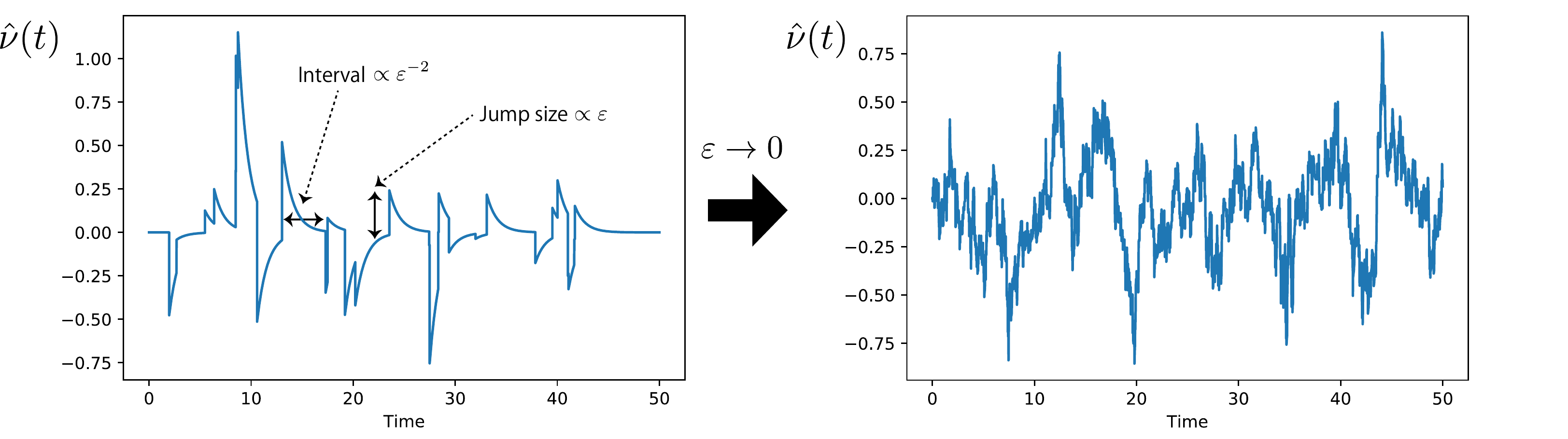}
			\caption{
						Schematic trajectory of the NLHawkes process in the diffusive limit. While the trajectory is composed of sparse jumps for large $\ve$ ($\ve=1.0$; left panel), the trajectory is composed of many small jumps for small $\ve$ and becomes approximately continuous ($\ve=0.1$; right panel). The trajectories were generated by assuming $h(t)=(\eta/\tau)e^{-t/\tau}$, $\rho_{\ve}(y)=e^{-y^2/(2\ve^2)}/\sqrt{2\pi\ve^2}$, and $g(\nu)=\lambda_0+k\nu^2$ with $\eta=0.5$, $\tau=1$, $k=1$, and $\lambda_0=0.5$. The discrete time step is $\Delta t = 10^{-4}$. 
					}	
			\label{fig:DiffusiveTrajectory}
		\end{figure*}
		
		We next consider the diffusive limit for the mark distribution according to the SSE, by assuming (i)~that the mark distribution is symmetric,
		\begin{equation}
			\rho(y) = \rho(-y),
		\end{equation}
		i.e., this is the case where inhibitory effects are as prevalent as excitatory effects. This situation will be further studied in detail in Sec.~\ref{sec:inhibitory_effects}. 
		This model is essentially different from the positive mark cases (i.e., $\rho(y)=0$ for $y\leq 0$) because both positive and negative feedback effects occur with the same probability. For instance, such assumption is natural for seismic models as the stress perturbations induced by earthquakes present indeed this symmetry (which has a complex tensorial spatial rendering, see for instance \cite{OuilSorstress06}). With this symmetry condition, all the odd-order KM coefficients are zero: $\alpha_{2k+1}=0$ for nonnegative integer $k$. 
		
		As the second assumption (ii), let us introduce a small parameter $\ve>0$ scaling the jump size in the original Hawkes process,
		\begin{equation}
			\hat{y}_i:= \ve \hat{Y}_i
		\end{equation}
		or equivalently, 
		\begin{equation}
			\hat{\lambda}(t) = g\left( \ve \sum_{i=1}^{\hN(t)} \hat{Y}_i h(t-\htt_i) \right). 
		\end{equation}
		In other words, each jump size $\hat{y}_i$ is assumed proportional to a small parameter $\ve$ and thus the rescaled jump size $\hat{Y}_i$ appears as the renormalised jump size independent of $\ve$ (see Fig.~\ref{fig:DiffusiveTrajectory}, left panel). For explicit clarification of the $\ve$ dependence, we denote below the original mark distribution $\rho(y)$ by $\rho_{\ve}(y)$. This assumption can be interpreted as a weak coupling limit between the system and the noise term. Considering the Jacobian relation (i.e., preservation of probability)
		\begin{equation}
			\rho_{\ve}(y)dy = \tl{\rho}(Y)dY
		\end{equation}
		with the scaled jump-size distribution $\tl{\rho}(Y)$, the above scaling assumption on the trajectory level is equivalent to that for the mark distribution
		\begin{equation}
			\rho_{\ve}(y): = \frac{1}{\ve}\tl{\rho}\left(\frac{y}{\ve}\right).
		\end{equation}
		We note that this scaling assumption is equivalent to the system size expansion (SSE, or often called the $\Omega$ expansion), which was originally introduced by Van Kampen for a systematic derivation of the Langevin equation within this kinetic theory (see the textbook by Van Kampen~\cite{VanKampen} and a review~\cite{KzBook} including recent extended SSEs~\cite{KzPRL2015,KzJSP2015}). With this assumption, the KM coefficients $\alpha_k$ have the following scaling
		\begin{equation}
			\alpha_{k} = 	\begin{cases}
								\ve^k \tl{\alpha}_{k} & (\mbox{even } k) \\
								0 & (\mbox{odd } k)
							\end{cases}
		\end{equation}
		with $\ve$-independent KM coefficient $\tl{\alpha}_k := \int_{-\infty}^\infty Y^k\tl{\rho}(Y)dY$.
		
		In the weak coupling limit $\ve \to 0$, each jump size is very small and thus the noise term becomes irrelevant if the intensity is constant. To keep the effect of the noise minimally relevant, let us take the diffusive limit by increasing the intensity as a function of $\ve$ (i.e., $g(\nu)$ is a function of $\ve$). As the third assumption (iii), therefore, we assume that the intensity function satisfies the diffusive scaling
		\begin{equation}
			g(\nu) = \frac{1}{\ve^2}\tl{g}(\nu)
		\end{equation}
		with $\ve$-independent intensity function $\tl{g}(\nu)$ (see Fig.~\ref{fig:DiffusiveTrajectory}, right panel). In other words, the model is explicitly written in the following form: 
		\begin{equation}
			\hat{\lambda}(t) = \frac{1}{\ve^2}\tl{g}\left( \ve \sum_{i=1}^{\hN(t)} \hat{Y}_i h(t-\htt_i) \right). 
		\end{equation}
		These three assumptions enable us to rewrite the field ME exactly in terms of the functional FPE in the diffusive limit $\ve\to 0$ (see Fig.~\ref{fig:DiffusiveTrajectory}, right panel) as we will elaborate in the following.  

		It is interesting to mention a report by Gao and Zhu~\cite{GaoZhu2018_NLHawkes}, where a similar but still different form of asymptotics is studied by assuming a one-sided mark distribution $\rho(y)=\delta(y-1)$ and a scaling for the tension-intensity map $g(\nu)=(1/\ve)\tl{g}(\nu)$ for a nonlinear version of the large baseline intensity regime for the LHawkes processes~\cite{GaoZhu2018_LHawkes}. For this setup, the trajectory fluctuates around a deterministic trajectory and thus shows quite different phenomenology.

		\subsubsection{Exponential memory case}
			To understand the main ingredients of our calculations, let us first focus on the exponential-memory case~\eqref{eq:exponential_memory}. The KM expansion can be rewritten as 
			\begin{equation}
				\frac{\partial P_t(\nu)}{\partial t} = \frac{1}{\tau}\frac{\partial }{\partial \nu}[\nu P_t(\nu)] + \sum_{k=1}^\infty \ve^{2k-2}\frac{\tl{\alpha}_{2k}}{(2k)!}\left(-\frac{\eta}{\tau}\right)^{2k} \frac{\partial^{2k}}{\partial \nu^{2k}} \tl{g}(\nu)P_t(\nu).
			\end{equation}
			By taking the diffusive limit $\ve \to 0$ (Fig.~\ref{fig:DiffusiveTrajectory}, right panel), we obtain the exact FPE
			\begin{equation}\label{eq:master_eq_diffusive}
				\frac{\partial P_t(\nu)}{\partial t} = \frac{1}{\tau}\frac{\partial }{\partial \nu}[\nu P_t(\nu)] + D\frac{\partial^{2}}{\partial \nu^{2}} \tl{g}(\nu)P_t(\nu), \>\>\>
				D := \frac{\tl{\alpha}_{2}\eta^{2}}{2\tau^{2}}. 
			\end{equation}
			We note that this FPE is equivalent to an It\^o process described by
			\begin{equation}\label{eq:SDE_diffusive-expon}
				\frac{d\hnu}{dt} = -\frac{\hnu}{\tau} + \sqrt{2D\tl{g}(\hnu)} \cdot \hxi^{\mathrm{G}}
			\end{equation}
			with the It\^o product $``\cdot"$ and the standard white Gaussian noise $\hxi^{\mathrm{G}}$, satisfying $\la\hxi^{\mathrm{G}}\ra=0$ and $\la \hxi^{\mathrm{G}}(t)\hxi^{\mathrm{G}}(t')\ra=\delta(t-t')$.

		\subsubsection{General cases}
			The above formulation can be extended for the field ME. Indeed, we obtain the following exact functional FPE
			\begin{equation}\label{eq:functional_FP_diffusive}
				\frac{\partial P_t[z]}{\partial t} = \int_0^\infty dx \frac{\delta}{\delta z(x)}\left(\frac{z(x)}{x}P_t[z]\right) + 
				\int_{0}^\infty dx\int_{0}^\infty dx' D(x,x')\frac{\delta^2 }{\delta z(x)\delta z(x')} \tl{G}\left[ z\right]P_t[z]
			\end{equation}
			with coefficient
			\begin{equation}
				D(x,x') := \frac{\alpha_{2}}{2}\tilh(x)\tilh(x').
			\end{equation}
			The functional FPE~\eqref{eq:functional_FP_diffusive} implies that the stochastic dynamics finally reduces to
			\begin{equation}
				\frac{\partial \hz(t,x)}{\partial t} = -\frac{\hz(t,x)}{x} + \sqrt{2\tl{G}\left[ \hz \right] }\cdot \hxi^{\mrG}(t;x)
			\end{equation}
			for the diffusive limit (Fig.~\ref{fig:DiffusiveTrajectory}, right panel) with the white Gaussian noise satisfying 
			\begin{equation}
				\la \hxi^{\mrG}(t;x) \ra = 0, \>\>\>
				\la \hxi^{\mrG}(t;x)\hxi^{\mrG}(t';x')\ra = D(x,x')\delta(t-t').
			\end{equation}

	\subsection{Laplace transformation}
		Here, we introduce the relevant notations for the Laplace transformation. We first define the $K$-dimensional Laplace transformation as
		\begin{equation}
			\mcL_K[f(\bm{z}); \bm{s}]:= \int_0^\infty d\bm{z} e^{-\bm{s}\cdot \bm{z}}f(\bm{z}), \>\>\> \bm{s} \in \mcR^+_K.
		\end{equation}
		In a parallel manner, the Laplace transformation in the function space can be defined as a straightforward generalisation as follows:
		\begin{equation}
			\mcL_{\mathrm{path}}[f[z]; s]:= \int_0^\infty \mcD z e^{-\int_0^\infty dxs(x)z(x)}f[z], \>\>\> s \in \mcS.
		\end{equation}
		We note that this Laplace transformation is a kind of path integral. 
		
\section{Solution 1: exponential memory kernel with one-sided mark distribution and with linear and ramp intensity maps}\label{sec:sol1_expon_memory_wo_inhibitory}
	In this section, we focus on exact solutions for the simplest case with the exponential memory kernel~\eqref{eq:exponential_memory}, whose dynamics is characterised by a simple ME~\eqref{eq:master_expon_kernel}. In particular, we here assume that all the marks are positive $\hat{y}>0$, implying the absence of inhibitory effects.

	\subsection{Exact solutions for one-sided exponential jump}
		\begin{figure}
			\centering
			\includegraphics[width=70mm]{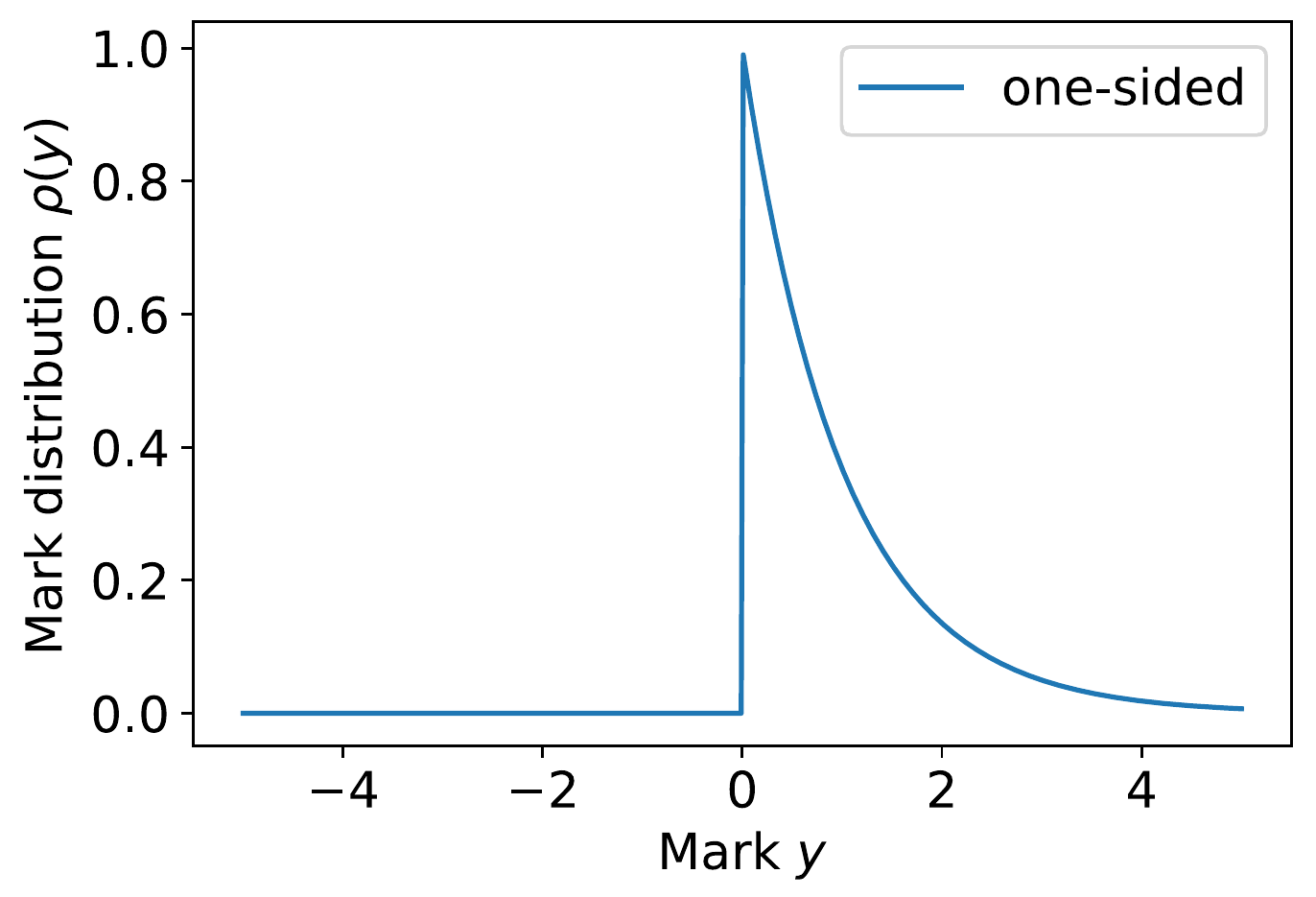}
			\caption{Schematic of the one-sided exponential mark distribution~\eqref{eq:exponential_mark_dist} with $y^*=1$}
			\label{fig:one_sided_markdist}
		\end{figure}

		Let us consider the case with the exponential memory kernel~\eqref{eq:exponential_memory} and with the one-sided exponential jump size\footnote{$\Theta(y)$ is the Heaviside function defined by $\Theta(y)=1$ for $y>0$, $\Theta(0)=1/2$, and $\Theta(y)=0$ for $y<0$.} (see Fig.~\ref{fig:one_sided_markdist})
		\begin{equation}
			\label{eq:exponential_mark_dist}
			\rho(y) := \frac{1}{y^*}e^{-y/y^*}\Theta(y),
		\end{equation}
		whose ME is known to be exactly tractable due to its special form~\cite{VanDenBroeck1983}. We assume $y^*=1$ without losing generality because the scale can be absorbed into the branching ratio $\eta$. Since both memory kernel and jump size are nonnegative, the inhibitory effects are absent in this model. Interestingly, even this simple model can exhibit nontrivial steady-state distribution functions of intensities resulting from the nonlinearity of the tension-intensity map $g(\hnu)$. This case is special because the exact steady solution to the ME~\eqref{eq:master_expon_kernel} is available. In the steady state, the exact steady solution is given by
		\begin{equation}
			P_{\mrss}(\nu) = \frac{\nu^{-1}}{Z}\exp \left[-c\nu + \tau\int \frac{g(\nu)}{\nu}d\nu\right]~, \>\>\> 
		\label{eq:steadyPDF_oneSidedExpon_ExponMem}
		\end{equation}
		with
		\begin{equation}
			c := \frac{\tau}{\eta}
		\label{wy5hnthbw8F}
		\end{equation}
		and with a normalisation constant given by 
		\begin{equation}
			Z:= \int_0^\infty d\nu \nu^{-1}\exp \left[-c\nu + \tau\int \frac{g(\nu)}{\nu} d\nu \right]~. 
 		\label{wy5hqetnthbw8F}
		\end{equation}
		
 		\subsubsection*{Derivation.}
		By utilising the following identity (see Appendix~\ref{sec:app:integral_identity} for the technical derivation),
		\begin{equation}\label{eq:identity_one_sided_expon}
			\left(1+\frac{1}{c}\frac{\partial}{\partial \nu}\right) \int_0^\infty dy e^{-y}g(\nu-\eta y/\tau)P_t(\nu-\eta y/\tau) = g(\nu)P_t(\nu),
		\end{equation}
		we can rewrite the ME as
		\begin{equation}
			\frac{\partial P_t(\nu)}{\partial t} = \frac{1}{\tau}\partial_\nu[\nu P_{t}(\nu)] - \frac{\partial_\nu/c}{1+\partial_\nu/c} g(\nu)P_{t}(\nu)
		\end{equation}
		with the differential operator $\partial_\nu:= \partial/\partial \nu$. We note that similar calculation technique can be found in Ref.~\cite{VanDenBroeck1983}. This ME can be rewritten as 
		\begin{equation}
			\frac{\partial P_t(\nu)}{\partial t} = -\frac{\pd}{\pd \nu} J_t(\nu), \>\>\> 
			J_t(\nu) := -\frac{1}{\tau}\nu P_{t}(\nu) + \frac{1/c}{1+\partial_\nu/c} g(\nu)P_{t}(\nu).
		\end{equation}
		Here we assume the natural boundary condition~\cite{GardinerB}:
		\begin{equation}
			\lim_{\nu \to \infty} J_t(\nu) = 0,
		\end{equation}
		ensuring that the mean probability ``velocity'' $J_t(\nu)/P_t(\nu)$ is zero at infinity. We reject the possibility of periodic boundary conditions which are non-physical. In the steady state, we thus obtain the exact steady solution~\eqref{eq:steadyPDF_oneSidedExpon_ExponMem}.

	\subsection{Example 1: linear Hawkes process.}
		For the linear intensity function
		\begin{equation}
			g(\nu) = \nu + \nu_0
			\label{eq:linear_Hawkes_intensity_map}
		\end{equation}
		with base intensity $\nu_0>0$, the model recovers the conventional LHawkes process.  In the subcritical case $\eta<1$, the exact steady-state solution is given by the gamma distribution
		\begin{equation}\label{eq:single_expon_linear_gamma_exact}
			P_{\mrss}(\nu) = \frac{1}{Z}\nu^{-1-a}e^{-\frac{\nu}{\nu_{\rm cut}}}, \>\>\> Z = \nu_{\rm cut}^{\tau \nu_0}\Gamma(\tau \nu_0), \>\>\> a:= -\tau\nu_0
		\end{equation}
		with the gamma function $\Gamma(x):= \int_0^\infty dt~t^{x-1}e^{-t}$. The characteristic tension for the exponential cutoff is defined as
		\begin{equation}
		\nu_{\rm cut}:= {\eta \over (1-\eta)} {1 \over \tau}~.
		\label{eq:linewtrh2tb}
		\end{equation}
		The PDF $P_{\mrss}(\nu)$ and thus the PDF of $\lambda$ is a power law with a non-universal negative exponent $a$ up to the cutoff tension $\nu_{\rm cut}$. Since the cutoff tension diverges near criticality, the power law ``tail" described by $\nu^{-1+\tau\nu_0}$ and $\lambda^{-1+\tau\nu_0}$ corresponds to an intermediate asymptotics~\cite{Barenblatt}, as reported in Ref.~\cite{KzDidier2019PRL}. 
			
	\subsection{Example 2: ramp tension-intensity map.}
		\begin{figure}
			\centering
			\includegraphics[width=70mm]{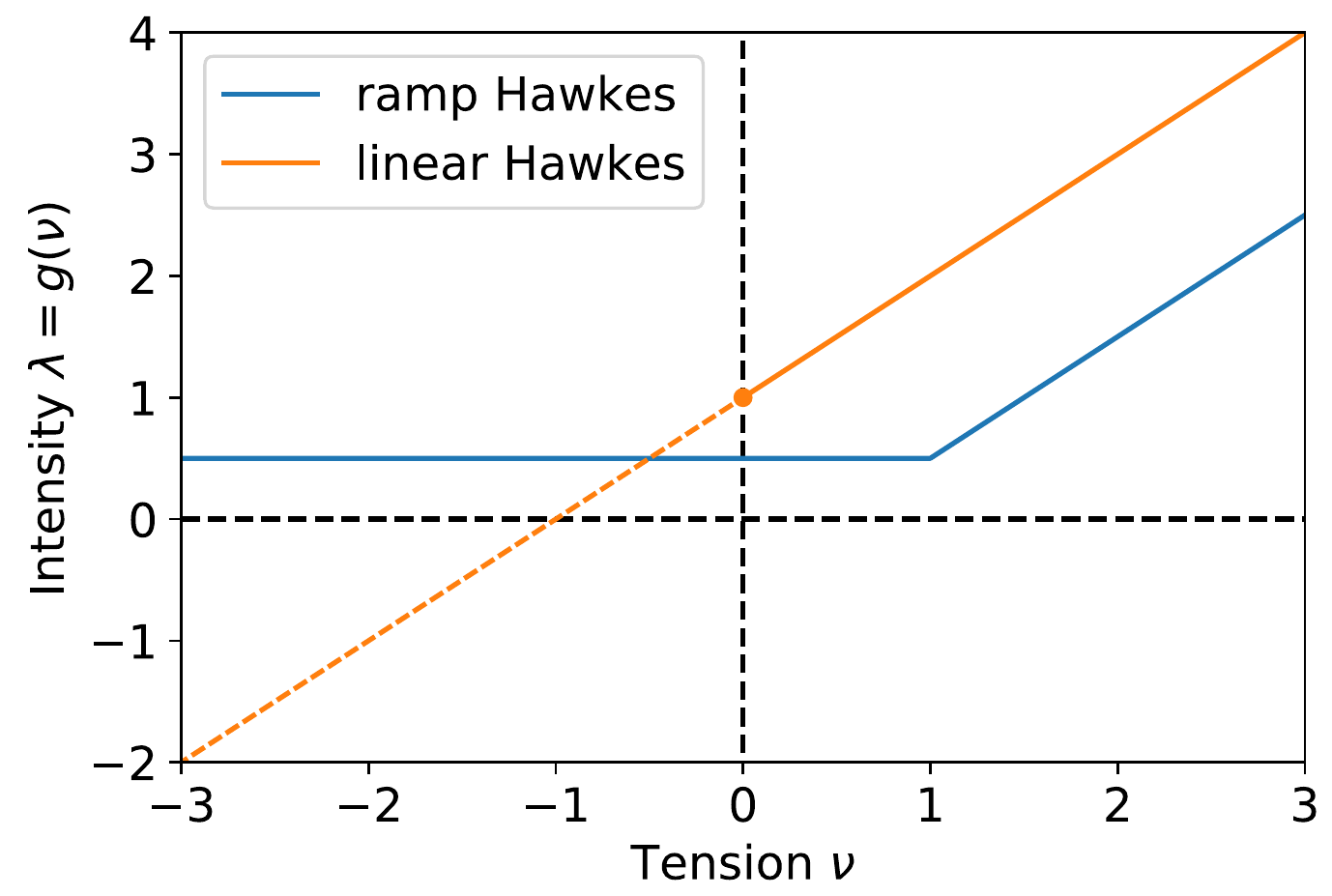}
			\caption{Schematic of the tension-intensity maps for the linear (i.e., $g(\nu)=\nu_0+\nu$ with $\nu_0=1$) and ramp (i.e., $g(\nu)=\max \{\nu_0, \nu-\nu_1\}$ with $\nu_0=1/2$ and $\nu_1=1/2$) Hawkes processes. While $\nu_0$ must be non-negative due to the non-negativity of the probability, $\nu_1$ can be either positive or non-positive.}
			\label{fig:RampVsLinear}
		\end{figure}

		Let us consider the ramp tension-intensity map (also called a rectified linear unit (ReLU) in the context of recent works in machine learning), 
		\begin{equation}\label{eq:ramp_tension-intensity_map}
			g(\nu) = \max\left\{\nu_0, \nu - \nu_1\right\}
		\end{equation}
		for positive $\nu_0$ and any real number $\nu_1$ (see Fig.~\ref{fig:RampVsLinear}). In this paper, the NLHawkes process with the ramp tension-intensity map~\eqref{eq:ramp_tension-intensity_map} is called the ramp Hawkes process. While the ramp Hawkes process is quite similar to the LHawkes process, its minimal nonlinearity leads to a genuine asymptotic power law tail, thus very different from the  LHawkes process. In the subcritical regime $\eta <1$, the exact steady solution is given by 
		\begin{equation}
			\label{eq:exact_sol_ramp_oneExpon_ExponMemory}
			P_{\mrss}(\nu) = 	\begin{cases}
							\displaystyle \frac{1}{Z}\nu^{-1-\tau\nu_1}e^{-\nu/\nu_{\rm cut}} & (\nu>\nu_0+\nu_1) \\
							\displaystyle \frac{(\nu_0+\nu_1)^{-\tau(\nu_0+\nu_1)}}{Z}\nu^{-1+\tau\nu_0}e^{\tau(\nu_0+\nu_1)-c\nu} & (\nu\leq \nu_0+\nu_1)
						\end{cases}.
		\end{equation}
		with exponential cutoff $\nu_{\rm cut}$ given by expression \eqref{eq:linewtrh2tb}, constant $c$ given by \eqref{wy5hnthbw8F} and normalisation constant $Z$ given by \eqref{wy5hqetnthbw8F}.		
		
		Interestingly, for $\nu_1>0$ and at criticality $\eta=1$,  for $\nu>\nu_0+\nu_1$, $P_{\mrss}(\nu)$ becomes a pure power law
		\begin{equation}\label{eq:true_power law_one-sided_exponential}
			P_{\mrss}(\nu) \propto \nu^{-1-a}, \>\>\> a:= \tau\nu_1,
		\end{equation}
		which is normalisable without truncation. Given the asymptotic linear relationship between $\nu$ and $\lambda$, the same power law behaviour holds for the PDF of $\lambda$. This power law is different from the intermediate asymptotic power law distribution~\eqref{eq:single_expon_linear_gamma_exact} for the LHawkes process. In this sense, the ramp Hawkes process can reproduce any power law relationship (including both true and intermediate asymptotics) at criticality, which may be useful to account for power law distributions observed empirically in various systems. It is remarkable that such a slight change from the affine structure~\eqref{eq:linear_Hawkes_intensity_map} to the rectified linear~\eqref{eq:ramp_tension-intensity_map} structure creates this large difference in the asymptotic intensity distribution. Note also that, since $\max\left\{\nu_0, \nu - \nu_1\right\} < \nu + \nu_0, \forall \nu_1>0$, the ramp tension-intensity map has a smaller intensity than that of the LHawkes process, which explains the thinner tail \eqref{eq:true_power law_one-sided_exponential} compared with \eqref{eq:single_expon_linear_gamma_exact} (this later becoming so heavy tailed close to criticality so as to become non-normalisable). Intuitively, the base tension $\nu_0$ in the ramp tension-intensity map~\eqref{eq:ramp_tension-intensity_map} acts as a replenishing engine that ensures a minimum activity, which can become the source of bursts. This structure of the  ramp tension-intensity map is somewhat reminiscent of the Kesten process~\cite{Kesten1973,SornettePhysA1998_Kesten,SornetteCont1997_Kesten}, which is well-known to produce power law distributions with tail exponent depending on the distribution of the multiplicative factors. It is interesting that the exponent $a=\tau \nu_1$ is independent of the ``resourcing'' term.

	\subsection{Existence of steady-state solutions}
		The exact solution~\eqref{eq:steadyPDF_oneSidedExpon_ExponMem} is useful in understanding the condition for the existence of a steady-state solution. For example, let us consider the case of the exponential tension-intensity map: 
		\begin{equation}
			g(\nu) = \lambda_0e^{\beta \nu}, \>\>\> \beta > 0,
		\end{equation}
		which has been used in the statistical calibration of neural spike time series in neural science~\cite{Truccolo2017}. The exact solution~\eqref{eq:steadyPDF_oneSidedExpon_ExponMem} predicts that this NLHawkes process has no steady-state solution. Indeed, 
		\begin{equation}
			P_{\mrss}(\nu)\propto \nu^{-1}\exp\left[-c\nu + \lambda_0 \tau \int \nu^{-1}e^{\beta \nu}d\nu\right] \propto \nu^{-1}\exp\left[-c\nu + \lambda_0 \tau {\rm Ei}(\beta \nu)\right]
			\simeq \nu^{-1}\exp\left[-c\nu + \lambda_0 \tau \frac{e^{\beta \nu}}{\beta \nu}\right]
		\end{equation}
		for large $\nu$ with the exponential integral ${\rm Ei}(x):=-\int_{-x}^\infty t^{-1}e^{-t}dt$. This PDF is not normalisable, implying that this NLHawkes process is always unstable independently of the model parameters. 

		To avoid this problem, one of the easiest solutions is to introduce an upper bound in the intensity function:
		\begin{equation}
			g(\nu) = \min\{\lambda_0e^{\beta \nu}, \lambda_{\max}\}
		\end{equation}
		with the finite upper boundary parameter $\lambda_{\max}>0$. Ref.~\cite{Truccolo2017} introduces a similar regularisation to guarantee the stability of their model. However, it is remarkable that this NLHawkes process is always unstable in the absence of the upper bound, and thus simulation results sensitively depend on the specific value of the cutoff $\lambda_{\max}$.  
		
		In general, if the tension-intensity map diverges faster than the linear (or ramp) function, there is no stationary solution. Indeed, for $g(\nu) \simeq \lambda_0\nu^{n}$ with $n>1$, we obtain
		\begin{equation}
			P_{\mrss}(\nu) \simeq \frac{\nu^{-1}}{Z}\exp\left[-c\nu + \lambda_0 \tau \int \nu^{n-1}d\nu\right] 
			\propto \nu^{-1}\exp\left[-c\nu + \frac{\lambda_0 \tau}{n} \nu^{n}\right],
		\end{equation}
		which is not normalisable. In this sense, the ramp Hawkes process is the boundary between the stationary and non-stationary Hawkes processes under the assumption of an exponential memory~\eqref{eq:exponential_memory} and one-sided exponential marks~\eqref{eq:exponential_mark_dist}.

		Thus, an NLHawkes process with one-sided positive marks is not so flexible, if we require its stationarity. However, this situation drastically changes if we allow for the coexistence of excitatory and inhibitory effects (i.e., marks can take both positive and negative values). Indeed, as will be shown in Sec.~\ref{sec:inhibitory_effects}, NLHawkes processes with two-sided marks are flexible enough to accommodate various nonlinearities without losing their stationarity. 		
		
	\subsection{Robust asymptotic results}
		The previous presentation of exact solutions for the ramp tension-intensity map  \eqref{eq:ramp_tension-intensity_map} for the special case of (a)~an exponential memory and (b)~an exponential jump-size distribution, allowed us highlighting the appearance of power law tails for the distribution of tensions near and at criticality. Here, we show that such a power law behaviour is asymptotically robust for general jump-size distributions, assuming that the memory is exponential. With the following notations
		\begin{align}
			h(t) = \frac{\eta}{\tau}e^{-t/\tau}, \>\>\> g(\nu) \simeq \nu - \nu_1 + o(\nu^0) \>\>\> &\mbox{for large }\nu, 
			\>\>\> \alpha_k := \int_0^\infty y^k \rho(y)dy < \infty \>\>\>\mbox{for any }k\geq 1, \>\>\> \alpha_1=1, \notag 
		\end{align}
		the steady-state intensity distribution $P_{\mrss}(\nu)$ is given by the following non-universal power law relation:
		\begin{equation}
		 P_{\mrss}(\nu) \propto \nu^{-1-a}, \>\>\> 
		 a := \frac{2\tau \nu_1}{\alpha_2} ~.
		 \label{yn3yhbwgq}
		\end{equation}
		Given the asymptotic linear relationship between $\nu$ and $\lambda$, the same power law behaviour holds for the PDF of $\lambda$. 
		
		We stress that $\nu_1$ can take any real value, either positive, negative, or zero.  If negative or zero, the derivation does not extend all the way to the limit $\eta=1$, and the power law \eqref{yn3yhbwgq} is truncated as in \eqref{eq:single_expon_linear_gamma_exact} by an exponential cut-off. This result implies a true power law tail for positive $\nu_1$ (i.e., normalisable without cutoff tail even at criticality) or intermediate asymptotic power law tail for non-positive $\nu_1$ (i.e., not normalisable without cutoff tail near criticality). Notably, this recovers Eq.~\eqref{eq:true_power law_one-sided_exponential} for the one-sided exponential mark distribution~\eqref{eq:exponential_mark_dist} for which $\alpha_2=2$.

		\subsubsection*{Derivation}
			Since we are interested only in the tail of the intensity PDF, let us focus on the asymptotic properties of the ME~\eqref{eq:master_expon_kernel} for large $\nu$. The ME~\eqref{eq:master_expon_kernel} has the asymptotic expression
			\begin{equation}
				\frac{1}{\tau}\frac{\partial }{\partial \nu}[\nu P_{\mrss}(\nu)] + 
				\int dy\rho(y)(\nu-\nu_1-\eta y/\tau)P_{\mrss}(\nu-\eta y/\tau)-(\nu-\nu_1)P_{\mrss}(\nu) \simeq 0 \>\>\>\> \mbox{for large }\nu
			\end{equation}
			in its steady state, obtained by replacing $g(\nu)$ by $\nu-\nu_1$ asymptotically. Applying the Laplace transform 
			\begin{equation}
				\tl{P}_{\mrss}(s) := \mcL_1\left[P_{\mrss}(\nu);s\right] = \int_{0}^\infty d\nu e^{-s\nu}P_{\mrss}(\nu),
			\end{equation}
			to this above equation yields
			\begin{equation}
				-\frac{s}{\tau}\frac{d }{ds}\tl{P}_{\mrss}(s) - \Phi(s) \left[\frac{d }{ds}\tl{P}_{\mrss}(s) 
				+ \nu_1\tl{P}_{\mrss}(s)\right] \simeq 0, \>\>\> \Phi(s) := \int_0^{\infty}dy(e^{-\eta y/\tau}-1)\rho(y).
			\end{equation}
			Its solution is given by
			\begin{equation}
				\log \tl{P}_{\mrss}(s) \simeq -\nu_1 s + \int_0^s \frac{\nu_1 s' ds'}{s'+ \tau\Phi(s')}
			\end{equation}
			with the normalisation condition $\tl{P}_{\mrss}(s=0)=1$. Considering the expansion
			\begin{equation}
				\Phi(s) = -\frac{\eta}{\tau}s + \frac{\eta^2\alpha_2}{2\tau^2}s^2+\dots, 
			\end{equation} 
			$\log\tl{P}_{\mrss}(s)$ has the following asymptotic form for small $s$ near criticality $1-\eta \ll 1$,
			\begin{align}
				\log \tl{P}_{\mrss}(s) \simeq \frac{2\tau \nu_1}{\alpha_2}\log s \>\>\> \mbox{ for small }s,
			\end{align}
			implying, by inverse Laplace transform, a power law asymptotics for the steady intensity PDF:
			\begin{equation}
				P_{\mrss}(\nu) \propto  \nu^{-1-2\tau \nu_1/\alpha_2}.
			\end{equation}
			For non-positive $\nu_1$, this PDF is not normalisable and thus requires a cutoff tail, such as the exponential given by Eq.~\eqref{eq:single_expon_linear_gamma_exact}.

\section{Solution 2: exponential memory kernel with two-sided symmetric mark distribution for linear to fast-accelerating intensity maps}\label{sec:inhibitory_effects}
	In this section, we study both exact and asymptotic solutions of the ME \eqref{eq:master_expon_kernel} for the PDF of the total tension $\nu$ valid for an exponential memory kernel~\eqref{eq:exponential_memory} and in the presence of inhibitory effects (i.e., marks $\hat{y}$ can be both positive and negative). The inhibitory effects imply that events can sometimes suppress or decrease the amplitude of bursts, which can lead to essentially different phenomena from those in the previous section. 

	\subsection{Exact solutions to two-sided symmetric exponential mark distribution}
	\begin{figure}
		\centering
		\includegraphics[width=75mm]{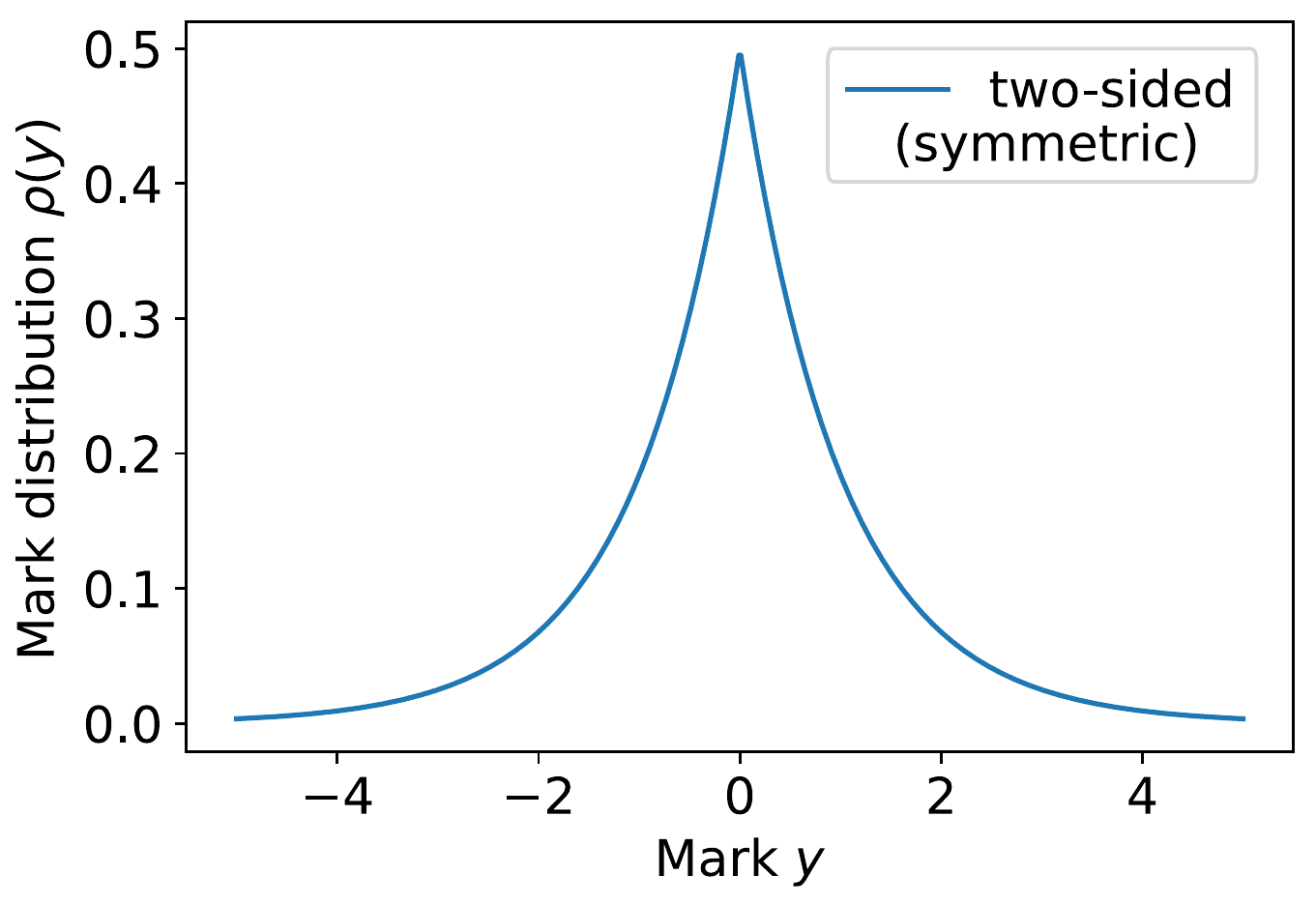}
		\caption{Schematic of the two-sided symmetric exponential mark distribution~\eqref{tht42gq} with $y^*=1$.}
		\label{fig:twosided_markdist}
	\end{figure}
		Let us focus on the case with the two-sided symmetric exponential mark distribution (see Fig.~\ref{fig:twosided_markdist}):
		\begin{equation}
			\rho(y) = \frac{1}{2y^*} e^{-|y|/y^*},
			\label{tht42gq}
		\end{equation}
		which corresponds to the existence of symmetric positive ($y>0$) and negative ($y<0$) feedback effects with zero mean. We again assume $y^*=1$, without loss of generality. 
		This negative feedback effect is called the inhibitory effect in Ref.~\cite{BouchaudBook} and is known to be difficult to deal with in analytical approaches. We present the exact solution of equation \eqref{eq:master_expon_kernel} with \eqref{tht42gq} for some specific 
		forms of $\lambda = g(\nu)$. 
		
		Let us recall the identity
		\begin{equation}
			\left(1-\frac{1}{c}\frac{\partial }{\partial \nu}\right)\int_{-\infty}^0 dy e^{-|y|}g(\nu-\eta y/\tau)P_t(\nu-\eta y/\tau) = g(\nu)P_t(\nu),
			\label{eq:identity_one_sided_expon2}
		\end{equation}
		where $c= \tau/\eta$ has been defined in \eqref{wy5hnthbw8F} (see Appendix~\ref{sec:app:integral_identity} for the derivation). This identity together with the other identity~\eqref{eq:identity_one_sided_expon} implies a third identity useful to solve the ME~\eqref{eq:master_expon_kernel}: 
		\begin{equation}
			\int_{-\infty}^{+\infty} dy\frac{e^{-|y|}}{2}g(\nu-\eta y/\tau)P_t(\nu-\eta y/\tau) = \frac{1}{2}\left[\frac{1}{1
			+\partial_{\nu}/c}+\frac{1}{1-\partial_{\nu}/c}\right]g(\nu)P_t(\nu) = \frac{1}{1-\partial_\nu^2/c^2}g(\nu)P_t(\nu).
		\end{equation}
		We thus obtain a simple representation of the ME~\eqref{eq:master_expon_kernel}: 
		\begin{equation}
			\frac{\partial P_t(\nu)}{\partial t} = \frac{1}{\tau}\partial_\nu[\nu P_t(\nu)] + \frac{\partial_\nu^2/c^2}{1-\partial_\nu^2/c^2}g(\nu)P_t(\nu). 
			\label{thrtgbrb}
		\end{equation}
		This ME can be written in the more familiar form
		\begin{equation}\label{eq:master_expon_kernelwb}
			\frac{\partial P_t(\nu)}{\partial t} = - {\partial J_t(\nu) \over \partial \nu}~,
		\end{equation}
		where the probability current is defined by
		\begin{equation}\label{eq:master_expon_kernel_current}
			J_t(\nu) :=   - \left[ \frac{\nu}{\tau}  +  \frac{\partial_\nu/c^2}{1-\partial_\nu^2/c^2}g(\nu)  \right] P_t(\nu). 
		\end{equation}
		This formulation makes more transparent the meaning of the boundary condition ${\rm lim}_{\nu \to \infty} J_t(\nu) = 0$ ensuring that the mean probability ``velocity'' $J_t(\nu)/P_t(\nu)$ is zero at infinity. We reject the possibility of periodic boundary conditions which are non-physical.
		
		Then, the steady-state solution satisfies the following second-order differential equation
		\begin{equation}\label{eq:master_exact_two-sided_expon}
			\frac{d^2}{d\nu^2}\left[\nu P_{\mrss}(\nu)\right] - \tau\frac{d}{d\nu}\left[g(\nu) P_{\mrss}(\nu)\right] -c^2 \nu P_{\mrss}(\nu) = 0~,
		\end{equation}
		This is obtained by putting $\frac{\partial P_t(\nu)}{\partial t} =0$ in \eqref{eq:master_expon_kernelwb}, and using ${\rm lim}_{\nu \to \infty} J_t(\nu) = 0$ that leads to $J_t(\nu) = 0, \forall \nu$, from which equation \eqref{eq:master_exact_two-sided_expon} derives.

		\subsubsection*{Example 1: ramp tension-intensity map}
			For the ramp tension-intensity map
			\begin{equation}
				\lambda = g(\nu) = \max\{\nu_0,| \nu | \} 	~,			
				\label{eq:ramp_tension-intensity_mapqt}
			\end{equation}
			which corresponds to setting $\nu_1=0$ in Eq.~\eqref{eq:ramp_tension-intensity_map} and adding the absolute value, the solution of \eqref{eq:master_exact_two-sided_expon} is a truncated-L\'evy-type intensity asymptotic tail: 
			\begin{equation}
				 P_{\mrss}(\lambda) \propto \lambda^{-1}e^{-\frac{\lambda}{\lambda_{\rm cut}}}, \>\>\>\lambda_{\rm cut} := \frac{2c^2}{\tau + \sqrt{4c^2+\tau^{2}}} ~,~~
				 {\rm for ~large}~\lambda~,
				\label{eq:ramp_tension-intensity_map_twosided}
			\end{equation}
			where $c= \tau/\eta$ has been defined in \eqref{wy5hnthbw8F}. For $\nu_1 \neq 0$ in Eq.~\eqref{eq:ramp_tension-intensity_map}, the exact form of the intensity distribution is also available. 
			
			Remarkably, this model has no critical point: the process is always stationary for all $\eta <\infty$, due to the stabilisation effect of the inhibitory component of the process, and thus $\lambda_{\rm cut}$ is always finite. This is in contrast to the ramp Hawkes process with one-sided exponential jumps (without inhibitory effect), and thus highlights the fact that the inhibitory effects can be crucial in understanding even the qualitative behaviour of the NLHawkes processes in general. 
			
			\paragraph*{Derivation of the exact solution.}
			The ME~\eqref{eq:master_exact_two-sided_expon} reduces to the following set of modified Bessel differential and constant-coefficient second-order differential equations: 
			\begin{align}
					\nu^2 \frac{d^2\phi(\nu)}{d\nu^2} + \nu \frac{d\phi(\nu)}{d\nu} - (c^2 \nu^2 + \gamma^2)\phi(\nu) &= 0, \>\>\> 
					\gamma := \frac{1-\tau\nu_0}{2}, \>\>\> 
					\phi(\nu):=\nu^{\gamma}P_{\mrss}(\nu) & (0\leq \nu \leq \nu_0), \\
					\frac{d^2}{d\nu^2}\psi(\nu) - \tau\frac{d}{d\nu}\psi(\nu) -c^2 \psi(\nu) &= 0, \>\>\> 
					\psi(\nu):= \nu P_{\mrss}(\nu) & (\nu_0\leq \nu ).
			\end{align}
			The exact solution is then given by
			\begin{equation}
				P_{\mrss}(\nu) = 	\begin{cases}
										\displaystyle \nu^{-\gamma}\left(C^{[1]}I_{\gamma}(|\nu|/c)+C^{[2]}K_{\gamma}(|\nu|/c)\right) & (|\nu|\leq \nu_0) \\
										\displaystyle C^{[3]}|\nu|^{-1}e^{-|\nu|/\nu_{\rm cut}} & (|\nu|>\nu_0)
									\end{cases}, \>\>\> 
				\nu_{\rm cut} := \frac{2c^2}{\tau + \sqrt{4c^2+\tau^{2}}} ~, \>\>\>    c := \frac{\tau}{\eta}
			\end{equation}
			with integral constants $C^{[1]},C^{[2]},C^{[3]}$ and modified Bessel functions of the first and second kinds (denoted by $I_{\gamma}(x)$ and $K_{\gamma}(x)$, respectively; see Appendix~\ref{app:specialFunctions:modiedBessel}). The integral constants are determined by the normalisation and continuity conditions: $\int_{-\infty}^{\infty}d\nu P_{\mrss}(\nu)=1$, $\lim_{\nu\uparrow \nu_0}P_{\mrss}(\nu) = \lim_{\nu\downarrow \nu_0}P_{\mrss}(\nu)$. We thus obtain that the intensity distribution is given by the sum of a $\delta$ function centred on $\nu_0$ and the truncated L\'evy distribution, 
			\begin{equation}
				P_{\mrss}(\lambda)  = \left(1-C^{[3]}\Gamma(0, \nu_0/\nu_{\rm cut})\right)\delta(\nu-\nu_0)	+ C^{[3]}\lambda^{-1}e^{-\lambda/\nu^*}\Theta(\nu-\nu_0). \label{eq:expon_tail_two_sided_ramp}
			\end{equation}
			where the incomplete gamma function is $\Gamma(a,x):=\int_{x}^\infty dt ~t^{a-1}e^{-t}$.

		\subsubsection*{Example 2: quadratic tension-intensity map}
			For the quadratic tension-intensity map corresponding to the ZHawkes process, see Sec.~\ref{subsec:QHawkes_review}, 
			\begin{equation}\label{eq:quadratic_tensiton_intensity_maprg1}
				\lambda = g(\nu) = k \nu^2 + \lambda_0 ~,
			\end{equation}
			the solution of \eqref{eq:master_exact_two-sided_expon} is a power law steady-state distribution\footnote{If $\lambda_0$ is zero, the steady-state distribution is singular at $\nu=0$ as $P_{\mrss}(\lambda)\propto \nu^{-1}$ and thus is not normalisable.}
			\begin{equation}\label{eq:quadratic_tensiton_intensity_map}
				P_{\mrss}(\lambda) \propto |\lambda|^{-1-a}~,~~~~ \>\>\> a:= \frac{1}{2}+\frac{c^2}{2k\tau}~,  ~~~  c := \frac{\tau}{\eta}
			\end{equation}
			with $\lambda_0>0$ and power law exponent $a>1/2$. The exact form of the intensity distribution is also available. We note that this non-universal power law scaling is consistent with Eq.~\eqref{eq:power-law_QHawkes_Bouchaud}, which was reported for the diffusive limit of the ZHawkes process in Ref.~\cite{QHawkesBouchaud}.

			\paragraph*{Derivation of the exact solution.}
			By the variable transformation $x=k\tau\nu^2/2$, the ME~\eqref{eq:master_exact_two-sided_expon} for $\nu>0$ reduces to
			\begin{equation}
				x\frac{d^2\phi(x)}{dx^2} + \left(\frac{3}{2}-\frac{\tau\lambda_0}{2}-x\right)\frac{d\phi(x)}{dx} - \left(1+\frac{c^2}{2k\tau}\right)\phi(x) = 0, \>\>\> \phi(x):=P(\nu(x)), \>\>\> \nu(x):= \sqrt{\frac{2x}{k\tau}}.
			\end{equation}
			This is the confluent hypergeometric differential equation and thus its exact solution is given by
			\begin{equation}
				\phi(x) = 	C^{[1]} {}_1F_1\left(1+\frac{c^2}{2k\tau}, \frac{3}{2}-\frac{\tau\lambda_0}{2}; x\right) + 
							C^{[2]} {}_1U_1\left(1+\frac{c^2}{2k\tau}, \frac{3}{2}-\frac{\tau\lambda_0}{2}; x\right) 
			\end{equation}
			with integral constants $C^{[1]}, C^{[2]}$ and the confluent hypergeometric functions of the first ${}_1F_1$ and second kind  ${}_1U_1$ (see Appendix~\ref{app:specialFunctions:confluentHyperGeometric}). The integral constants are determined by the normalisation condition $\int_{-\infty}^{\infty}d\nu P_{\mrss}(\nu) = 1$. Interestingly, this solution has the following asymptotic form for large $x$ 
			\begin{equation}
				\phi(x) \propto x^{-1-c^2/(2k\tau)}.
			\end{equation}
			The steady distribution of the intensity $\lambda$ is then given by
			\begin{equation}\label{eq:quadratic_intensity_powerlaw_twosidedPoisson}
				P_{\mrss}(\lambda) = \left|\frac{d\nu}{d\lambda}\right|P_{\mrss}(\nu)\propto \lambda^{-1-a}, \>\>\> a:= \frac{1}{2}+\frac{c^2}{2k\tau}~, 
				~~~c := \frac{\tau}{\eta}
			\end{equation}
			for the tail $\lambda \to \infty$. This is a power law asymptotic distribution with a non-universal exponent $a$ without truncation.

		\subsubsection*{Example 3: exponential intensity map}
			For the exponential tension-intensity map 
			\begin{equation}
				\lambda = g(\nu) =  \lambda_0 |\nu|e^{\beta \nu} ~,
			\end{equation}
			the solution of \eqref{eq:master_exact_two-sided_expon} is Zipf's law for the intensity distribution:
			\begin{equation}
				P_{\mrss}(\lambda) \propto \lambda^{-2}~,
				\label{eq:exact_Zipf_expIntensity}
			\end{equation}
			up to a logarithmic factor $\log \lambda$, for large $\lambda$ with positive constants $\lambda_0$ and $\beta$. This intensity map is inspired by the MSA model \cite{MSA_PRL,MSA_Geophys}, where the dominant contribution comes from the exponential factor originating from the Arrhenius law (see Sec.~\ref{sec:MSA_diffusive} for more detail).  
			
			\paragraph*{Derivation of the exact solution.}
			The exact steady-state solution of \eqref{eq:master_exact_two-sided_expon} is given by
			\begin{equation}
				P_{\mrss}(\nu) = 	\begin{cases}
										\displaystyle
										\frac{C^{[1]}}{\nu}e^{c \nu} {}_1U_1\left(1+\frac{c}{\beta},1+\frac{2c}{\beta}; \frac{\lambda_0 \tau}{\beta}e^{\beta \nu}\right) + \frac{C^{[2]}}{\nu}e^{c\nu}L_{-1-c/\beta}^{2c/\beta}\left(\frac{\lambda_0 \tau}{\beta}e^{\beta \nu}\right) & (\nu \geq 0) \\
										\displaystyle
										\frac{C^{[3]}}{\nu}\exp\left[c \nu-\frac{\lambda_0 \tau}{\beta}e^{\beta \nu}\right] {}_1U_1\left(\frac{c}{\beta},1+\frac{2c}{\beta}; \frac{\lambda_0 \tau}{\beta}e^{\beta \nu}\right) 
										+ \frac{C^{[4]}}{\nu}\exp\left[c \nu-\frac{\lambda_0 \tau}{\beta}e^{\beta \nu}\right]L_{-c/\beta}^{2c/\beta}\left(\frac{\lambda_0 \tau}{\beta}e^{\beta \nu}\right) & (\nu<0)
									\end{cases}
			\end{equation}
			with integral coefficients $C^{[1]}$, $C^{[2]}$, $C^{[3]}$, and $C^{[4]}$, and the generalised Laguerre function $L_a^b(x)$ (see Appendix~\ref{app:specialFunctions:gLaguerre}). Considering the asymptotic formulas~\eqref{eq:special_CH_asymp} and~\eqref{eq:special_LG_asymp}, $C^{[2]}$ must be zero since $P_{\mrss}(\nu)\to 0$ for $\nu \to \infty$. This means that the asymptotic tail is given by
			\begin{equation}
				P_{\mrss}(\nu) \propto \nu^{-1}e^{-\beta \nu},
			\end{equation}
			which leads to the Zipf law \eqref{eq:exact_Zipf_expIntensity} for the steady intensity distribution, by using the Jacobian relation $d\nu P_{\mrss}(\nu) = d\lambda P_{\mrss}(\lambda) \Longleftrightarrow P_{\mrss}(\lambda) =|d\nu/d\lambda|P_{\mrss}(\nu)$. As shown in Secs.~\ref{sec:MSA_diffusive} and~\ref{sec:ZipfForMSA_exponMemory}, this asymptotic Zipf law~\eqref{eq:exact_Zipf_expIntensity} is robust for exponential-type tension-intensity maps under general symmetric mark distribution (or more generally when the mark average is zero), on the condition that the memory kernel is exponential and the mark distribution has its moments at all orders being finite.

	\subsection{Exact solutions in the diffusive limit}
		Let us now consider the diffusive limit formulated in Sec.~\ref{sec:system_size_expansion}, and assume an exponential memory kernel~\eqref{eq:exponential_memory}:
		\begin{equation}\label{eq:sol_diffisivetrb}
			h(t) = \frac{\eta}{\tau} e^{-t/\tau}, \>\>\>
			\rho_{\ve}(y)=\frac{1}{\ve}\tl{\rho}\left(\frac{y}{\ve}\right), \>\>\> g(\nu) = \frac{1}{\ve^2}\tl{g}(\nu)~.
		\end{equation}
		For this case, by solving the FPE~\eqref{eq:master_eq_diffusive} in the steady state, we obtain the explicit solution
		\begin{equation}\label{eq:sol_diffisive}
			P_{\mrss}(\nu) \propto_{ \ve \to 0} \frac{1}{\tl{g}(\nu)}\exp\left[-\frac{1}{\tau D}\int \frac{\nu d\nu}{\tl{g}(\nu)}\right], \>\>\>D := \frac{\tl{\alpha}_{2}\eta^{2}}{2\tau^{2}}
		\end{equation}
		for any $\tl{g}(\nu)$, assuming that $\tl{\rho}(y)$ and $\tl{g}(\nu)$ are independent of $\ve$ and that all the integrals appropriately converge.

		\subsubsection*{Example 1: ramp tension-intensity map}
			Let us first consider the example of the ramp tension-intensity map
			\begin{equation}
				g(\nu) = \max\{\nu_0, |\nu - \nu_1|\}
			\end{equation}
			with positive real number $\nu_0>0$ and arbitrary real number\footnote{Here $\nu_1$ can be either positive, zero, or negative.} $\nu_1$. For $\nu>\nu_0+\nu_1$, the solution of the FPE~\eqref{eq:master_eq_diffusive} is the truncated L\'evy distribution for the tension (and thus for the intensity)
			\begin{equation}
				P_{\mrss}(\nu) = \frac{1}{Z}\frac{e^{-\nu/\nu_{\mathrm{cut}}}}{(\nu-\nu_1)^{1+a}}, \>\>\>\nu_{\mathrm{cut}}:= D\tau, \>\>\> a := \frac{\nu_1}{D\tau}, 
			\end{equation}
			with an exponential tail tapering the intermediate power law tail. We note that this model has no critical point due to the inhibitory effects leading to the characteristic intensity for the exponential cutoff $\nu_{\mathrm{cut}}$ to be always finite.
	
		\subsubsection*{Example 2: quadratic tension-intensity map}
			We next study the exact solution of the quadratic intensity \eqref{eq:quadratic_tensiton_intensity_maprg1}, corresponding to the ZHawkes process, see Sec.~\ref{subsec:QHawkes_review}, in the presence of inhibitory effects. Using formula~\eqref{eq:sol_diffisive}, the exact solution of the FPE~\eqref{eq:master_eq_diffusive} in the steady-state regime is given by
			\begin{equation}
				P_{\mrss}(\nu) = \frac{1}{Z}  \left(k\nu^2+\lambda_0\right)^{-1-1/(2kD\tau)}~,
			\end{equation}
			which is equivalent to
			\begin{equation}\label{eq:quadratic_intensity_powerlaw_Gaussian}
				P_{\mrss}(\lambda) = \frac{1}{Z'}\frac{\lambda^{-1-1/(2kD\tau)}}{\sqrt{\lambda - \lambda_0}} \propto \lambda^{-1-a} \>\>\> (\lambda\to \infty), \>\>\> a := \frac{1}{2}+\frac{1}{2kD\tau}
			\end{equation}
			with $Z':= 2\sqrt{k}Z$. This is a power law distribution without truncation and with a non-universal exponent $a$. We note that this non-universal power law scaling is essentially identical to Eq.~\eqref{eq:power-law_QHawkes_Bouchaud} for the diffusive limit of the ZHawkes process reported in Ref.~\cite{QHawkesBouchaud}.

		\subsubsection*{Example 3: polynomial tension-intensity map}
			Let us consider the case of the polynomial intensity given by
			\begin{equation}\label{eq:poly_intensity_diffusive}
				\tl{g}(\nu) = k|\nu|^n + \nu_0 \>\>\> (n>2)
			\end{equation}
			with positive constant $\nu_0>0$. Using formula~\eqref{eq:sol_diffisive}, we obtain the exact steady-state distribution, solution of the FPE~\eqref{eq:master_eq_diffusive},
			\begin{equation}
				P_{\mrss}(\nu) \propto \frac{1}{k|\nu|^n + \nu_0}\exp\left[-\frac{\nu^2}{2\tau D\nu_0} \>{}_2F_1\left(1,\frac{2}{n}, 1+\frac{2}{n}; -\frac{k|\nu|^n}{\nu_0}\right)\right]
				\label{eq:exact_diff_poly_inten}
			\end{equation}
			with the hypergeometric function ${}_2F_1$ (see Appendix~\ref{app:specialFunctions:hyperGeometric}). By considering the following asymptotic expansion 
			\begin{equation}\label{eq:hyperGeometricAsymptotics_poly}
				\nu^2 {}_2F_1\left(1,\frac{2}{n},1+\frac{2}{n}; -\frac{k\nu^n}{\nu_0}\right) 
				=
				\begin{cases}
					\displaystyle
					\underbrace{\frac{\frac{2\pi}{n}}{\sin \frac{2\pi}{n}}
					\left(\frac{\nu_0}{k}\right)^{\frac{2}{n}}}_{\mbox{const.}} + o(\nu^0)
					& (n>2) \\ 
					\displaystyle
					\frac{\nu_0}{k}\log \left(\frac{k\nu^2}{\nu_0}\right) + o(\nu^0)
					& (n=2) 
				\end{cases}
			\end{equation}
			for large $\nu$'s, we obtain the asymptotic form of the steady PDF for $|\nu|\to \infty \Longrightarrow \lambda \to \infty$ as 
			\begin{equation}\label{eq:poly_diffusive_sol_powerlaw}
				P_{\mrss}(\nu) \propto \frac{1}{k|\nu|^n + \nu_0} \>\>\> \Longrightarrow \>\>\> P_{\mrss}(\lambda) \propto \lambda^{-1-a}, \>\>\> a:= 1-\frac{1}{n}.
			\end{equation}
			Note that the limit $n \to +\infty$ recovers Zipf's law. The manner with which the exact solution~\eqref{eq:exact_diff_poly_inten} 
			recovers Eq.~\eqref{eq:quadratic_intensity_powerlaw_Gaussian} for $n\to 2$ is now elaborated.

		\paragraph*{Crossover between $n=2$ and $n>2$.}
			Remarkably, the solution~\eqref{eq:quadratic_intensity_powerlaw_Gaussian} for the QHawkes (i.e., $n=2$) and the one~\eqref{eq:exact_diff_poly_inten} for the polynomial Hawkes with $n>2$ are slightly different. This qualitative difference can be seen from the analytical singularity of the hypergeometric function ${}_2F_1$ at $n=2$ and suggests a crossover between two power law regimes. Here we explicitly estimate the crossover point.
			
			Let us introduce a small positive parameter $\eps$ as 
			\begin{equation}
				n:= \frac{2}{1-\eps} > 2
			\end{equation}
			and consider the limit $\eps \downarrow 0$. We focus on the discontinuous switching in Eq.~\eqref{eq:hyperGeometricAsymptotics_poly} between $n>2$ and $n=2$. To estimate the crossover point, it is necessary to evaluate their higher-order asymptotic behaviour for large $\nu$ with nonzero $\eps$ as given by Eq.~\eqref{eq:app:hyperGeometricAsymptotics_poly:c}. As summarised in Appendix~\ref{app:specialFunctions:hyperGeometric:crossover}, the threshold intensity is estimated to be 
			\begin{equation}
				\label{eq:threshold_crossover_nonzero_eps}
				\lambda^* := \nu_0 e^{2/\eps},
			\end{equation}
			which characterises the crossover between the two regimes. We thus obtain the explicit crossover formula as  
			\begin{equation}
				P_{\mrss}(\lambda) \propto 
				\begin{cases}
					\lambda^{-1-a_1} & (\lambda \ll \lambda^*, a_1:=\frac{1}{2}+\frac{1}{2kD\tau}) \\
					\lambda^{-1-a_2} & (\lambda \gg \lambda^*, a_2:=\frac{1}{2})
				\end{cases}.
			\end{equation}

			The existence of this crossover point can be intuitively understood as follows: let us go back to the SDE representation~\eqref{eq:SDE_diffusive-expon}. Remarkably, the cases $n=2$ and $n>2$ are critically different in the sense that the relaxation term $-\nu/\tau$ is the same order as the fluctuation $\sqrt{2D\tl{g}(\nu)} \xi^{\mrG}$ for $n=2$, whereas it is negligible for $n>2$
			\begin{equation}
				\left|-\frac{\nu}{\tau}\right| \ll \sqrt{2D\tl{g}(\nu)}\xi^{\mrG}
			\end{equation}
			for a sufficiently large $\nu \gg \nu^*$. Such a crossover point can be roughly estimated by the relationship $\nu^*/\tau = \sqrt{2D\tl{g}(\nu^*)}$, suggesting $\nu^* = C^{1/\eps}$ with some constant $C$. We therefore obtain $\log{\lambda^*}\propto \eps^{-1}$ consistently with Eq.~\eqref{eq:threshold_crossover_nonzero_eps}. 
						
		\subsubsection*{Example 4: Multifractal stress activation model}\label{sec:MSA_diffusive}
			An interesting example is the MSA model for earthquake triggering proposed in Refs.~\cite{MSA_PRL,MSA_Geophys} and summarised for our purpose in section \ref{rtwjhyrbgq}, which corresponds to
			\begin{equation}  
				\tl{g}(\nu) = \lambda_0\exp(\beta \nu)~,
				\label{eq:MSA_intensity_expon}
			\end{equation}
			with base intensity $\lambda_0>0$ and effective inverse temperature $\beta>0$. From the steady-state solution~\eqref{eq:sol_diffisive} of the FPE~\eqref{eq:master_eq_diffusive}, we obtain the steady solution $P_{\mrss}(\nu)$ for the tension $\nu$ and $P_{\mrss}(\lambda)$ of the intensity $\lambda = \lambda_0e^{\beta \nu}$:
			\begin{equation}\label{MSA_nu_diffusive}
				P_{\mrss}(\nu) = \frac{1}{Z}\frac{e^{-\beta \nu}}{\lambda_0}\exp\left[\frac{e^{-\beta \nu}(1+\beta\nu)}{\lambda_0 \beta^2 \tau D}\right]
				\>\>\> \Longleftrightarrow \>\>\>
				P_{\mrss}(\lambda) = \frac{1}{\beta Z}\lambda^{-2}\exp\left[\frac{\lambda^{-1}\{1+\log(\lambda/\lambda_0)\}}{\beta^2 \tau D}\right]
			\end{equation}
			The derivation of $P_{\mrss}(\lambda)$  from  $P_{\mrss}(\nu)$ uses the Jacobian relation $d\nu P_{\mrss}(\nu) = d\lambda P_{\mrss}(\lambda)  \Longleftrightarrow  P_{\mrss}(\nu) = \beta\lambda P_{\mrss}(\lambda)$. This steady intensity distribution exhibits Zipf's law similarly to the aforementioned result~\eqref{eq:exact_Zipf_expIntensity}:
			\begin{equation}
				P_{\mrss}(\lambda) \propto \lambda^{-2} ~~~~~\mbox{   (for large } \lambda). 
			\end{equation}
		
		\subsubsection*{Example 5: fast-accelerating intensity}
			Let us focus on a large class of intensity map $g(\nu)$ satisfying 
			\begin{equation}
				\tl{g}(\nu) \gg \nu^2 \>\>\>~~~~ \mbox{for large } \nu,
				\label{eq:RII_intensity}
			\end{equation}
			which we refer to as a fast-accelerating intensity (FAI) map. For example, the polynomial intensity~\eqref{eq:poly_intensity_diffusive} beyond second order and the MSA intensity~\eqref{eq:MSA_intensity_expon} belong to this class. FAI maps are special in the sense that the asymptotic PDF of $\nu$, which is solution of the FPE~\eqref{eq:master_eq_diffusive},
			 is given by
			\begin{equation}
				P_{\mrss}(\nu) \propto \frac{1}{\tl{g}(\nu)}\exp\left [-\frac{1}{\tau D}\int \frac{\nu d\nu}{\tl{g}(\nu)} \right] = \exp\left [-\log{\tl{g}(\nu)} - o(\nu^{-c}) \right]
			\end{equation}
			with some positive constant $c>0$. This expression is derived from Eq.~\eqref{eq:sol_diffisive},  considering that $\nu/\tl{g}(\nu) = o(\nu^{-1})$. We thus obtain a general asymptotic form 
			\begin{equation}
				\label{eq:RII_PDF_diffusive}
				P_{\mrss}(\lambda) \propto \lambda^{-1} \left|\frac{d\tl{g}(\nu)}{d\nu}\right|_{\nu=\tl{g}^{-1}(\lambda)}^{-1}. 
			\end{equation}

	\subsection{Robust asymptotic solutions}  
		\subsubsection{Robust exponential tail for the ramp intensity}\label{sec:expMem_rampInt_robustAsymp}
			Here we show that the exponential tail \eqref{eq:ramp_tension-intensity_map_twosided} and \eqref{eq:expon_tail_two_sided_ramp} for the ramp intensity~\eqref{eq:ramp_tension-intensity_mapqt} of the steady-state solution~\eqref{eq:sol_diffisive} of the FPE~\eqref{eq:master_eq_diffusive} remains valid for general symmetric mark distributions, assuming appropriate convergence of the moment-generating function and with an exponential memory function:
			\begin{equation}\label{eq:tLevyAsymptotics_ramp_symmetric}
				h(t)=\frac{\eta}{\tau}e^{-t/\tau},\>\>\> \rho(y) = \rho(-y), \>\>\> g(\nu) \simeq \nu + \nu_0 \>\>\>\mbox{for large $\nu$} \>\>\> \Longrightarrow \>\>\> 
				P_{\mrss}(\lambda) \propto e^{-\lambda/\lambda_{\rm cut}} \>\>\> \mbox{ for large }\lambda,
			\end{equation}
			up to a sub-leading contribution in the form of a truncated power law. The parameter $\lambda_{\rm cut}$ is given by the self-consistent relation
			\begin{equation}\label{eq:tLevyAsymptoticsSelConsisntent}
				\frac{1}{\tau\lambda_{\rm cut}} = \Phi\left(\frac{\eta}{\tau\lambda_{\rm cut}}\right), \>\>\> 
				\Phi(x) := \int_{-\infty}^\infty dy\rho(y)(e^{xy} -1) = \int_{-\infty}^\infty dy\rho(y)(\cosh{yx} -1),
			\end{equation}
			where $\Phi(x)$ is the moment-generating function. The equation for $\lambda_{\rm cut}$ has a single positive solution (see Appendix~\ref{sec:app:phi_symmetric}).

			\paragraph*{Derivation.}
				The solution \eqref{eq:tLevyAsymptotics_ramp_symmetric} can be derived by direct substitution into the ME~\eqref{eq:master_expon_kernel} as follows. Let us make an ansatz that the solution is given by
				\begin{equation}
					P_{\mrss}(\nu) \propto e^{-\nu/\lambda_{\rm cut}} \>\>\> \mbox{for large $\nu$}.
				\end{equation} 
				By considering the relations for large $\nu$
				\begin{align}
					&\frac{1}{\tau}\frac{d}{d \nu}\left[\nu e^{-\frac{\nu}{\lambda_{\rm cut}}}\right] 
					+ \int_{-\infty}^\infty dy\rho(y)(\nu+\nu_0-\eta y/\tau)e^{-\frac{\nu}{\lambda_{\rm cut}} + \frac{\eta y}{\tau \lambda_{\rm cut}}} 
					- (\nu+\nu_0) e^{-\frac{\nu}{\lambda_{\rm cut}}} \notag \\
					= &\nu e^{-\frac{\nu}{\lambda_{\rm cut}}}\left[
							-\frac{1}{\tau \lambda_{\rm cut}}
							+ \int_{-\infty}^\infty dy\rho(y) e^{\frac{\eta y}{\tau \lambda_{\rm cut}}} - 1
					 \right]
					+ o\left(\nu e^{-\frac{\nu}{\lambda_{\rm cut}}}\right) \notag \\
					= &\nu e^{-\frac{\nu}{\lambda_{\rm cut}}} \left[-\frac{1}{\tau \lambda_{\rm cut}} + \Phi\left(\frac{\eta}{\tau \lambda_{\rm cut}}\right)\right]
					+ o\left(\nu e^{-\frac{\nu}{\lambda_{\rm cut}}}\right),
				\end{align}
				the ME~\eqref{eq:master_expon_kernel} in the steady state reads 
				\begin{align}
					\nu e^{-\frac{\nu}{\lambda_{\rm cut}}} \left[-\frac{1}{\tau \lambda_{\rm cut}} + \Phi\left(\frac{\eta}{\tau \lambda_{\rm cut}}\right)\right]
					+ o\left(\nu e^{-\frac{\nu}{\lambda_{\rm cut}}}\right) = 0
				\end{align}
				for large $\nu$. This relation is equivalent to the self-consistent relation~\eqref{eq:tLevyAsymptoticsSelConsisntent}.

		\subsubsection{Robust power law tail for quadratic intensity}
			We show that the power law tail for the quadratic intensity, such as Eq.~\eqref{eq:quadratic_intensity_powerlaw_twosidedPoisson} and Eq.~\eqref{eq:quadratic_intensity_powerlaw_Gaussian}, of the steady-state solution~\eqref{eq:sol_diffisive} of the FPE~\eqref{eq:master_eq_diffusive} is generally valid for general symmetric mark size distributions:
			\begin{equation}
				h(t)=\frac{\eta}{\tau}e^{-t/\tau},\>\>\> 
				\rho(y) = \rho(-y), \>\>\> 
				g(\nu) \simeq k\nu^2 + \lambda_0 \>\>\>\mbox{for large $\nu$} \>\>\>
				\Longrightarrow \>\>\> 
				P_{\mrss}(\lambda) \propto \lambda^{-1-a}
			\end{equation} 
			for large $\lambda$ with $a>1/2$, assuming appropriate convergence of the KM coefficients.
			
			Note that the authors of Ref.~\cite{QHawkesBouchaud} conjectured that the PDF of the intensity of various ZHawkes processes should be a power law with a non-universal exponent. Our results confirm this conjecture, as least for an exponential memory kernel, in the sense that the power law asymptotics with a non-universal exponent $a>1/2$ is a robust property of ZHawkes processes, independently of the shape of the mark distribution, as long as it is symmetric with finite moments.
			
			\paragraph*{Derivation.}
				Let us go back to the ME in the steady state,
				\begin{equation}
					\frac{1}{\tau}\frac{\partial }{\partial \nu}[\nu P_{\mrss}(\nu)] + \int_{-\infty}^\infty dy\rho(y)\left(k (\nu-\eta y/\tau)^2+\lambda_0\right)P_{\mrss}(\nu-\eta y/\tau) - (k \nu^2+\lambda_0)P_{\mrss}(\nu) = 0.
				\end{equation}
				We make the anzatz that the asymptotic solution is given by
				\begin{equation}
					P_{\mrss}(\nu) \simeq C \nu^{-\kappa} + o(\nu^{-\kappa}) \>\>\>\>\> (\mbox{for large }\nu)
				\end{equation}
				with a positive $\kappa$ and a certain constant $C$. This implies that
				\begin{align}
					&C\left[    \frac{1-\kappa}{\tau}\nu^{-\kappa} +  \int_{-\infty}^\infty dy \rho(y) \left\{k\left(\nu-\frac{\eta y}{\tau}\right)^2+\lambda_0\right\}\left(\nu-\frac{\eta y}{\tau}\right)^{-\kappa} - (k\nu^2 + \lambda_0)\nu^{-\kappa}+ o(\nu^{-\kappa})  \right] \notag \\
					= &C\left[    \frac{1-\kappa}{\tau}\nu^{-\kappa} +  
					\int_{-\infty}^\infty dy \rho(y) \left\{
						k\nu^{-\kappa+2}\left(1-\frac{\eta y}{\tau\nu}\right)^{-\kappa+2}
						+\lambda_0\nu^{-\kappa}\left(1-\frac{\eta y}{\tau\nu}\right)^{-\kappa} \right\}
						- (k\nu^2 + \lambda_0)\nu^{-\kappa}+ o(\nu^{-\kappa})  \right] \notag \\
					= &C\left[    \frac{1-\kappa}{\tau}\nu^{-\kappa} +  \int_{-\infty}^\infty dy \rho(y) \left\{ 
						k\nu^{-\kappa+2}\left(1+\frac{(\kappa-2)\eta y}{\tau \nu} + \frac{\eta^2y^2(\kappa-1)(\kappa-2)}{2\tau^2 \nu^2}\right) + \lambda_0\nu^{-\kappa}
					\right\} - (k\nu^2 + \lambda_0)\nu^{-\kappa}+ o(\nu^{-\kappa})  \right] \notag \\
					= &C\left[  
											\frac{1-\kappa}{\tau}\nu^{-\kappa} 
										+  \frac{k \alpha_2 \eta^2}{2\tau^2}(\kappa-1)(\kappa-2)\nu^{-\kappa} +o(\nu^{-\kappa})			
							\right] \simeq 0,
				\end{align}
				which leads to the self-consistent relation 
				\begin{equation}
					\kappa = 2 + \frac{1}{kD\tau}, \>\>\> D:= \frac{\alpha_2 \eta^{2}}{2\tau^{2}}.
				\end{equation}
				We thus obtain the power law tail of the intensity distribution:
				\begin{equation}
					P_{\mrss}(\lambda) \propto \lambda^{-1-a}, \>\>\> a := \frac{1}{2} + \frac{1}{2kD\tau}. 
				\end{equation}

		\subsubsection{Robust Zipf's law for the multifractal stress activation model}\label{sec:ZipfForMSA_exponMemory}
			We have shown that the exact steady-state solution of the FPE~\eqref{eq:master_eq_diffusive} exhibits Zipf's law for the MSA model~\eqref{eq:MSA_intensity_expon} with exponential memory kernel in the diffusive limit. Here, we show that Zipf's law universally and robustly appears for the MSA model with any general symmetric mark distribution, on the condition that the memory is exponential and the appropriate integrals converge. In other words,
			\begin{equation}
				h(t) = \frac{\eta}{\tau} e^{-t/\tau}, \>\>\> 
				g(\nu)=\lambda_0 e^{\beta \nu}, \>\>\> \rho(y) = \rho(-y) \>\>\> \Longrightarrow \>\>\> P_{\mrss}(\lambda) \propto \lambda^{-2} \>\>\> (\mbox{for large } \lambda).
			\end{equation}

			\paragraph*{Derivation.}
				By defining $\phi(\nu) := g(\nu)P_{\mrss}(\nu)$, the steady-state ME is given by
				\begin{equation}
					\label{eq:MSA_exponMemory_masterEq.trans1}
					\frac{1}{\tau }\frac{\partial}{\partial \nu}[\nu e^{-\beta \nu} \phi(\nu)] + \int_{-\infty}^{\infty} dy\rho(y)\phi(\nu-\eta y/\tau) - \phi(\nu) = 0.
				\end{equation}
				For large $\nu$, the first term in the left-hand side is negligible due to the exponential factor $e^{-\beta \nu}$, implying 
				\begin{equation}
					\int_{-\infty}^{\infty} dy\rho(y)\phi(\nu-\eta y/\tau) - \phi(\nu)\simeq 0 \>\>\>\mbox{for large }\nu.
				\end{equation}
				Assuming that $\phi(\nu)$ is nonnegative, this integral equation has a general solution 
				\begin{equation}
					\phi(\nu)=C_0 + C_1\nu
				\end{equation}
				with constants $C_0$ and $C_1$ (see Appendix~\ref{sec:app:phi_symmetric}). By imposing the natural boundary condition, $C_1$ must be zero as shown later, and, therefore, the general solution is given by $\phi(\nu)=C_0$. This implies the following asymptotic form of the steady-state PDF
				\begin{equation}
					P_{\mrss}(\nu) = \frac{\phi(\nu)}{g(\nu)} \propto e^{-\beta \nu}.
				\end{equation}
				We thus obtain Zipf's law for the intensity PDF, from the Jacobian relation $P_{\mrss}(\lambda)=P_{\mrss}(\nu)|d\nu/d\lambda|$. 

			\paragraph*{Natural boundary condition.}
				Here we impose the natural boundary condition to remove $C_1$. Let us use the KM expansion~\eqref{eq:KM_expon} to define the probability current as
				\begin{align}
					\frac{\pd P_t(\nu)}{\pd t} = - \frac{\pd}{\pd \nu}J_t(\nu), \>\>\> 
					J_t(\nu) := -\frac{1}{\tau}[\nu P_t(\nu)] - \sum_{k=1}^\infty \frac{\alpha_{2k}}{(2k)!}\frac{\eta^{2k}}{\tau^{2k}} \frac{\partial^{2k-1}}{\partial \nu^{2k-1}} g(\nu)P_t(\nu).
				\end{align}
				For the steady-state distribution, let us ignore the first term in $J_t(\nu)$ for large $\nu$ to obtain
				\begin{equation}
					J_{\mrss}(\nu) \simeq - \sum_{k=1}^\infty \frac{\alpha_{2k}}{(2k)!}\frac{\eta^{2k}}{\tau^{2k}} \frac{\partial^{2k-1}}{\partial \nu^{2k-1}} g(\nu)P_{\mrss}(\nu)
					\>\>\> \mbox{ for large $\nu$}. 
				\end{equation}
				By direct substitution of the general solution $g(\nu)P_{\mrss}(\nu)=\phi(\nu)=C_0+C_1\nu$, we obtain 
				\begin{align}
					J_{\mrss}(\nu) \simeq - \sum_{k=1}^\infty \frac{\alpha_{2k}}{(2k)!}\frac{\eta^{2k}}{\tau^{2k}} \frac{\partial^{2k-1}}{\partial \nu^{2k-1}} \left(C_0+C_1\nu\right)
					= -\frac{\eta^2\alpha_2}{2\tau^2}C_1. 
				\end{align}
				Since the natural boundary condition implies $\lim_{\nu\to\infty}J_t(\nu)=0$ for any $t$, we obtain $C_1=0$.

		\subsubsection{Robust asymptotic form for fast-accelerating intensity maps}\label{sec:RII_exponMemory}
			We now show that the asymptotic form~\eqref{eq:RII_PDF_diffusive} of the steady-state solution~\eqref{eq:sol_diffisive} of the FPE~\eqref{eq:master_eq_diffusive} is robust even for general mark distribution for any FAI map: 
			\begin{equation}
				\label{eq:RII_robust_singleExpon}
				h(t) = \frac{\eta}{\tau} e^{-t/\tau}, \>\>\> 
				g(\nu) \gg \nu^2 \mbox{ (for large } \nu), \>\>\> \rho(y) = \rho(-y) \>\>\> \Longrightarrow \>\>\> P_{\mrss}(\lambda) \propto \lambda^{-1} \left|\frac{dg(\nu)}{d\nu}\right|_{\nu=g^{-1}(\lambda)}^{-1} \mbox{ (for large } \lambda).
			\end{equation}

			\paragraph*{Derivation.} 
				By defining $\phi(\nu) := g(\nu)P_{\mrss}(\nu)$, the steady-state ME is given by
				\begin{equation}
					\label{eq:RII_exponMemory_masterEq.trans1}
					\frac{1}{\tau }\frac{\partial}{\partial \nu}\left\{ \frac{\nu}{g(\nu)}  \phi(\nu)\right\} + \int_{-\infty}^{\infty} dy\rho(y)\phi(\nu-\eta y/\tau) - \phi(\nu) = 0.
				\end{equation}
				For large $\nu$, the first term in the left-hand side is negligible because $g(\nu)$ is a FAI. The self-consistency of this assumption will be confirmed later. This implies
				\begin{equation}
					\int_{-\infty}^{\infty} dy\rho(y)\phi(\nu-\eta y/\tau) - \phi(\nu)\simeq 0 \>\>\>\mbox{for large }\nu.
				\end{equation}
				Assuming the nonnegativity of $\phi(\nu)$ and the natural boundary condition, this integral equation has a single solution $\phi(\nu)=C_0$ with a constant $C_0$ in the same logic to that in Sec.~\ref{sec:ZipfForMSA_exponMemory}. Finally this implies the following asymptotic form of the steady-state PDF of $\nu$:
				\begin{equation}
					P_{\mrss}(\nu) \propto \frac{1}{g(\nu)}. 
				\end{equation}
				Formula~\eqref{eq:RII_robust_singleExpon} for the intensity PDF then derives from the Jacobian relation $P_{\mrss}(\lambda)=P_{\mrss}(\nu)|d\nu/d\lambda|$.

			\paragraph*{Self-consistency of the assumption.}
				Let us check whether this solution is consistent with the assumption that the first term in Eq.~\eqref{eq:RII_exponMemory_masterEq.trans1} is irrelevant for large $\nu$. For simplicity, we focus on the case of $g(\nu)=\nu^n$ with integer $n\geq 2$. We first assume the expansion of the solution 
				\begin{equation}
					\phi(\nu)=\phi_0(\nu) + \phi_1(\nu) + \dots, \>\>\> 
					\phi_0(\nu)=C^{[0]}, \>\>\> |\phi_0(\nu)| \gg |\phi_1(\nu)|
				\end{equation}
				for large $\nu$ with a constant $C^{[0]}$. By assuming the the first term in the left-hand side of Eq.~\eqref{eq:RII_exponMemory_masterEq.trans1} is subleading, we obtain 
				\begin{equation}
					\frac{(1-n)C^{[0]}}{\tau }\nu^{-n} + \int_{-\infty}^{\infty} dy \rho(y)\phi_1(\nu-\eta y/\tau) -\phi_1(y)\simeq 0.
					\label{eq:self-consisntency_symmetric_rho_robust_exponeMem}
				\end{equation}
				We solve this non-homogeneous integral equation by assuming a solution anzatz: 
				\begin{equation}
					\phi_1(\nu) \simeq C^{[1]}\nu^{-\kappa}, \>\>\> \kappa > 0.
				\end{equation}
				By using $\int_{-\infty}^\infty y\rho(y)dy=0$, we obtain 
				\begin{align}
					\int_{-\infty}^\infty dy \rho(y)\phi_1(\nu-\eta y/\tau) 
					&= \int_{-\infty}^\infty dy \rho(y)C^{[1]}\left(\nu-\eta y/\tau\right)^{-\kappa} \notag \\
					&= \int_{-\infty}^\infty dy \rho(y)C^{[1]}\nu^{-\kappa}\left(1-\frac{\eta y}{\tau}\nu^{-1}\right)^{-\kappa} \notag \\
					&= \int_{-\infty}^\infty dy \rho(y)C^{[1]}\nu^{-\kappa}\left(1+\frac{\kappa \eta y}{\tau}\nu^{-1}+\frac{\eta^2 \kappa (\kappa+1) y^2}{2\tau^2}\nu^{-2} + O(\nu^{-3})\right) \notag \\
					&= C^{[1]}\nu^{-\kappa}\left(1 + C^{[2]}\nu^{-2} + O(\nu^{-3})\right)
				\end{align}
				with the constant $C^{[2]}$ defined by 
				\begin{equation}
					C^{[2]} := \frac{\eta^2\kappa(\kappa+1)}{2\tau^2}\int_{-\infty}^\infty y^2\rho(y) dy. 
				\end{equation}
				Equation~\eqref{eq:self-consisntency_symmetric_rho_robust_exponeMem} is thus equivalent to 
				\begin{equation}
					C^{[1]}C^{[2]}\nu^{-\kappa-2} \simeq \frac{(n-1)C^{[0]}}{\tau }\nu^{-n}. 
				\end{equation}
				This implies that the exponent $\kappa$ must satisfy
				\begin{equation}
					\kappa = n - 2.  
				\end{equation}
				This means that the subleading term $\phi_{1}(\nu)$ is actually negligible when $\kappa >0 \Longleftrightarrow n > 2$. We thus confirm that the first term in Eq.~\eqref{eq:RII_exponMemory_masterEq.trans1} can be dropped for FAI maps with dependence on $\nu$ faster than $\nu^2$ for large $\nu$'s.


\section{Solution 3: exponential memory kernel with two-sided asymmetric mark distribution for linear to fast-accelerating intensity maps}\label{sec:inhibitory_effects_AS}
		We here study both exact and asymptotic results for the case with the exponential memory kernel, and two-sided asymmetric mark distribution with negative mean mark
		\begin{equation}
			h(t) = \frac{\eta}{\tau}e^{-t/\tau}, \>\>\> p_+ := \int_{0}^\infty dy \rho(y) >0, \>\>\> p_- := \int_{-\infty}^0 dy \rho(y) >0, \>\>\> 
			m := \int_{-\infty}^\infty y\rho(y)dy < 0.
		\end{equation}
		and consider FAI maps $g(\nu) \gg \nu^2$. 

		\subsection{Exact solution for two-sided asymmetric exponential mark distribution}
			\begin{figure}
				\centering
				\includegraphics[width=75mm]{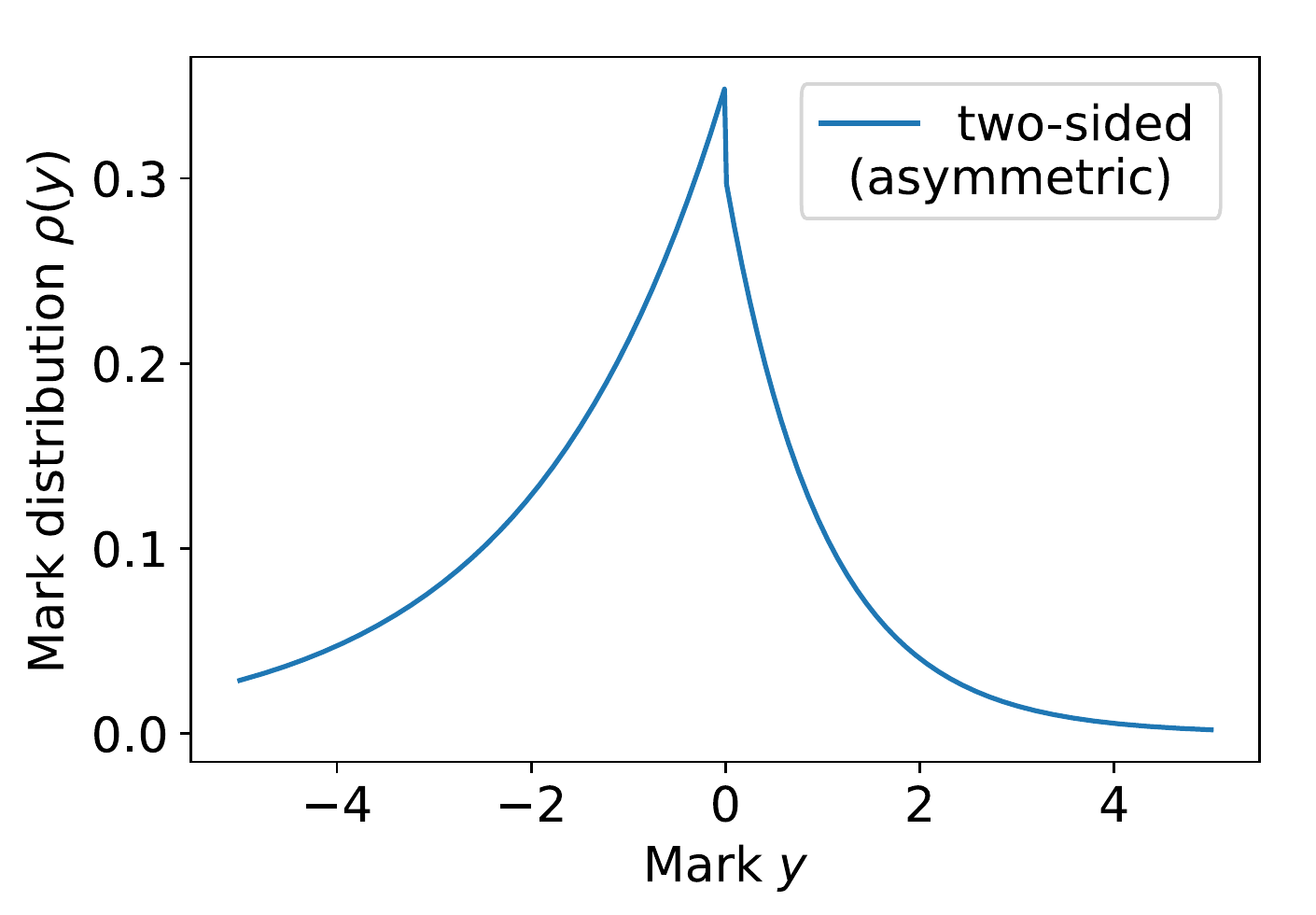}
				\caption{Schematic of the two-sided asymmetric exponential mark distribution~\eqref{eq:markdist_two-sided_A} with $(p_+,p_-,y^*_+,y^*_-)=(0.3,0.7,1,2)$.}
				\label{fig:twosided_markdist_A}
			\end{figure}
			Let us focus on the case with the two-sided asymmetric exponential mark distribution: 
			\begin{equation}
				\rho(y) = \begin{cases}
					\displaystyle \frac{p_+}{y^*_+} e^{-y/y^*_+} & (y \geq 0) \\
					\displaystyle \frac{p_-}{y^*_-} e^{y/y^*_-} & (y < 0) 
				\end{cases},
				\label{eq:markdist_two-sided_A}
			\end{equation}
			where $p_+ + p_- = 1$, $y_+^* >0$, and $y_-^* >0$. The mean mark is given by
			\begin{equation}
				m:= y^*_+ p_+ - y^*_- p_- < 0.
			\end{equation}
			By using the identities~\eqref{eq:identity_one_sided_expon} and~\eqref{eq:identity_one_sided_expon2}, the ME reads
			\begin{equation}
				\frac{\pd P_t(\nu)}{\pd t} = \frac{1}{\tau}\pd_\nu [\nu P_t(\nu)] 
				+ \pd_\nu \frac{(\frac{p_-}{c_-} - \frac{p_+}{c_+}) + \frac{\pd_\nu}{c_+c_-}}{(1+\frac{\pd_\nu}{c_+})(1-\frac{\pd_\nu}{c_-})}g(\nu)P_t(\nu), \>\>\> 
				c_\pm := \frac{\tau }{\eta y^*_{\pm}}.
			\end{equation}
			This means that the ME expresses the condition of probability conservation,
			\begin{equation}
				\frac{\pd P_t(\nu)}{\pd t} = - \frac{\pd J_t(\nu)}{\pd \nu}, \>\>\> 
				J_t(\nu) := -\frac{\nu}{\tau}P_t(\nu) - \frac{(\frac{p_-}{c_-} - \frac{p_+}{c_+}) + \frac{\pd_\nu}{c_+c_-}}{(1+\frac{\pd_\nu}{c_+})(1-\frac{\pd_\nu}{c_-})}g(\nu)P_t(\nu). 
			\end{equation}
			By requiring the natural boundary condition $\lim_{\nu\to \infty}J_t(\nu) = 0$, we obtain the second-order partial differential equation that the steady-state PDF $P_{\mrss}(\nu)$ satisfies: 
			\begin{equation}
				\tau \left[A + B \frac{d}{d\nu}\right] \phi(\nu) + \left[1 + C\frac{d}{d\nu} - B\frac{d^2}{d\nu^2}\right]\left\{\frac{\nu}{g(\nu)}\phi(\nu)\right\} = 0, \>\>\> \phi(\nu):= g(\nu)P_{\mrss}(\nu)
			\end{equation}
			with 
			\begin{equation}
				A := \frac{p_-}{c_-} - \frac{p_+}{c_+} := -\frac{\eta}{\tau}m > 0, \>\>\> 
				B:= \frac{1}{c_+c_-} > 0, \>\>\> 
				C:= \frac{1}{c_+} - \frac{1}{c_-}.
			\end{equation}
			We note that the coefficients are simplified for the symmetric mark distribution $y^*_+=y^*_-$ and $p_+=p_-=1/2$, such that $A=C=0$.

			\subsubsection*{Example 1: ramp tension-intensity map}
				For the ramp tension-intensity map 
				\begin{equation}
					\lambda = g(\nu) = \max \{\nu_0,|\nu|\},
				\end{equation}
				we obtain the exact solution for $\lambda > |\nu_0|$
				\begin{equation}
					P_{\mrss}(\lambda) \propto \lambda^{-1} e^{-\lambda/\lambda_{\cut}}, \>\>\> 
					\lambda_{\cut} := \frac{C + \tau B + \sqrt{(C+\tau B)^2 + 4B(1+\tau B)}}{2(1+\tau A)} > 0.
				\end{equation}
				under the natural boundary condition. This implies that an exponential tail is observed for this model. 

			\subsubsection*{Example 2: exponential tension-intensity map}
				For the exponential tension-intensity map
				\begin{equation}
					\lambda = g(\nu) = \lambda_0 |\nu| e^{\beta \nu},
				\end{equation}
				we obtain the exact solution for $\nu > 0$ as 
				\begin{equation}
					\phi(\nu) = C^{[1]} y^{\gamma_1} {}_1F_1 \left(q_1,r_1;y\right)
										+ C^{[2]} y^{\gamma_2} {}_1F_1 \left(q_2,r_2;y\right),\>\>\> 
					y(\nu) = \frac{\tau e^{\beta \nu}}{\beta} 
				\end{equation}
				where
				\begin{subequations}
				\begin{equation}
					\gamma_1: = \frac{C+2 \beta  B-\sqrt{4 B+C^2}}{2 \beta  B}, \>\>\>
					q_1:= 1 + \frac{C+2A-\sqrt{4 B+C^2}}{2 \beta  B}, \>\>\> 
					r_1:= 1-\frac{\sqrt{4 B+C^2}}{\beta  B},
				\end{equation}
				\begin{equation}
					\gamma_2: = \frac{C+2 \beta  B+\sqrt{4 B+C^2}}{2 \beta  B},\>\>\> 
					q_2:= 1 + \frac{C+2A+\sqrt{4 B+C^2}}{2 \beta  B}, \>\>\> 
					r_2:= 1+\frac{\sqrt{4 B+C^2}}{\beta  B}~,
				\end{equation}
				\end{subequations}
				and  $C^{[1]}, C^{[2]}$ are two integral constants.
				For large $\nu\to \infty$, assuming the boundary condition $\lim_{\nu\to \infty}P_{\mrss}(\nu)= 0$, we obtain the asymptotic formula
				\begin{equation}
					\phi(\nu) \propto y^{-A/(\beta B)},
				\end{equation}
				leading to the PDF tail of the intensity:
				\begin{equation}
					P_{\mrss}(\lambda) \propto \lambda^{-2-\beta^{-1} u},\>\>\> 
					u = \frac{A}{B} = -\frac{\tau m}{\eta y^*_+ y^*_-}.
					\label{eq:power-law_exact_twosidedAsymExpon_ExponMem}
				\end{equation}
				This result implies that the power-law scaling deviates from Zipf's law in proportion to the amplitude $m$ of the asymmetry of the mark distribution.

		\subsection{Robust asymptotic solutions}
			In this subsection, we generalise the above exact results in the form of robust asymptotic results under a wide range of two-sided mark distributions with negative mean, in the presence of an exponential memory kernel. 

			\subsubsection{Robust exponential tail for the ramp intensity}
				Let us assume that the ramp intensity is asymptotically
				\begin{subequations}
					\label{eq:tLevyAsymptotics_ramp_AS}
					\begin{equation}
						g(\nu) \simeq \nu - \nu_1 \>\>\>\mbox{for large $\nu$}
					\end{equation}
					with an exponential memory with two-sided mark distribution of negative mean:
					\begin{equation}
						h(t) = \frac{\eta}{\tau}e^{-t/\tau}, \>\>\> 
						p_+ := \int_{0}^\infty \rho(y)dy > 0,\>\>\> 
						p_- := \int_{-\infty}^0 \rho(y)dy > 0, \>\>\> 
						m := \int_{-\infty}^\infty y\rho(y)dy < 0.
					\end{equation}
					Here $\nu_1$ is an arbitrary real number, either positive or nonpositive in contrast to the LHawkes process. Under this assumption, we obtain 
					\begin{equation}
						P_{\mrss}(\lambda) \propto e^{-\lambda/\lambda_{\rm cut}} \>\>\> \mbox{ for large }\lambda.
					\end{equation}
				\end{subequations}
				The parameter $\lambda_{\rm cut}$ is given by the self-consistent relation
				\begin{equation}
					\frac{1}{\tau\lambda_{\rm cut}} = \Phi\left(\frac{\eta}{\tau\lambda_{\rm cut}}\right), \>\>\> 
					\Phi(x) := \int_{-\infty}^\infty dy\rho(y)(e^{xy} -1),
				\end{equation}
				where $\Phi(x)$ is the moment-generating function. The equation for $\lambda_{\rm cut}$ has a single positive solution (see Appendix~\ref{sec:app:phi_symmetric}). This relation can be derived by a straightforward generalisation of the derivation in Sec.~\ref{sec:expMem_rampInt_robustAsymp}. 		

			\subsubsection{Robust power law tail for fast-accelerating intensity maps}
				We show that, under the following general assumptions
				\begin{equation}
					h(t) = \frac{\eta}{\tau}e^{-t/\tau}, \>\>\> 
					g(\nu) \gg \nu^2 \>\> \mbox{ (for large $\nu$)}, \>\>\> 
					p_+ := \int_{0}^\infty \rho(y)dy > 0,\>\>\> 
					p_- := \int_{-\infty}^0 \rho(y)dy > 0, \>\>\> 
					m := \int_{-\infty}^\infty y\rho(y)dy < 0,
					\label{eq:robust_asymp_exponMem_genMark_condition}
				\end{equation}
				we obtain the robust asymptotic relationship
				\begin{equation}
					P_{\mrss}(\lambda) \propto \lambda^{-1}\left[\left|\frac{dg(\nu)}{d\nu}\right|^{-1} e^{-u \nu}\right]_{\nu = g^{-1}(\lambda)}, \>\>\>
					u := \frac{\tau c^*}{\eta},
					\label{eq:robust_asymp_exponMem_genMark}
				\end{equation}
				where $c^*$ is the unique positive root of $\Phi(c^*)=0$, where the moment-generating function is defined by $\Phi(x):=\int_{-\infty}^\infty dy \rho(y)\left(e^{xy}-1\right)$. 
				
				\paragraph*{Examples.}
					From this formula, we readily deduces the power law PDF for the exponential intensity
					\begin{equation}
						g(\nu) \simeq \lambda_0 e^{\beta }\>\>\>
						\Longrightarrow \>\>\> 
						P_{\mrss}(\lambda) \propto \lambda^{-2- \beta^{-1} u}.
					\end{equation} 
					We note that this result is consistent with the aforementioned exact result~\eqref{eq:power-law_exact_twosidedAsymExpon_ExponMem} by considering Eq.~\eqref{eq:app:root_c*_two-sided_exponential} in Appendix \ref{sec:app:phi_symmetric} for the case with the exponential intensity and the two-sided asymmetric exponential mark distribution. 
					
					In addition, we obtain the truncated power law PDF for the polynomial intensity
					\begin{equation}
						g(\nu) \simeq \lambda_0 \nu^{n},\>\>\> 
						n > 2 \>\>\>
						\Longrightarrow \>\>\> 
						P_{\mrss}(\lambda) \propto \lambda^{-2+\frac{1}{n}}e^{-u \left(\frac{\lambda}{\lambda_0}\right)^{\frac{1}{n}}},
					\end{equation}
					where the cutoff length appears due to the asymmetry of the mark distribution. For the zero mean mark limit $m\uparrow 0$, the cutoff disappears as $u \downarrow 0$.

				\paragraph*{Derivation.}
					By defining $\phi(\nu)=g(\nu)P_{\mrss}(\nu)$, the ME is given by 
					\begin{equation}
						\frac{1}{\tau}\frac{\pd}{\pd \nu}\left\{\frac{\nu}{g(\nu)}\phi(\nu)\right\} + 
						\int_{-\infty}^\infty dy \rho(y)\phi(\nu-\eta y/\tau) - \phi(y) = 0.
						\label{eq:robust_powerlaw_asymRho_FAI_trans1}
					\end{equation}
					As an asymptotic assumption for the solution, let us first neglect the first term of Eq.~\eqref{eq:robust_powerlaw_asymRho_FAI_trans1} to obtain 
					\begin{equation}
						\int_{-\infty}^\infty dy \rho(y)\phi(\nu-\eta y/\tau) - \phi(y) \simeq 0 \>\>\> \mbox{ for large }\nu
					\end{equation}
					for case~\eqref{eq:robust_asymp_exponMem_genMark_condition}. The self-consistency of this assumption will be confirmed later. According to Appendix~\ref{sec:app:sol_integralSolution}, the general solution is given by the superposition of exponentials, 
					\begin{equation}
						\phi(\nu) \simeq \sum_{i} C_i e^{-(\tau c_i/\eta)\nu},
					\end{equation}
					where the $c_i$'s are the  roots of the moment-generating function $\Phi(x)=0$. The moment-generating function is defined by
					\begin{equation}
						\Phi(c) = 0, \>\>\> \Phi(x):= \int_{-\infty}^\infty dy \rho(y)\left(e^{xy}-1\right),
					\end{equation}
					whose analytical characters are summarised in Appendix~\ref{sec:app:phi_symmetric}. According to Appendix~\ref{sec:app:phi_symmetric}, $\Phi(x)=0$ has only two roots at $x=0$ and $x=c^*>0$. This means that the general asymptotic solution is given by 
					\begin{equation}
						\phi(\nu) \simeq C_0 e^{-(\tau c^*/\eta)\nu} + C_1
					\end{equation}
					with integral constants $C_0$ and $C_1$. By imposing the natural boundary condition, $C_1$ must be zero (see below for the natural boundary condition). We thus have the solution 
					\begin{equation}
						\phi(\nu) \simeq C_0 e^{-(\tau c^*/\eta)\nu}. 
					\end{equation}
					This implies that the steady-state intensity PDF has the following asymptotic form 
					\begin{equation}
						P_{\mrss}(\nu) = \frac{\phi(\nu)}{g(\nu)} \propto \frac{1}{g(\nu)} e^{-(\tau c^*/\eta)\nu},
					\end{equation}
					which implies Eq.~\eqref{eq:robust_asymp_exponMem_genMark} from the Jacobian relation $P_{\mrss}(\lambda)=P_{\mrss}|d\nu/d\lambda|$.

				\paragraph*{Self-consistency of the assumption.}
					Finally, we here confirm the self-consistency of the ansatz for the solution under the assumption of FAI maps. Let us assume that the solution is given by the following expansion 
					\begin{equation}
						\phi(\nu) = \phi_0(\nu) + \phi_1(\nu) + \dots , \>\>\> \phi_0(\nu) := C^{[0]} e^{-(\tau c^*/\eta)\nu}, \>\>\> |\phi_0(\nu)| \gg |\phi_1(\nu)| \>\>\> \mbox{ for large }\nu
					\end{equation}
					with an integral constant $C^{[0]}$. For simplicity, let us focus on the case $g(\nu)=\nu^n$ with integer $n>2$. By assuming that the first term in Eq.~\eqref{eq:robust_powerlaw_asymRho_FAI_trans1} is subleading, we substitute this expansion into Eq.~\eqref{eq:robust_powerlaw_asymRho_FAI_trans1} to obtain 
					\begin{align}
						&\int_{-\infty}^\infty dy \rho(y)\phi_1(\nu-\eta y/\tau) - \phi_1(y) \simeq 
						- \frac{1}{\tau}\frac{\pd}{\pd \nu}\left\{\nu^{1-n}\phi_0(\nu)\right\} \notag \\
						\Longrightarrow 
						&\int_{-\infty}^\infty dy \rho(y)\phi_1(\nu-\eta y/\tau) - \phi_1(y) \simeq 
						C^{[0]}\frac{c^*}{\eta}\nu^{1-n}e^{-(\tau c^*/\eta)\nu}.
						\label{eq:robutsAsymp_asymrho_perturbation}
					\end{align}
					We make the anzatz for the solution in the form
					\begin{equation}
						\phi_1(\nu) \simeq C^{[1]} \nu^{-\kappa}e^{-(\tau c^*/\eta)\nu}, \>\>\> \kappa > 0
					\end{equation}
					to obtain the special solution with a constant $C^{[1]}$. Here the condition $\kappa>0$ is essential; otherwise the consistency relationship $|\phi_0(\nu)| \gg |\phi_1(\nu)|$ does not hold. By direct substitution, Eq.~\eqref{eq:robutsAsymp_asymrho_perturbation} is equivalent to 
					\begin{align}
						\int_{-\infty}^\infty dy \rho(y)\phi_1(\nu-\eta y/\tau) &\simeq C^{[1]}\int_{-\infty}^\infty dy \rho(y)(\nu-\eta y/\tau)^{-\kappa} e^{c^*y-(\tau c^*/\eta)\nu}  \notag \\
						&=  C^{[1]}\int_{-\infty}^\infty dy \rho(y)\nu^{-\kappa}\left(1-\frac{\eta y}{\tau}\nu^{-1}\right)^{-\kappa} e^{c^*y}e^{-(\tau c^*/\eta)\nu}  \notag \\
						&= C^{[1]}\int_{-\infty}^\infty dy \rho(y)\nu^{-\kappa}\left(1+\frac{\kappa \eta y}{\tau}\nu^{-1} + O(\nu^{-2})\right) e^{c^*y}e^{-(\tau c^*/\eta)\nu} \notag \\
						&= C^{[1]}\nu^{-\kappa}e^{-(\tau c^*/\eta)\nu} \left(\Phi(c^*)+1 + C^{[2]}\nu^{-1}+O(\nu^{-2})\right)
					\end{align}
					with 
					\begin{equation}
						C^{[2]} := \frac{\kappa \eta}{\tau}\int_{-\infty}^\infty ye^{c^*y}\rho(y)dy.
					\end{equation}
					By using $\Phi(c^*)=0$, we thus obtain 
					\begin{equation}
						C^{[1]}C^{[2]} \nu^{-\kappa-1}e^{-(\tau c^*/\eta)\nu} \simeq C^{[0]}\frac{c^*}{n}\nu^{1-n}e^{-(\tau c^*/\eta)\nu}. 
					\end{equation}
					This implies that the power law exponent $\kappa$ must satisfy the relationship
					\begin{equation}
						\kappa = n - 2.
					\end{equation}
					Because of the assumption $\kappa > 0$, we obtain the self-consistency condition 
					\begin{equation}
						n > 2,
					\end{equation}
					which is equivalent to the assumption that the tension-intensity maps must be FAI ($g(\nu)\gg \nu^2$).

				\paragraph*{Natural boundary condition. }
					Here we impose the natural boundary condition to remove $C_1$. Let us use the KM expansion~\eqref{eq:KM_expon} to define the probability current as
					\begin{align}
						\frac{\pd P_t(\nu)}{\pd t} = - \frac{\pd}{\pd \nu}J_t(\nu), \>\>\> 
						J_t(\nu) := -\frac{1}{\tau}[\nu P_t(\nu)] - \sum_{k=1}^\infty \frac{(-1)^k\alpha_{k}}{k!}\frac{\eta^{k}}{\tau^{k}} \frac{\partial^{k-1}}{\partial \nu^{k-1}} g(\nu)P_t(\nu).
					\end{align}
					For the steady-state distribution, let us ignore the first term in $J_t(\nu)$ for large $\nu$ to obtain
					\begin{equation}
						J_{\mrss}(\nu) \simeq - \sum_{k=1}^\infty \frac{(-1)^k\alpha_{k}}{k!}\frac{\eta^{k}}{\tau^{k}} \frac{\partial^{k-1}}{\partial \nu^{k-1}} g(\nu)P_{\mrss}(\nu)
						\>\>\> \mbox{ for large $\nu$}. 
					\end{equation}
					By direct substitution of the general solution $g(\nu)P_{\mrss}(\nu)=\phi(\nu)=C_0+C_1 e^{-(\tau c^*/\eta)\nu}$, we obtain 
					\begin{align}
						J_{\mrss}(\nu) &\simeq - \sum_{k=1}^\infty \frac{(-1)^k\alpha_{k}}{k!}\frac{\eta^{k}}{\tau^{k}} \frac{\partial^{k-1}}{\partial \nu^{k-1}} \left(C_1+C_0 e^{-(\tau c^*/\eta)\nu}\right) \notag \\
						&= \frac{\eta m}{\tau}C_1 + \frac{\eta C_0}{c^*\tau}e^{-(\tau c^*/\eta)\nu}\sum_{k=1}^\infty \frac{\alpha_k}{k!}c^{*k} \notag \\
						&= \frac{\eta m}{\tau}C_1 + \frac{\eta C_0}{c^*\tau}e^{-(\tau c^*/\eta)\nu}\Phi(c^*)\notag \\
						&= \frac{\eta m}{\tau}C_1,
					\end{align}
					where we have used $\Phi(x)=\sum_{k=1}^\infty (\alpha_k/k!)x^k$ and $\Phi(c^*)=0$.
					Since the natural boundary condition implies $\lim_{\nu\to\infty}J_t(\nu)=0$ for any $t$, we obtain $C_1=0$.


\section{Solution 4: general memory kernel for one-sided mark distribution  for the ramp Hawkes process}\label{sec:sol3_arbitrary_memory}
		We have studied the exact solution for the NLHawkes process assuming that (i) the memory kernel is exponential and (ii) the jump size obeys the one-sided exponential distributions. In particular, we derived the power law tail~\eqref{eq:true_power law_one-sided_exponential} without truncation, at the critical point for the ramp Hawkes process \eqref{eq:ramp_tension-intensity_map}. As shown in the following in this section, this exact power law relation is robust for general ramp Hawkes processes with any memory kernel and jump-size distribution, only assuming the finiteness of 
		\begin{equation}
			\la \tau \ra := \int_0^\infty t~h(t)dt = \int_{0}^{\infty}x^2 ~\tilh(x) dx <\infty~,
		\end{equation}
		where $\tilh(x)$ has been defined in \eqref{eq:decomposition}. 
			Note that the critical condition is characterised by 
			\begin{equation}
				\eta := \int_0^\infty h(t)dt = \int_0^\infty x \tilh(x)dx = 1.
			\end{equation}

		\subsection{Discrete sum of exponentials} 
			Let us first consider the case of a discrete sum of exponentials. In this case, we find a power law asymptotics at the critical point $\eta=1$
			\begin{align}
				h(t) = \sum_{k=1}^K \tilh_ke^{-t/\tau_k}, \>\>\> & \lambda = g(\nu) \simeq \nu - \nu_1 + o(\nu^0) \> \mbox{ for large }\nu, \>\>\> \rho(y)=0 \>  \mbox{ for negative }y, \>\>\> \int_0^\infty y\rho(y)dy = 1\notag\\
				\Longrightarrow \>\>\>
				&P_{\mrss}(\lambda) \propto \lambda^{-1-a}, \>\>\> a:= \frac{2\nu_1\la\tau \ra}{\alpha_2}, \>\>\> \la\tau \ra:= \sum_{k=1}^K\tau_k^2 ~\tilh_k
				\label{eq:power-law_gen_memory_K_one-sidedMark_ramp}
			\end{align}
			for either negative or non-negative $\nu_1$. This relation is a true power law for positive $\nu_1$ (i.e. normalizable even without cutoff), while it is an intermediate asymptotics for non-positive $\nu_1$ (i.e., not normalizable without cutoff). 
			Note that the critical condition is given by 
			\begin{equation}
				\eta := \sum_{k=1}^K \tau_k\tilh_k =1. 
			\end{equation}

			\subsubsection*{Derivation.}	
				Let us first write the asymptotic form of $P_{\mrss}(\bm{z})$ as $S(\bm{z})$:
				\begin{equation}
					P_{\mrss}(\bm{z}) = S(\bm{z}) + R(\bm{z}), \>\>\> S(\bm{z}) \gg R(\bm{z}) \mbox{ for large }\bm{z}
				\end{equation}
				with a correction term $R(\bm{z})$ for small $\bm{z}$. Here $S(\bm{z})$ is assumed to have a fat tail represented by a power law, while $R(\bm{z})$ is assumed to have a thinner tail. The ME~\eqref{eq:master_eq_discrete_K} in the steady state reduces asymptotically to
				\begin{equation}
					\sum_{k=1}^K \frac{\partial}{\partial z_k}\frac{z_k}{\tau_k}S(\bm{z}) 
					+\int_{0}^\infty dy\rho(y)\left[ \sum_{k=1}^K (z_k-y\tilh_k)-\nu_1\right]S(\bm{z}-y\tilbh)
					 - \left[\sum_{k=1}^Kz_k-\nu_1 \right]S(\bm{z}) \simeq 0 \>\>\> \mbox{ for large }\nu.
				\label{wthgtm1vf}
				\end{equation}
				
				Since the asymptotic form of this ME is the same as that for the LHawkes process presented in Ref.~\cite{KzDidier2019PRL} except that $\nu_1$ can be either negative or non-negative, its asymptotic solution for large $\nu$ can be obtained from a similar calculation to that presented in Ref.~\cite{KzDidier2019PRL}. While we refer the reader to Ref.~\cite{KzDidier2019PRL} for an elementary introduction to the calculations, let us sketch the main steps of the derivation. We first define the Laplace transformations,
				\begin{equation}
					\tl{P}_{\mrss}(\bm{s}):= \mcL_K[P_{\mrss}(\bm{z});\bm{s}], \>\>\> \tl{S}(\bm{s}):= \mcL_K[S(\bm{z});\bm{s}], \>\>\> \tl{R}(\bm{s}):= \mcL_K[R(\bm{z});\bm{s}].
				\end{equation}
				Since $P_{\mrss(\bm{z})}$ is a PDF, the normalisation implies $\int d\bm{z}P_{\mrss}(\bm{z}) = \tl{P}_{\mrss}(\bm{s}=\bm{0}) = 1$. However, $S(\bm{z})$ is just an asymptotic form of the PDF, and there is no guarantee that $\tl{S}(\bm{s}=\bm{0})=1$. For example, assuming that $\tl{S}(s\bm{1}) = As^{a} + o(s^a)$ with any non-integer number $a$ and indicator vector $\bm{1}:= (1,1,...,1)$, we can expand $\tl{P}_{\mrss}(s):= \tl{P}_{\mrss}(s\bm{1})$ as 
				\begin{equation}
					\tl{P}_{\mrss}(s) = \tl{S}(s\bm{1}) + \tl{R}(s\bm{1}) \simeq As^{a} +
					 \sum_{k=0}^{m}c_ks^{k} + o(s^{a}), \>\>\> \tl{R}(s\bm{1}) = \sum_{k=0}^{m}c_ks^{k} + o(s^{a})
				\end{equation}
				for small $\bm{s}$ with $m:= \max(\lfloor a\rfloor ,0)$ and the floor function $\lfloor x \rfloor = \max\{k \in \bm{Z} \>|\> k \leq x\}$ with the set of integers $\bm{Z}$. The normalisation condition requires $\tl{P}(s=0)=c_0=1$. By applying the Laplace transformation to the steady-state ME \eqref{wthgtm1vf}, we obtain
				\begin{equation}
					-\sum_{k=1}^K \frac{s_k}{\tau_k}\frac{\partial \tl{S}(\bm{s})}{\partial s_k} 
					- (\Phi(\bm{s})-1)\left(\nu_1+\sum_{k=1}^K\frac{\partial }{\partial s_k}\right)\tl{S}(\bm{s}) \simeq 0,
					 \>\>\> \Phi(\bm{s}):= \int_{0}^\infty dy\rho(y)e^{-y\tilbh\cdot \bm{s}}, 
				\end{equation}
				which is valid for small $\bm{s}$. Considering $(1/\tl{S}(\bm{s}))\partial \tl{S}(\bm{s})/\partial s_k = (\partial/\partial s_k)\log |\tl{S}(\bm{s})|$, this equation can be rewritten as 
				\begin{equation}
					\sum_{k=1}^K\left (1 - \Phi(\bm{s}) - \frac{s_k}{\tau_k}\right) \frac{\partial \Psi (\bm{s})}{\partial s_k} \simeq \nu_1\left(\Phi(\bm{s})-1\right), 
					\>\>\> \Psi(\bm{s}) := \log |\tl{S}(\bm{s})|.
				\end{equation}
				
				\begin{figure*}
					\centering
					\includegraphics[width=175mm]{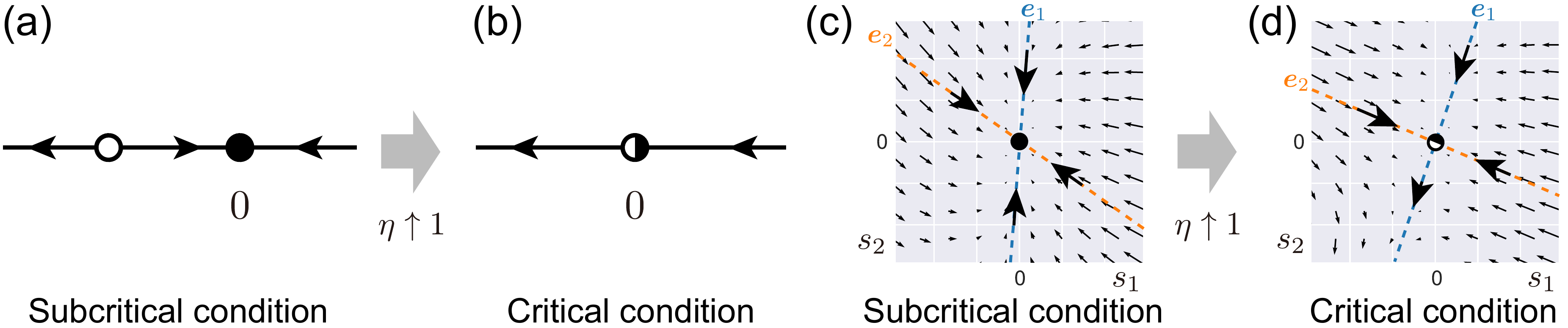}
					\caption{
						Schematics of the ``vector field'' $\bm{V}(\bm{s}):=d\bm{s}/dl$ for the special cases $K=1$ (Figs.~a and b below and at criticality, respectively) and $K=2$ (Figs.~c and d below and at criticality, respectively). The origin $\bm{s}=\bm{0}$ is the stable attractor below criticality $\eta < 1$. An unstable attractor merges into $\bm{s}=0$ at criticality $\eta \uparrow 1$, which is consistent with the standard picture of the transcritical bifurcation. 
					}
					\label{fig:transcritical}
				\end{figure*}
				Since this equation belongs to the class of first-order partial differential equations, it can be solved by the method of characteristics. Let us thus consider the corresponding Lagrange-Charpit equations:
				\begin{equation}
					\frac{ds_k}{dl} = 1-\frac{s_k}{\tau_k} - \Phi(\bm{s}), \>\>\> \frac{d\Psi}{dl} = \nu_1\left(\Phi(\bm{s})-1\right)
				\end{equation}
				with a parameter $l$ describing the position on characteristic curves. By regarding $l$ as an imaginary ``time" of this system, we can apply the standard bifurcation theory of dynamical systems. Since we are interested in the regime of small $\bm{s}$, let us consider the ``long time" asymptotic limit $l\to \infty$, where $\bm{s}(l)$ relaxes to the attractor at $\bm{s}=\bm{0}$ (see the schematic figures~\ref{fig:transcritical}a and c for the vector field $\bm{V}(\bm{s}):=d\bm{s}/dl$ for the cases $K=1$ and $K=2$ below criticality $\eta < 1$, respectively), such that $\lim_{l\to \infty}\bm{s}(l)=\bm{0}$. Let us expand the equations for small $s$:
				\begin{subequations}
				\begin{align}
					\frac{ds_k}{dl} &\simeq - \bm{H}\bm{s} - \frac{\alpha_2}{2}\left(\sum_{j=1}^K \tilh_j s_j\right)^2+ o(\bm{s}^2),  
					\label{eq:LC_RampHawkes_gen_K_LC1} \\
					\frac{d\Psi}{dl} &\simeq -\nu_1\bm{K} \bm{s} + o(\bm{s}),
					\label{eq:LC_RampHawkes_gen_K_LC2}
				\end{align}
				\end{subequations}
				with
				\begin{equation}	
					\bm{H} :=	 	\begin{pmatrix}
									\frac{1}{\tau_1}-\tilh_1,& -\tilh_2,& \dots& -\tilh_K \\
									-\tilh_1,& \frac{1}{\tau_2}-\tilh_2,& \dots& -\tilh_K \\
									\vdots& \vdots& \ddots& \vdots \\
									-\tilh_1,& -\tilh_2,& \dots& \frac{1}{\tau_K}-\tilh_K
								\end{pmatrix}, \>\>\>
					\bm{K} := \left(\tilh_1, \dots, \tilh_K\right). 
					\label{wrnhmnnh3}
				\end{equation}
				Note that the matrix $\bm{H}$ is the same as that in Ref.~\cite{KzDidier2019PRL,KzDidier2019PRR}. Defining its eigenvalues $\{\lambda_k\}_{k=1,\dots,K}$ and corresponding eigenvectors $\{\bm{e}_{k}\}_{k=1,\dots,K}$ by
					\begin{equation}
						\bm{H}\bm{e}_k = \lambda_k \bm{e}_k~,
					\end{equation}
				$\bm{H}$  has the following mathematical properties (see Ref.~\cite{KzDidier2019PRR} and Appendix~\ref{sec:app:proof_of_H} for details): 
				\begin{screen}
					\begin{enumerate}
						\item 	All the eigenvalues are real: $\lambda_k \in \bm{R}^1$. Accordingly, we assume that $\lambda_i \leq \lambda_j$ for $i<j$. 
						\item 	The determinant of $\bm{H}$ is given by
								\begin{equation}
									\det \bm{H} = \frac{1-\sum_{k=1}^K \tau_k\tilh_k}{\prod_{k=1}^K\tau_k}.
								\end{equation}
								This means that the zero eigenvalue appears at criticality $\eta:=\sum_{k=1}^K \tau_k\tilh_k = 1$. 
						\item 	$\bm{H}$ can be diagonalised by $\bm{P}$ as follows:
								\begin{equation}
									\bm{P}:= (\bm{e}_1,\dots,\bm{e}_K), \>\>\> \bm{P}^{-1}\bm{H}\bm{P} = 
									\begin{pmatrix}
										\lambda_1,& 0, & \dots& 0 \\
										0,& \lambda_2,& \dots& 0 \\
										\vdots& \vdots& \ddots& \vdots \\
										0,& 0,& \dots& \lambda_K
									\end{pmatrix}.
								\end{equation}
						\item 	Let us introduce a new representation based on the eigenvectors:
								\begin{equation}
									\bm{X} = 	\begin{pmatrix}
												X_1 \\
												X_2 \\
												\vdots \\
												X_K
											\end{pmatrix}
									:= \bm{P}^{-1}\bm{s}, \>\>\>
									\bm{s} = 	\begin{pmatrix}
										s_1 \\
										s_2 \\
										\vdots \\
										s_K
									\end{pmatrix}, \>\>\>
									\bm{P}^{-1} =  \begin{pmatrix}
												\bm{g}_1 \\
												\bm{g}_2 \\
												\vdots \\
												\bm{g}_K
											\end{pmatrix}.
								\end{equation}
								At criticality $\eta=1$, the smallest eigenvalue is zero, $\lambda_1=0$, and its eigenvector is given by
								\begin{equation}
									\bm{e}_1 = \begin{pmatrix}
												\tau_1 \\
												\tau_2 \\
												\vdots \\
												\tau_K
											\end{pmatrix}.
								\end{equation}
								In addition, $X_1$ is represented by
								\begin{equation}
									X_1 = \bm{g}_1 \cdot \bm{s} = \frac{1}{\la \tau \ra}\sum_{k=1}^K \tau_k\tilh_k s_k, \>\>\> \bm{g}_1 = \left(\frac{\tau_1\tilh_1}{\la\tau \ra},\>\dots\>, \frac{\tau_K\tilh_K}{\la\tau \ra}\right), \>\>\> \la \tau \ra:= \sum_{k=1}^K \tau_k^2\tilh_k.
								\end{equation}
					\end{enumerate}	
				\end{screen}
				
				Given these properties, let us consider the Lagrange-Charpit equation~\eqref{eq:LC_RampHawkes_gen_K_LC1} in the representation $\bm{X}:= (X_1,.\dots,X_K)^T$. At criticality $\eta=1$, the leading order contribution in the Lagrange-Charpit Eq.~\eqref{eq:LC_RampHawkes_gen_K_LC1} is given by
				\begin{equation}
					\frac{dX_1}{dl} \simeq 0 -\frac{\alpha_2}{2\la\tau \ra}\left(\sum_{k=1}^K \tilh_k s_k\right)^2 
					+ o(\bm{X}^2), \>\>\> \frac{dX_j}{dl} = - \lambda_j X_j +o (\bm{X})\mbox{  for } j\geq 2.
				\end{equation}
				Since the leading order contribution will come from the $X_1$ direction, we can assume that $|X_1| \gg |X_j|$ for $j\geq 2$ for large $l$. We thus ignore contribution other than $X_1$ by assuming $X_j \simeq 0$ for $j\geq 2$: 
				\begin{equation}
					\bm{s} 	= \bm{P}\bm{X} \simeq (\bm{e}_1,\dots,\bm{e}_K) 
											\begin{pmatrix}
												X_1 \\
												0 \\
												\vdots \\
												0
											\end{pmatrix}
							= X_1 \bm{e}_1.
				\end{equation} 
				We thus obtain 
				\begin{equation}
					\frac{dX_1}{dl} \simeq 0 -\frac{\alpha_2}{2\la\tau \ra}X_1^2 + o(\bm{X}^2), \>\>\> \frac{dX_j}{dl} = - \lambda_j X_j + o(\bm{X})\mbox{  for } j\geq 2.
				\end{equation}
				This is the standard normal form of the transcritical bifurcation when regarding $l$ as a physical time {(see the schematic figures~\ref{fig:transcritical}b and d for $K=1$ and $K=2$ at criticality $\eta\uparrow 1$, respectively). The solution is given by
				\begin{equation}
					X_1(l) \simeq \frac{2\la\tau\ra}{\alpha_2} \frac{1}{l-l_0}, \>\>\> X_j(l) \simeq C_je^{-\lambda_j(l-l_0)}
				\end{equation}
				with integral constants $l_0$ and $C_j$ for $j\geq 2$. We can assume $l_0=0$ as the initial point of the characteristic curve without losing generality. From expanding $\Phi(s)$, we obtain the solution
				\begin{equation}
					\Psi(l) \simeq -\nu_1\int dl \sum_{k=1}^K \tilh_k s_k(l) \simeq \frac{2\nu_1\la \tau\ra}{\alpha_2}\log |X|_1 + O(\bm{X}) + C_0
				\end{equation}
				with an integral constant $C_0$. According to the method of characteristics, the general solution is given by 
				\begin{equation}
					\mcH(C_2, ,\dots, C_K) = C_0
				\end{equation}
				with a function $\mcH$ which needs to be determined by the initial condition. The constants $C_j$ with $j\geq 2$ are related to each other, such that
				\begin{align}
					l = \frac{2\la \tau\ra}{\alpha_2 X_1}, \>\>\> C_j = X_j\exp\left\{\frac{2\la\tau \ra\lambda_j}{\alpha_2X_1} \right\}.
				\end{align}
				This means that the explicit form of the general solution is given by
				\begin{equation}
					\Psi(\bm{X}) \simeq \frac{2\nu_1\la \tau\ra}{\alpha_2}\log |X_1| + O(\bm{X}) 
					+ \mcH\left(X_2\exp\left\{\frac{2\la\tau \ra\lambda_2}{\alpha_2X_1}\right\},\dots, 
					X_K\exp\left\{\frac{2\la\tau \ra\lambda_K}{\alpha_2X_1}\right\} \right)~.
				\end{equation}
				Note the existence of the divergent term $\log|X_1|$ resulting from neglecting the UV cutoff. Since $\Psi(\bm{X})$ must be constant for $\bm{s}\to \bm{0}$, except for the artificial log divergence, we obtain
				\begin{equation}
					\lim_{\bm{X}\to 0}\mcH\left(X_2\exp\left\{\frac{2\la\tau \ra\lambda_2}{\alpha_2X_1}\right\},
					\dots, X_K\exp\left\{\frac{2\la\tau \ra\lambda_K}{\alpha_2X_1}\right\} \right) = \mbox{const.}
					\label{eq:Gen_Kmem_woInhibitory_trans1}
				\end{equation}
				
				Let us now consider the specific limit $X_1 \to 0$, by writing
				\begin{equation}
					X_j = Z_j\exp\left\{-\frac{2\la\tau \ra\lambda_j}{\alpha_2X_1}\right\}
				\end{equation}
				with any positive number $Z_j$ for $j\geq 2$. This specific limit satisfies the relation, 
				\begin{equation}
					\lim_{X_1\to 0}\bm{X} = \bm{0}.
				\end{equation}
				Since Eq.~\eqref{eq:Gen_Kmem_woInhibitory_trans1} should hold for any path taken to reach the limit $\bm{X}\to 0$, we obtain the relation even for the specific limit
				\begin{equation}
					\lim_{X_1\to 0} \mcH (Z_2,\dots,Z_K) = \mbox{const.}
				\end{equation}
				for any positive $\{Z_j\}_{j=2,\dots K}$, implying that $\mcH$ is a constant function. We thus obtain 
				\begin{equation}
					|\tl{S}(\bm{z})| = \exp \Psi (\bm{X}) \simeq \tl{A}s^{a}, \>\>\> a:= \frac{2\nu_1\la \tau\ra}{\alpha_2} 
				\end{equation}
				with some positive number $\tl{A}$. 
				
				\vskip 0.5cm
				
				\paragraph{Case with negative $a<0$.}
				When $\nu_1$ is negative, we have
				\begin{equation}
					\tl{P}_{\mrss}(s) \simeq As^{a} + o(s^{a})
				\end{equation}
				for small $s$ with some constant $A$ satisfying $|A|=\tl{A}$ and negative value $a<0$. By applying the inverse Laplace transform (see Appendix~\ref{sec:app:Laplace_0}), we obtain the power law asymptotic form~\eqref{eq:power-law_gen_memory_K_one-sidedMark_ramp}. In this case, the sign of $A$ is determined to be positive (i.e., $A=\tl{A}$) for the consistency with the probability interpretation. 
				
				\vskip 0.5cm
				\paragraph{Case with $0<a<1$.}
				This case is equivalent to $m:=\lfloor a \rfloor = 0$. We obtain
				\begin{equation}
					\tl{P}_{\mrss}(s) \simeq 1 - As^{a} + o(s^{a})
				\end{equation}
				with some constant $A$ for small $s$. Assuming that $\tl{A}$ is a positive real number, we obtain the power law asymptotic form~\eqref{eq:power-law_gen_memory_K_one-sidedMark_ramp} (see Appendix~\ref{sec:app:Laplace_1}). 

				\vskip 0.5cm
				\paragraph{Case with positive non-integer $a$.}
				Let us define $m:= \lfloor a \rfloor$ in order to classify the solutions. Since the asymptotic series of the Laplace transformation is given by
				\begin{equation}
					\tl{P}_{\mrss}(s) \simeq As^{a} + \sum_{k=0}^{m}c_ks^{k} + o(s^{a}),
				\end{equation}
				we obtain the power law asymptotic form~\eqref{eq:power-law_gen_memory_K_one-sidedMark_ramp} (see Appendix~\ref{sec:app:Laplace_2}), by setting $A$ to a positive (negative) number for even (odd) $m$, for consistency with the probability interpretation. 

				\vskip 0.5cm
				\paragraph{Case with positive integer $a$.}
				Technically, the positive integer case requires a special treatment since the gamma function in the Laplace transformation formula~\eqref{eq:app:Laplace_gen} diverges: $\Gamma(-a)= \infty$. However, since the power law asymptotics~\eqref{eq:power-law_gen_memory_K_one-sidedMark_ramp} is valid for any non-integer $a$, it is straightforward to obtain the power law asymptotics~\eqref{eq:power-law_gen_memory_K_one-sidedMark_ramp} for positive integer $a$, assuming that the power law exponent $a$ is a continuous function in terms of $\nu_1$:
				\begin{equation}
					a(\nu_1) = \lim_{x\to \nu_1}a(x).\label{eq:continuity_power-law-exponent}
				\end{equation}
				While we have numerically checked the validity of this result~\eqref{eq:power-law_gen_memory_K_one-sidedMark_ramp} for some specific cases (see Sec.~\ref{sec:RampIntensity_gen_numerical} for the numerical results), a rigorous proof of the continuity assumption~\eqref{eq:continuity_power-law-exponent} is beyond the scope of this paper, as it requires further technical investigation while the continuity assumption~\eqref{eq:continuity_power-law-exponent} is physically reasonable.
				
				In summary, we obtain the power law asymptotics~\eqref{eq:power-law_gen_memory_K_one-sidedMark_ramp} for general $a$.

	\subsection{General memory kernel}
		Since the power law asymptotics~\eqref{eq:power-law_gen_memory_K_one-sidedMark_ramp} holds for general discrete sums of exponentials, as a straightforward generalisation, we find a power law asymptotics at the critical point $n=1$ for general memory kernel $h(t)$, such that 
		\begin{align}
			h(t) = \int_0^\infty dx\tilh(x) e^{-t/x}, \>\>\> & \lambda = g(\nu) \simeq \nu - \nu_1 + o(\nu^0) \> \mbox{ for large }\nu, \>\>\> \rho(y)=0 \>  \mbox{ for negative }y, \>\>\> \int_0^\infty y\rho(y)dy = 1\notag\\
			\Longrightarrow \>\>\>
			&P_{\mrss}(\lambda) \propto \lambda^{-1-a}, \>\>\> a:= \frac{2\nu_1\la\tau \ra}{\alpha_2}, \>\>\> \la \tau\ra:= \int_0^\infty x^2~\tilh(x)dx
			\label{eq:power-law_gen_memory_one-sidedMark_ramp}
		\end{align}
		for either negative or non-negative $\nu_1$. This relation is a true power law for positive $\nu_1$ (i.e. normalisable even without cutoff), while it is an intermediate asymptotics for non-positive $\nu_1$ (i.e., not normalisable without cutoff). 

	\subsection{Numerical confirmation}\label{sec:RampIntensity_gen_numerical}
		\begin{figure}
			\centering
			\includegraphics[width=170mm]{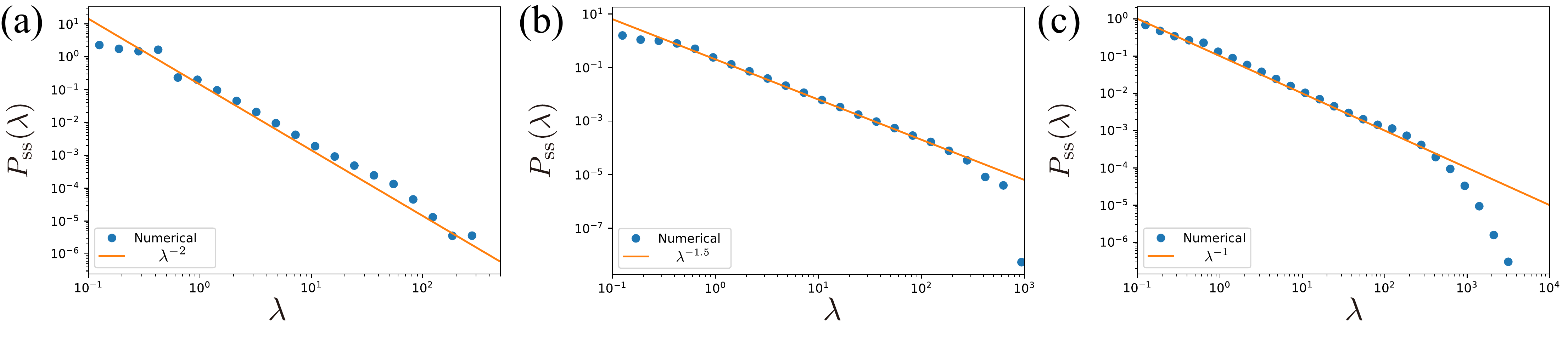}
			\caption{
						Numerical confirmation of our theoretical prediction on the power law exponents~\eqref{eq:power-law_gen_memory_K_one-sidedMark_ramp}.
						(a)~Simulation based on $K=2$, $(\tau_1, \tau_2)=(1, 2)$, $(\tilh_1,\tilh_2)=(0.7,0.14995)$, $\eta=0.9999$, $\nu_0=0.01$, and $\nu_1\simeq 0.385$, predicting $a\simeq 1.0$ (i.e., a true power law; Zipf's law). 
						(b)~Simulation based on $K=3$, $(\tau_1, \tau_2, \tau_3)=(1, 2, 3)$, $(\tilh_1,\tilh_2,\tilh_3)=(0.5, 0.15, 0.1999/3)$, $\eta=0.9999$, $\nu_0=0.01$, and $\nu_1\simeq 0.147$, predicting $a\simeq 0.5$ (i.e., a true power law). 
						(c)~Simulation based on $K=3$, $(\tau_1, \tau_2, \tau_3)=(1, 2, 3)$, $(\tilh_1,\tilh_2,\tilh_3)=(0.5, 0.15, 0.1999/3)$, $\eta=0.9999$, $\nu_0=0.01$, and $\nu_1=0$, predicting $a=0$ (i.e., an intermediate power law asymptotic). 
					}
			\label{fig:numerical_ramp}
		\end{figure}
		Figure~\ref{fig:numerical_ramp} shows the numerical results based on the Monte Carlo simulation of the SDE~\eqref{def:NL_Hawkes_gen} obtained for a memory function constructed as a discrete sum of exponentials, for the ramp intensity \eqref{eq:ramp_tension-intensity_map}, and a mark distribution reducing to the Dirac function centred on $y=1$:
		\begin{equation}
			h(t) = \sum_{k=1}^K\tilh_k e^{-t/\tau_k}, \>\>\> \lambda = g(\nu) = \max\{\nu_0, \nu-\nu_1\}, \>\>\> \rho(y)=\delta (y-1)~.
		\end{equation}
		The simulations are performed by using an adaptive time discretisation scheme (see Appendix~\ref{sec:app:numerical_implementation} for the detailed numerical scheme). All panels in Fig.~\ref{fig:numerical_ramp} exhibit the predicted power law tail of the intensity distribution, in excellent agreement with our theoretical prediction~\eqref{eq:power-law_gen_memory_K_one-sidedMark_ramp}. Notably, the power law exponents varies continuously as a function of $\nu_1$ and the power law formula~\eqref{eq:power-law_gen_memory_K_one-sidedMark_ramp} is found to be valid even for integer exponents such as $a=0$ and $a=1$.  

\section{Solution 5: general memory kernel for fast-accelerating intensity maps and two-sided mark distribution with nonpositive mean mark}\label{sec:sol4_arbitrary_memory}
		In Secs.~\ref{sec:inhibitory_effects} and \ref{sec:inhibitory_effects_AS}, we have shown that a general asymptotic formula is available for the exponential memory and the two-sided mark distributions with non-positive mean. Here we show that, by solving the corresponding MEs, the asymptotic formula is valid for a wider class of memory kernels with FAI.
				
		\subsection{Discrete sum of exponentials}\label{sec:MSA_gen_discrete_der}
			We first show that the power law tail of the PDF of intensities is robust for various memory kernel $h(t)$ for the MSA intensity function in the presence of a two-sided mark distribution with non-positive mean mark. Specifically, we make the following assumptions (i.e., discrete-sum of exponentials, MSA intensity, two-sided mark distribution with nonpositive mean mark)
			\begin{equation}
					h(t) = \sum_{k=1}^K \tilh_k e^{-t/\tau_k}, \>\>\> 
					g(\nu) = \lambda_0 e^{\beta \nu}, \>\>\> 
					p_+ := \int_{0}^\infty \rho(y)dy > 0,\>\>\> 
					p_- := \int_{-\infty}^0 \rho(y)dy > 0, \>\>\> 
					m := \int_{-\infty}^\infty y\rho(y)dy \leq 0.
			\end{equation}
			Under these conditions, we obtain the power law intensity PDF
			\begin{equation}
				P_{\mrss}(\lambda) \propto \lambda^{-2-\beta^{-1} u} \>\>\> \mbox{ for large }\lambda, \>\>\>
				 u:= \frac{c^*}{h(0)},
				\label{eq:MSA_gen_intensity_Zipf}
			\end{equation}
			where $c^*$ is the positive root of $\Phi(x)$ for $m<0$ (or $c^*=0$ for $m=0$). Remarkably, we recover Zipf's law exactly for the zero mean mark case $m=0$.

			\subsubsection*{Derivation}
				From Eq.~\eqref{eq:master_eq_discrete_K}, the steady-state ME is given by
				\begin{equation}
					\label{eq:MSA_master_transient_der_genK}
					\sum_{k=1}^K\frac{1}{\tau_k \lambda_0}\frac{\partial}{\partial z_k}\left( z_ke^{-\sum_{k'=1}^Kz_{k'}} \phi(\bm{z}) \right) + \int dy \rho(y)\phi\left(\bm{z}-y\tilbh\right) - \phi(\bm{z}) = 0,
				\end{equation}
				where we have defined $\phi(\bm{z}):= G(\bm{z})P_{\mrss}(\bm{z})$. For large $\bm{z}$, the first term in the r.h.s. is negligible due to the exponential factor, leading to
				\begin{equation}
					\label{eq:integral_eq_MSA_steady_genK}
					\int dy \rho(y)\phi\left(\bm{z}-y\tilbh\right) - \phi(\bm{z}) \simeq 0 \>\>\> \mbox{for large } \bm{z}. 
				\end{equation}
				We then apply the transformation from $\bm{z} = (z_1,\dots, z_K)$ to $\bm{Z}:= (W, Z_2,\dots, Z_{K})$: 
				\begin{equation}
					z_1 = \tilh_1 W, \>\>\> 
					z_2 = \tilh_2 W + Z_2, \>\>\> 
					z_3 = \tilh_3 W + Z_3, \>\>\> 
					\>\>\> \dots, \>\>\>
					z_K = \tilh_K W + Z_K.
					\label{eq:genMem_FAIHawkes_var_trans_1}
				\end{equation}
				Using this variable set, we can rewrite
				\begin{equation}
					\psi(W-y;  Z_2, \dots, Z_K) := \phi(\bm{z}-y\tilbh) = \phi\left(\tilh_1(W-y), \tilh_2(W-y) + Z_2, \dots,  \tilh_K(W-y) + Z_K\right). 
				\end{equation}
				The integral equation~\eqref{eq:integral_eq_MSA_steady_genK} is then reduced to
				\begin{equation}
					\label{eq:integral_eq_MSA_steady_genK_2}
					\int dy \rho(y)\psi\left(W-y; \bm{Z}'\right) - \psi(W; \bm{Z}') \simeq 0 \>\>\> \mbox{for large } \bm{Z}
				\end{equation}
				with $\bm{Z}':= (Z_2, \dots, Z_K)$. This variable transformation is useful because Eq.~\eqref{eq:integral_eq_MSA_steady_genK} is an effectively one-dimensional integral equation. Since the variable subset $Z'$ is irrelevant in this integral equation, its solution is given by
				\begin{equation}
					\psi (W; \bm{Z}') = C_0(\bm{Z}')e^{-c^* W} + C_1(\bm{Z}')
					\label{eq:genMem_FAIHawkes_solutionAnzatz}
				\end{equation}
				with arbitrary nonnegative functions $C_0(\bm{Z}')$ and $C_1(\bm{Z}')$ without the variable $W$ (see Appendix~\ref{sec:app:sol_integralSolution}). In addition, by defining the moment-generating function $\Phi(x):=\int_{-\infty}^\infty dy\rho(y)(e^{xy}-1)$, the constant $c^*$ is the positive root of $\Phi(c^*)=0$ for the case of negative mean mark $m<0$ or $c^*=0$ for the case of zero mean mark $m=0$ (see Appendix~\ref{sec:app:phi_symmetric} for the detailed properties of $\Phi(x)$). Assuming the natural boundary condition, $C_1(\bm{Z}')$ must be set zero as shown later. We then derive the steady distribution $P_{\mrss}(\nu)$ as
				\begin{align}
					P_{\mrss}(\nu) &:= \int_{-\infty}^\infty d\bm{z} P_{\mrss}(\bm{z})\delta \left(\nu - \sum_{k=1}^K z_k\right)\notag\\
					&\simeq \frac{1}{\lambda_0}\int_{-\infty}^\infty d\bm{z} e^{-\beta\sum_{k=1}^Kz_k} C_0(\bm{Z}')e^{-c^* W}\delta \left(\nu - \sum_{k=1}^K z_k\right) \notag\\
					&= \frac{e^{-\beta \nu}}{\lambda_0}\int_{-\infty}^\infty dz_1\int_{-\infty}^\infty \left(\prod_{j=2}^K dz_j\right)  C_0\left(z_2- \frac{\tilh_2}{\tilh_1}z_1, \dots, z_K- \frac{\tilh_K}{\tilh_1}z_1\right)\exp\left(-\frac{c^*}{\tilh_1}z_1\right)\delta \left(\nu - \sum_{k=1}^K z_k\right)
				\end{align}
				Applying the transformation 
				\begin{equation}
					z'_j:=z_j - \frac{\tilh_j}{\tilh_1}z_1~, ~~~{\rm for} ~j=2,\dots, K~,
				\end{equation}
				we obtain 
				\begin{align}
					P_{\mrss}(\nu) 
					&\simeq \frac{e^{-\beta \nu}}{\lambda_0}\int_{-\infty}^\infty dz_1\int_{-\infty}^\infty \left(\prod_{j=2}^K dz'_j\right)  C_0\left(z'_2, \dots, z'_K\right)\exp\left(-\frac{c^*}{\tilh_1}z_1\right)\delta \left(\nu - rz_1 - \sum_{k=2}^K z'_k\right) \notag \\
					&= \frac{e^{-\beta \nu}}{\lambda_0}\int_{-\infty}^\infty 
					\left(\prod_{j=2}^K dz'_j\right) C_0\left(z'_2, \dots, z'_K\right) \int_{-\infty}^\infty dz_1 \exp\left(-\frac{c^*}{\tilh_1}z_1\right)\delta \left(\nu - rz_1 - \sum_{k=2}^K z'_k\right) \notag \\
					&= \frac{e^{-(\beta-\eu) \nu}}{r \lambda_0}\int_{-\infty}^\infty \left(\prod_{j=2}^K dz'_j\right) C_0\left(z'_2, \dots, z'_K\right)\exp\left(\frac{c^*}{h(0)}\sum_{k=2}^K z'_k\right),
				\end{align}
				where we have used 
				\begin{equation}
					\delta\left(\nu - rz_1 - \sum_{k=2}^K z'_k\right) = \frac{1}{r}\delta\left(z_1 - \frac{\nu-\sum_{k=2}^K z'_k}{r}\right)
				\end{equation}
				with
				\begin{equation}
					r:= \frac{1}{\tilh_1}\sum_{k=1}^K \tilh_k = \frac{h(0)}{\tilh_1}, \>\>\> 
					\eu:=	\frac{c^*}{h(0)}.
				\end{equation}
				Assuming that 
				\begin{equation}
					\frac{1}{r}\int_{-\infty}^\infty \left(\prod_{j=2}^K dz'_j\right) C_0\left(z'_2, \dots, z'_K\right)\exp\left(\frac{c^*}{h(0)}\sum_{k=2}^K z'_k\right) < \infty,
				\end{equation}
				we find that the asymptotic PDF for large $\nu$ is given by
				\begin{equation}
					P_{\mrss}(\nu) \propto e^{-(\beta- u) \nu} \>\>\> \mbox{for large }\nu. 
				\end{equation}
				This asymptotic form implies the power law~\eqref{eq:MSA_gen_intensity_Zipf} for the intensity $\lambda:= g(\nu)$. 
				\subsubsection*{Natural boundary condition}
					By neglecting the first term in the ME~\eqref{eq:master_eq_discrete_K} and by applying the variable transformation~\eqref{eq:genMem_FAIHawkes_var_trans_1}, we obtain an approximate ME
					\begin{equation}
						\frac{\pd P_t(W;\bm{Z}')}{\pd t} \simeq \int_{-\infty}^\infty dy\left\{G(W-y;\bm{Z}')P_t(W-y;\bm{Z}')- G(W;\bm{Z}')P_t(W;\bm{Z}')\right\}. 
					\end{equation}
					By applying the KM expansion, we obtain the conservation of probability: 
					\begin{equation}
						\frac{\pd P_t(W;\bm{Z}')}{\pd t} \simeq -\frac{\pd J_t(W;\bm{Z}')}{\pd W}
					\end{equation}
					with the probability current 
					\begin{equation}
						J_t(W;\bm{Z}'):= \sum_{n=1}^\infty \frac{(-1)^{n-1}\alpha_n}{n!}\frac{\pd^{n-1}}{\pd W^{n-1}}G(W;\bm{Z}')P_t(W;\bm{Z}').
					\end{equation}
					By substituting the solution~\eqref{eq:genMem_FAIHawkes_solutionAnzatz}, we obtain 
					\begin{equation}
						J_{\mrss}(W;\bm{Z}') = \sum_{n=1}^\infty \frac{(-1)^{n-1}\alpha_n}{n!}\frac{\pd^{n-1}}{\pd W^{n-1}}\left(C_1(\bm{Z}')+e^{-c^*W}C_0(\bm{Z}')\right)
						= m C_1(\bm{Z}') + \frac{C_0(\bm{Z}')}{c^*}\Phi(c^*)e^{-c^*W},
					\end{equation}
					where we have used $\Phi(x)=\sum_{n=1}^\infty (\alpha_n/n!)x^n$. Since $\Phi(c^*)=0$ by definition, we obtain $J_{\mrss}(W;\bm{Z}')=m C_1(\bm{Z}')$. The natural boundary condition requires 
					\begin{equation}
						\lim_{W\to \infty}J_{\mrss}(W;\bm{Z}')=0
					\end{equation}
					for any $\bm{Z}'$, implying $C_1(\bm{Z}')=0$. 
			
	\subsection{General memory kernel}
		As done before, any memory kernel can be approximated by a sum of exponentials, such that 
		\begin{equation}\label{eq:MSA_approx_genMemToConMem}
			h(t) = \int_0^\infty dx \tilh(x)e^{-t/x} \approx \sum_{k=1}^K \tilh_k e^{-t/\tau_k}. 
		\end{equation}
		Since the power law tail for the PDF of the intensity is found for any discrete sum of exponentials, it remains valid for general superpositions of exponential memory. Under the assumption 
		\begin{equation}
			\forall ~h(t), \>\>\> 
			g(\nu) = \lambda_0 e^{\beta \nu}, \>\>\> 
			p_+ := \int_{0}^\infty \rho(y)dy > 0,\>\>\> 
			p_- := \int_{-\infty}^0 \rho(y)dy > 0, \>\>\> 
			m := \int_{-\infty}^\infty y\rho(y)dy \leq 0.
		\end{equation}
		we obtain 
		\begin{equation}
			P_{\mrss}(\lambda) \propto \lambda^{-2-\beta^{-1}u} \>\>\> \mbox{ for large }\lambda, \>\>\> 
			u:= \frac{c^*}{h(0)},
		\end{equation}
		where $c^*$ is the positive root of $\Phi(c^*)=0$ for $m<0$ or $c^*=0$ for $m=0$. For the zero mean mark case $m=0$, the PDF obeys Zipf's law exactly. 
		
	\subsection{Numerical confirmation}
		\begin{figure}
			\centering
			\includegraphics[width=150mm]{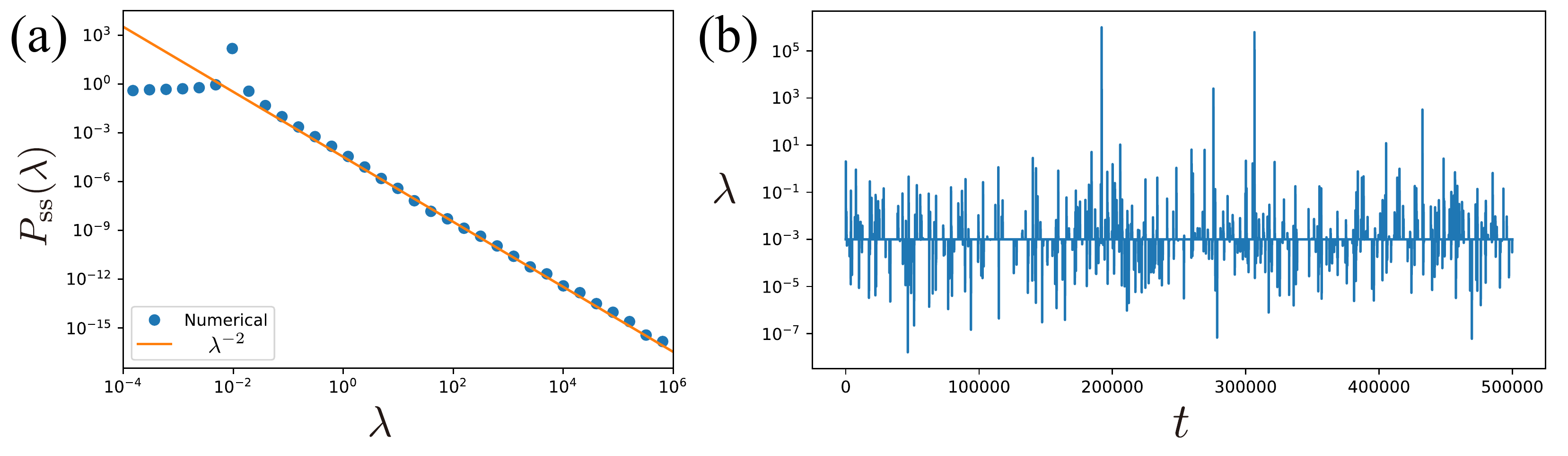}
			\caption{
						(a)~Numerical confirmation of our theoretical prediction~\eqref{eq:MSA_gen_intensity_Zipf} of the Zipf law for the intensity distribution.
						This simulation corresponds to the following set of parameters: $K=3$, $(\tau_1,\tau_2,\tau_3)=(1,0.5,2)$, $(\tilh_1,\tilh_2,\tilh_3)=(0.5,0.6,0.1)$, $\lambda_0=0.01$, $\beta=6$, $\lambda_{\max}=10^6$, and $\sigma=0.3$. The predicted power law exponent is given by $a=1.0$ (i.e., a true power law; the Zipf law) and is shown as the red straight line.
						(b)~Sample trajectory of the intensity for $K=3$, $(\tau_1, \tau_2, \tau_3)=(1, 0.5, 2)$, $(\tilh_1,\tilh_2,\tilh_3)=(0.5,0.6,0.1)$, $\lambda_0=0.001$, $\beta=10$, $\lambda_{\max}=10^6$, and $\sigma=0.3$. Note that the horizontal reference line is determined by $\lambda_0=0.001$.
					}
			\label{fig_numerical_MSA}
		\end{figure}	
		
		Figure~\ref{fig_numerical_MSA} shows numerical results obtained by Monte Carlo simulations of the SDE~\eqref{def:NL_Hawkes_gen} for the memory function made of a discrete sum of exponential functions, for the exponential intensity MSA with finite cutoff (to ensure convergence of the numerical scheme), and a zero-mean Gaussian mark distribution 
		\begin{equation}
			h(t) = \sum_{k=1}^K\tilh_k e^{-t/\tau_k}, \>\>\> \lambda = g(\nu) = \min\left \{ \lambda_0e^{\beta \nu}, \lambda_{\max}\right\}, \>\>\> \rho(y)=\frac{1}{\sqrt{2\pi \sigma^2}}e^{-\frac{y^2}{2\sigma^2}} ~.
		\end{equation}
		We use an adaptive time discretisation scheme (see Appendix~\ref{sec:app:numerical_implementation} for the detailed numerical scheme). Here $\lambda_{\max}$ is a cutoff parameter to control numerical rounding errors. Figure~\ref{fig_numerical_MSA}a exhibits the Zipf law in the intensity distribution up to the cutoff $\lambda_{\max}$, showing agreement with our theoretical prediction~\eqref{eq:MSA_gen_intensity_Zipf}. 

		Figure~\ref{fig_numerical_MSA}b shows a sample trajectory of the intensity obtained for the parameter set $K=3$, $(\tau_1, \tau_2, \tau_3)=(1, 0.5, 2)$, $(\tilh_1,\tilh_2,\tilh_3)=(0.5,0.6,0.1)$, $\lambda_0=0.001$, $\beta=10$, $\lambda_{\max}=10^6$, and $\sigma=0.3$. 	This semi-log plot illustrates that the NLHawkes model exhibits an intermittent behaviour in terms of its intensity, which is qualitatively consistent with observed phenomena in various complex systems, such as seismic activity.

	\subsection{Generalisation to fast-accelerating intensity maps}
		The above framework can be readily generalised to FAI maps defined by $g(\nu) \gg \nu^2$. Our general result can be formulated as follows. Under the assumptions
		\begin{subequations}
			\label{eq:RII_gen_intensity_power-law}
		\begin{equation}
			h(t) = \sum_{k=1}^K \tilh_k e^{-t/\tau_k}, \>\>\> 
			g(\nu) \gg \nu^2 \>\>\> \mbox{for large $\nu$}, \>\>\> 
			p_+ := \int_{0}^\infty \rho(y)dy > 0,\>\>\> 
			p_- := \int_{-\infty}^0 \rho(y)dy > 0, \>\>\> 
			m := \int_{-\infty}^\infty y\rho(y)dy \leq 0,
		\end{equation}
		we obtain 
		\begin{equation}
			P_{\mrss}(\lambda) \propto \lambda^{-1}\left[e^{-u\nu}\left\{\frac{dg(\nu)}{d\nu}\right\}^{-1}\right]_{\nu=g^{-1}(\lambda)} \>\>\> \mbox{for large }\lambda, \>\>\> 
			u := \frac{c^*}{h(0)}
		\end{equation}
		\end{subequations}
		with $c^*$ being the positive root of $\Phi(c^*)=0$ for $m<0$ or zero for $m=0$, where $\Phi(x):=\int_{-\infty}^\infty dy\rho(y)(e^{xy}-1)$. The derivation of this result is essentially the same as that in Sec.~\ref{sec:MSA_gen_discrete_der}, by replacing the MSA intensity map with the general FAI map. 

		Since any memory kernel can be approximated by a discrete sum of exponentials (see Eq.~\eqref{eq:MSA_approx_genMemToConMem}),  the continuous version of the statement~\eqref{eq:RII_gen_intensity_power-law} also holds: 
		\begin{align}
			\forall ~h(t), \>\>\> 
			g(\nu) \gg \nu^2, \>\>\> 
			p_+ > 0,\>\>\> 
			p_- > 0, \>\>\> 
			m \leq 0 \>\>\> \Longrightarrow \>\>\>
			P_{\mrss}(\lambda) \propto \lambda^{-1}\left[e^{-u\nu}\left\{\frac{dg(\nu)}{d\nu}\right\}^{-1}\right]_{\nu=g^{-1}(\lambda)}, \>\>\> 
			u:= \frac{c^*}{h(0)}
		\end{align}
		with $c^*$ being the positive root of $\Phi(c^*)=0$ for $m<0$ or zero for $m=0$, where $\Phi(x):=\int_{-\infty}^\infty dy\rho(y)(e^{xy}-1)$. Remarkably, this result implies that the power law tail for the steady-state PDF of intensities holds robustly for super-polynomial intensity maps, such as $g(\nu)\propto e^{\beta \nu}$ (i.e., the MSA case) and $g(\nu)\propto e^{\beta \nu^2}$.

\section{Discussion}\label{sec:discussion}
	\subsection{Relationship to nonlinear Kesten processes}
		We have shown that power law asymptotics robustly appears for the quadratic and FAI cases. Here we provide another derivation based on more heuristic arguments, by removing inessential technicalities, using the viewpoint of Kesten processes \cite{Kesten1973}. Let us focus on the case with exponential memory $h(t)=(\eta/\tau)e^{-t/\tau}$ and in the diffusive limit (i.e., for the symmetric mark distribution $\rho(y)=\rho(-y)$) described by the FPE~\eqref{eq:master_eq_diffusive} and the corresponding SDE~\eqref{eq:SDE_diffusive-expon}. 
	
		\subsubsection{Case with quadratic intensity map: $\tl{g}(\nu) =k \nu^2 +\nu_0$}
			Inserting $\tl{g}(\nu) = k \nu^2 +\nu_0$ in the SDE~\eqref{eq:SDE_diffusive-expon}, we obtain
			\begin{equation}
				\frac{d\hnu}{dt} = -\frac{\hnu}{\tau} + \sqrt{k \nu^2 +\nu_0} \sqrt{2D} \cdot \hxi^{\mrG}_t~.
				\label{wrth2bhg}
			\end{equation}
			Here, $\hxi^{\mathrm{G}}_t$ is the standard white Gaussian noise satisfying 
			$\la\hxi^{\mathrm{G}}\ra=0$ and $\la \hxi^{\mathrm{G}}(t)\hxi^{\mathrm{G}}(t')\ra=\delta(t-t')$. 
			With respect to the power law structure of the tail of the PDF of $\nu$,  by using the discretisation $d\nu/dt = (\nu(t+dt) - \nu(t))/dt$, this SDE can be regarded as a continuous version of the discrete-time Kesten process \cite{Kesten1973}
			\begin{equation}
				\nu(t+dt) = a_t \nu(t) + b_t  ~,~~~{\rm with}~a_t = 1 + (2Dk)^{1/2} \hxi^{\mrG}_t dt - (dt/\tau)  ~~{\rm and}~~   b_t =\nu_0  (2k)^{1/2} \hxi^{\mrG}_t dt~.
				\label{h2btbg1bq}
			\end{equation}
			The first term $a_t \nu(t)$ controls the intermittent excursions of $\nu(t)$ to large values, for which $\nu_0$ can be neglected in the last term of the r.h.s. of Eq.~\eqref{wrth2bhg}. The second term $b_t$ in the Kesten map \eqref{h2btbg1bq} is the ``reinjecting term'' obtained when $\nu$ becomes smaller than $\nu_0$. As shown in Ref.~\cite{SornetteCont1997_Kesten}, the detailed shape of this ``reinjecting term'' has no impact on the existence of a power law tail and on the value of its exponent. The only important point is that the ``reinjecting term'' exists to prevent $\nu$ from being too small. Remaining no less than a stochastic variable proportional to $\nu_0$, intermittent runs of exponential growth occur when there is a succession of positive realisations of $\hxi^{\mrG}_t$ for several consecutive times such that the multiplicative factor $a_t$ is larger than $1$ over this run \cite{SornetteCont1997_Kesten,SornettePhysA1998_Kesten}.
			
			The condition for the existence of a steady-state PDF for the Kesten process \eqref{h2btbg1bq} is that $\langle \ln a_t \rangle  <0$ \cite{Kesten1973,SornetteCont1997_Kesten}. For infinitesimal $dt$, $\ln a_t$ can be expanded as $\ln a_t  = (2Dk)^{1/2} \hxi^{\mrG}_t dt - (dt/\tau)$ and its mean is then $\langle \ln a_t \rangle  = - (dt/\tau)$ since $\langle \hxi^{\mrG}_t \rangle =0$ by definition. Hence, the condition for a stationarity process holds true. It is then easy to show by explicitly writing the self-consistent equation for the steady-state PDF of $\nu$ that it is a power law with exponent $a$ given as the solution of the equation
			\begin{equation}
				\langle |a_t|^a \rangle =1~.
				\label{3j3unhbgqgb}
			\end{equation}
			Using the fact that $\hxi^{\mrG}_t dt = dW$ is a Gaussian random variable with zero mean and variance $dt$ (i.e. it is the infinitesimal increment of the Wiener process), the average in \eqref{3j3unhbgqgb} is obtained by using the saddle-node approximation, and we find that the corresponding solution recovers exactly expression \eqref{eq:quadratic_tensiton_intensity_map}, namely $a:= 1/2+c^2/(2k\tau)$ with $c := \tau/\eta$. This confirms that our treatment in terms of the diffusive limit gives equations in the general class of Kesten processes. Other forms of the tension-intensity map $\lambda=g(\nu)$ can thus be interpreted as continuous nonlinear extension of the Kesten process.
			
 			Therefore, from an intuitive point of view, the power law tails of the PDFs of $\nu$ and $\lambda$ can be qualitatively related to an underlying multiplicative structure together with additional ingredients to ensure the existence and stationarity of the process.

		\subsubsection{Case with fast-accelerating intensity map: $\tl{g}(\nu) \gg \nu^2$}
			We here consider the case of FAI maps satisfying
			\begin{equation}
				\tl{g}(\nu) \gg \nu^2 \mbox{ (for positive large } \nu), \>\>\> \tl{g}(\nu) \simeq \mbox{const.} \mbox{ (for negative large } \nu).
			\end{equation}
			Let us take $D=1/2$ to simplify notations, so that the corresponding SDE is given by
			\begin{equation}
				\label{eq:discuss_RII_Kesten}
				\frac{d\hnu}{dt} = -\frac{\hnu}{\tau} + \sqrt{\tl{g}(\hnu)} \cdot \hxi^{\mrG}. 
			\end{equation}
			Since $\tl{g}(\nu)\simeq \mbox{const.}$ for negative large $\nu$, $\hnu$ cannot go to $-\infty$ due to the relaxation term $-\hnu/\tau$. On the other hand, for positive large $\hnu$, the dynamics is approximated by
			\begin{equation}
				\frac{d\hnu}{dt} \simeq \sqrt{\tl{g}(\hnu)} \cdot \hxi^{\mrG},
			\end{equation}
			because $\sqrt{\tl{g}(\hnu)} \gg \hnu/\tau$ for FAI maps. 

			This model is thus similar to a Brownian motion with a position dependent variance or diffusion coefficient. Interestingly, such a Brownian model has a well-defined steady-state PDF for FAI maps. Let us thus consider a Brownian motion obeying the following SDE
			\begin{equation}\label{eq:discuss_Brownian_positionDependence}
				\frac{d\hnu}{dt} = \sqrt{\tl{g}(\hnu)} \cdot \hxi^{\mrG} \mbox{ (for }\nu \geq 0)~,
			\end{equation}
			which is complemented by the condition of a repulsive hard wall at $\hnu=0$, which prohibits the ``Brownian particle" from going to $-\infty$. This condition is a simplification to and ensures a similar result as in the presence of the relaxation effect in the original model~\eqref{eq:discuss_RII_Kesten}. The steady FPE is given by
			\begin{equation}
				\frac{d^2}{d\nu^2}\{ \tl{g}(\nu)P_{\mrss}(\nu) \} = 0 \mbox{ (for }\nu > 0), \>\>\> P_{\mrss}(\nu) = 0 \mbox{ (for }\nu < 0). 
				\label{nbhg2wvf}
			\end{equation}
			If $\tl{g}(\nu)$ were not a FAI map, this steady FPE might not have a normalisable steady solution. For example, if $\tl{g}(\nu)=\mbox{const.}$, the general solution of the FPE is given by a non-normalisable steady solution $\tl{P}_{\mrss}(\nu)=c_0 + c_1\nu$, satisfying $\int_0^\infty d\nu P_{\mrss}(\nu) = \infty$. This model is therefore non-stationary. In contrast, when $\tl{g}(\nu)$ is a FAI map, the FPE (\ref{nbhg2wvf}) has a normalisable steady solution for any FAI map, with
			\begin{equation}
				P_{\mrss}(\nu) \propto \begin{cases}
					\frac{1}{\tl{g}(\nu)} & \mbox{ (for }\nu > 0)\\
					0 & \mbox{ (for }\nu < 0) 
				\end{cases}
				\>\>\> \Longrightarrow \>\>\>
				\int_{-\infty}^\infty d\nu P_{\mrss}(\nu) < \infty. 
			\end{equation}
			We then obtain the robust expression of the steady-state PDF of the intensity $\lambda := \tl{g}(\nu)$: 
			\begin{equation}
				P_{\mrss}(\lambda) \propto \lambda^{-1}\left| \frac{d\tl{g}(\nu)}{d\nu}\right|_{\nu=\tl{g}^{-1}(\lambda)}^{-1} \mbox{ (for large }\lambda). 
			\end{equation}
			This result readily implies that the Zipf law 
			\begin{equation}
				\label{eq:discuss_MSA_Kesten_Zipf}
				P_{\mrss}(\lambda) \propto \lambda^{-2} \mbox{ (for large }\lambda)
			\end{equation}
			is observed for a wide class of superpolynomial intensity maps, such as $\tl{g}(\nu)=e^{\beta \nu}$ and $\tl{g}(\nu)=e^{\beta \nu^2}$. 
			
			Let us complete this discussion by mentioning that the rigorous mathematical demonstration of the existence of steady-state solutions of the SDE \eqref{eq:discuss_Brownian_positionDependence} is obtained from the theorems presented in Ref.~\cite{bookCherny_sin05}. In particular, we refer to the theorems in section 5.2 in Ref.~\cite{bookCherny_sin05}.

		\paragraph{Numerical simulation.}
		
			\begin{figure}
				\centering
				\includegraphics[width=70mm]{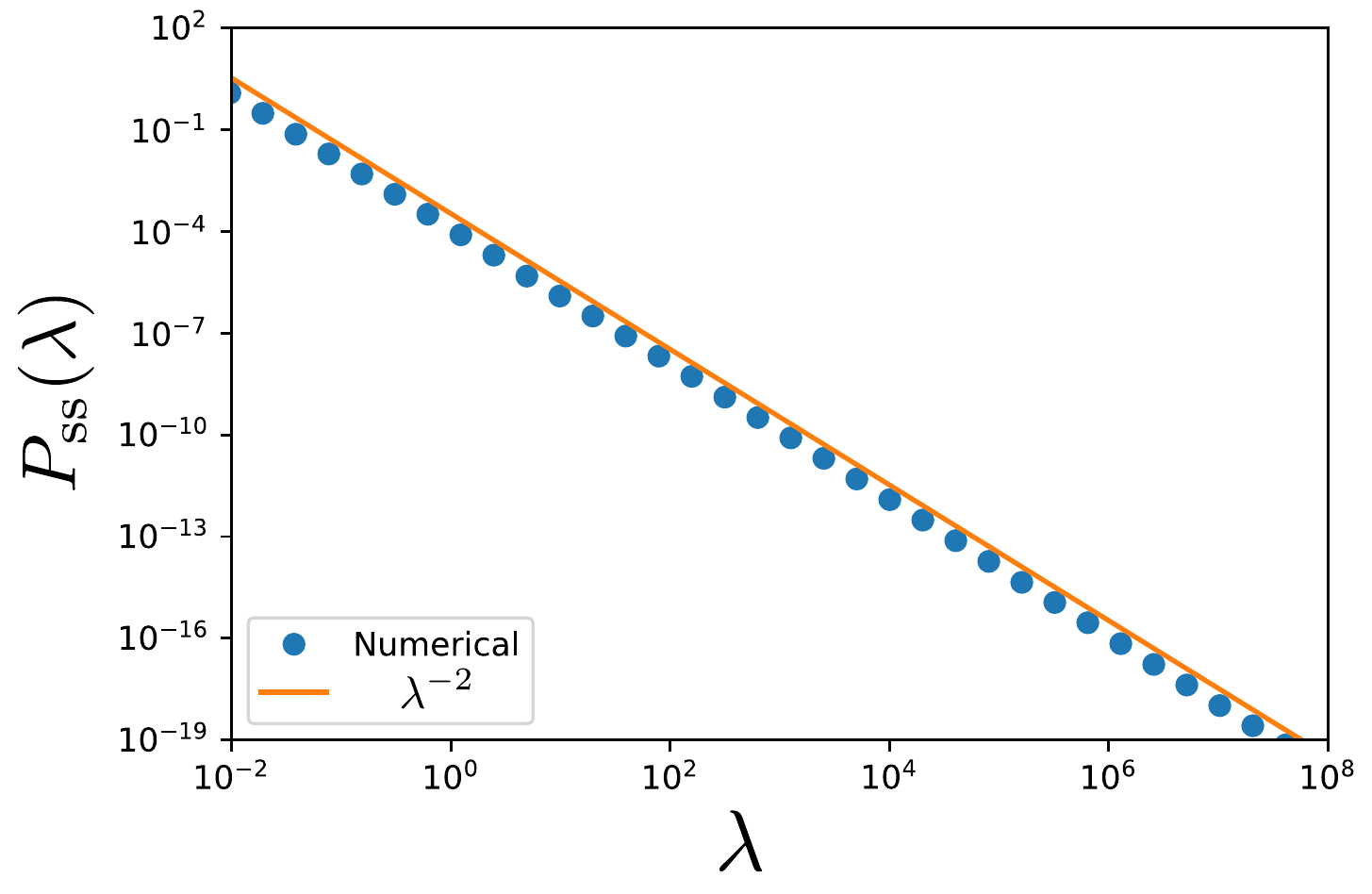}
				\caption{
							Numerical simulations confirming our theoretical prediction on the Zipf law~\eqref{eq:discuss_MSA_Kesten_Zipf} for the intensity distribution. The simulated SDE describes a Brownian motion with the $\hnu$-dependent diffusion constant $\tl{g}(\nu)/2$ and a reflecting barrier at $\hnu=0$. 
						}
				\label{fig_numerical_BrownianMSA}
			\end{figure}	
			
			Figure~\ref{fig_numerical_BrownianMSA} presents the PDF of the intensity obtained from the numerical solution of the SDE~\eqref{eq:discuss_Brownian_positionDependence} with an exponential intensity $\tl{g}(\nu)=\lambda_0e^{\beta \nu}$. The Zipf law~\eqref{eq:discuss_MSA_Kesten_Zipf} is obtained for the steady-state intensity distribution of the SDE describing a Brownian motion with $\hnu$-dependent diffusion constant $\tl{g}(\nu)/2$ in the presence of a reflecting barrier at $\hnu=0$. The detailed numerical implementation is described in Appendix~\ref{sec:app:numerical_implementation}. 

		\paragraph{Intuitive discussion.}
			Why is the Brownian model with position dependent variance stationary for FAI maps? This might be understandable from the viewpoint of step-size explosion for large $\hnu$. As $\hnu$ grows to very large values, the steps of the random walk explode even faster and thus it is very likely that a negative step occurs of huge size which brings back the Brownian particle to $0$ or even pushes it to negative values if the repelling boundary was absent. In the presence of the repelling boundary, the huge negative steps bring the Brownian particle close to $0$, for which the random step sizes become small, which implies that $\hnu$ remains quite a long time in the vicinity of the origin. Eventually, $\hnu$ escapes again to large values of $\hnu$ but then the huge random walk step sizes, when negative, bring it back again to a neighbourhood of $0$. This process occurs repeatedly and leads to a stationary PDF decaying rapidly as $1/\tl{g}(\hnu)$, due to the effect of the negative random steps that push back $\hnu$ to the left and the boundary somehow traps the process in its neighbourhood. In summary, this picture implies that the explosive step size leads to an effective strong ``trapping potential", which might be counter-intuitive at the first glance.

			It might be interesting to rephrase the above argument from the viewpoint of the recurrence time of one-dimensional Brownian motion. It is well-known that the recurrence probability of one-dimensional Brownian motion with constant variance $\tl{g}(\nu)=\mbox{const}.$ is unity, while the average recurrence time is infinity. In other words, a Brownian particle will surely come back to the origin after a long time, but this waiting time may be too large so that repeated recurrence events cannot be expected within a finite observation time. In the case of Eq.~\eqref{eq:discuss_Brownian_positionDependence}, the variance of the step lengths depends on the distance from the origin and becomes larger and larger for large $\nu$. Thus, the time evolution of this Brownian particle becomes faster and faster for large $\hnu$, such that the particle can come back to the origin much sooner than with a constant diffusion coefficient and repeated recurrence events can be expected in a finite time. 

			Last, it is useful to stress the difference between the mechanism underlying the existence of a steady-state power law distribution in the conventional linear Kesten process \cite{Kesten1973} and that of our FAI Hawkes model. In usual linear Kesten processes, the underlying mechanism is ``proportional growth'' or multiplicative proportional growth in order for Zipf's law and related power laws to occur, in the presence of an average contraction, i.e., the mean growth rate is negative, together with a reinjection mechanism. In contrast, our FAI Hawkes models are characterised by explosive expansions in the presence of a reflecting or bounded boundary condition. This is in stark contrast to the conventional approaches based on the Kesten type models, proportional growth type, and preferential attachment type mechanisms.

	\subsection{Implication to financial data analyses}
			From a broader perspective, our results have significant implications for financial modelling. Recall that one of the motivations for introducing the NLHawkes family is to reproduce empirical stylised facts, in particular the power law distribution of returns (see Sec.~\ref{subsec:QHawkes_review} for a brief review). Indeed, according to Ref.~\cite{QHawkesBouchaud}, one of the advantages of the QHawkes process lies in the fact that it can reproduce a power law intensity distribution with non universal exponents (see Eq.~\eqref{eq:power-law_QHawkes_Bouchaud}), from which the power law distribution of price changes derives. 
			
			From this point of view, our results summarised in Table~\ref{table:summary_classification_discussion} show that various NLHawkes processes can reproduce power law intensity PDFs, and not just the QHawkes processes. Even the ramp Hawkes processes with one-sided marks (which is arguably a minor modification of the LHawkes process) can reproduce a power law distribution with arbitrary exponent, when near criticality. If one focuses only on power law intensity PDFs, various Hawkes models can be suitable candidate models. Moreover, by assuming symmetric marks, note that QHawkes is the model at the boundary between a class of models with universal power law exponents and a class of models with non-universal exponents. 
			
			Another debatable point is whether the exponent should be universal or non-universal. The linear and ramp Hawkes processes with one-sided marks, the symmetric-mark QHawkes processes, and the two-sided FAI Hawkes process with nonpositive mean mark have non-universal exponents for the power law intensity PDF. In contrast, the symmetric-mark FAI Hawkes processes have universal exponents. The non-universal exponents are useful for flexible data calibration, while one would like to have strong justifications of why such parameters are selected in empirical data analyses. On the other hand, universal exponents are useful if the empirical exponent seems robust and universal, but this removes flexible data calibration. It might be necessary to construct a suitable framework for model selection with the goal of developing practical reverse engineering approaches. This is beyond the scope of the present paper.

	\subsection{Mathematical relation to quantum field theories}
		We have studied a wide variety of generalised Hawkes processes. Here we discuss their mathematical relation with quantum field theories. Let us rewrite $z(x)\to \phi(x)$ and introduce a ``momentum'' operator $\pi(x)$
		\begin{equation}
			\pi(x) := -i\frac{\delta}{\delta \phi(x)},
		\end{equation}
		which satisfies the canonical commutative relation
		\begin{equation}
			[\phi(x), \pi(x')] = i\delta (x-x').
		\end{equation}
		By introducing the state vector
		\begin{equation}
			|P_t \ra := \int \mathcal{D}\phi P_t[\phi] |\phi \ra,
		\end{equation}
		the ME~\eqref{eq:functional_master_gen} becomes a Schr\"{o}dinger-like equation for the field $\{\phi(x)\}_x$ as 
		\begin{equation}
			\frac{\partial }{\partial t}|P_t \ra = H |P_t\ra, \>\>\> H := i\int_0^{\infty}\frac{dx}{x}\pi(x)\phi(x) + \int_{-\infty}^{\infty}dy\rho(y)\left(T\left[y\tilh \right]-1\right)G[\phi]
		\end{equation}
		with the non-Hermitian Hamiltonian $H$ and the translation operator $T[y]$, defined by
		\begin{equation}
			T[y]:= \exp\left[-i\int_{0}^{\infty} dxy(x)\pi(x)\right]
		\end{equation}
		satisfying $T[y]P_t[z]=P_t[z-y]$. This equation is non-local, since the Hamiltonian includes infinite-order of ``momentum'' operators.
		
		In the diffusive limit~\eqref{eq:functional_FP_diffusive}, the Hamiltonian reduces to local version as a result of the SSE: 
		\begin{equation}
			\frac{\partial }{\partial t}|P_t \ra = H |P_t\ra, \>\>\> H := i\int_0^{\infty}\frac{dx}{x}\pi(x)\phi(x) - \int_0^\infty dx\int_0^\infty dx' D(x,x') \pi(x)\pi(x')G[\phi], \>\>\> D(x,x'):= \frac{\alpha_2}{2}\tilh(x)\tilh(x'), 
		\end{equation}
		where the ``momentum'' operator appears via a quadratic form. In this sense, the SSE for the field ME can be regarded as a mathematical procedure to obtain local forms of non-Hermitian field quantum theories in an appropriate limit.

	\subsection{Future application: field master equation for the general QHawkes processes}
	
		Our current formulation covers a part of the ZHawkes processes, but not yet the whole class of general QHawkes processes~\eqref{eq:QHawkes_review}. Including the general QHawkes processes as special cases of our formalism can in principle be easily performed as follows. The linear and quadratic kernels $L(t-s)$ and $K(t-s,t-u)$ can be decomposed according to a Laplace representation: 
		\begin{equation}
			L(t-s) = \int_0^\infty dx ~e^{-\frac{t-s}{x}}\tl{L}(x) ,\>\>\>
			K(t-s,t-u) = \int_0^\infty dx \int_0^\infty dx' ~e^{-\frac{t-s}{x}-\frac{t-u}{x'}}\tl{K}(x,x').
		\end{equation}
		The QHawkes process then reads
		\begin{align}
			\hat{\lambda}(t) = &\lambda_0 + 
							\int_{-\infty}^t ds\hxi^{\mrP}_{\rho(y);\hat{\lambda}(s)}(s)
							\left(\int_0^\infty dx e^{-\frac{t-s}{x}}\tl{L}(x) \right) \notag\\
							&+
							\int_{-\infty}^t ds ~\hxi^{\mrP}_{\rho(y);\hat{\lambda}(s)}(s)\int_{-\infty}^t du~\hxi^{\mrP}_{\rho(y);\hat{\lambda}(u)}(u)
							\left(\int_0^\infty dx~ \int_0^\infty dx' e^{-\frac{t-u}{x'}-\frac{t-s}{x}}\tl{K}(x,x')\right)\notag \\
							=  &\lambda_0 + 
							\int_0^\infty dx~ \tl{L}(x) 
							\left(\int_{-\infty}^t ds~e^{-\frac{t-s}{x}}\hxi^{\mrP}_{\rho(y);\hat{\lambda}(s)}(s)\right) \notag\\
							&+
							\int_0^\infty dx \int_0^\infty dx' \tl{K}(x,x') 
							\left(\int_{-\infty}^t ds~e^{-\frac{t-s}{x}}\hxi^{\mrP}_{\rho(y);\hat{\lambda}(s)}(s)\right) \left(\int_{-\infty}^t du~e^{-\frac{t-u}{x'}}\hxi^{\mrP}_{\rho(y);\hat{\lambda}(u)}(u)\right), 
		\end{align}
		where we have exchanged the order of integration. By considering the Markovian SPDE
		\begin{subequations}
			\label{eq:Markov_embedding_QHawkes_set}
			\begin{equation}
				\frac{\partial \hz(t,x)}{\partial t} = -\frac{\hz(t,x)}{x} + \hxi^{\mrP}_{\rho(y);\hat{\lambda}(t)}(t),
			\end{equation}
			whose explicit solution is given by
			\begin{equation}
				\hz(t,x) = \int_{-\infty}^t dse^{-\frac{t-s}{x}}\hxi^{\mrP}_{\rho(y);\hat{\lambda}(s)}(s),
			\end{equation}
			we obtain 
			\begin{equation}
				\hat{\lambda}(t) = \lambda_0 + \int_0^\infty dx \tl{L}(x) \hz(t,x) + \int_0^\infty dx \int_0^\infty dx' \tl{K}(x,x') \hz(t,x)\hz(t,x'),
			\end{equation}
		\end{subequations}
		which is equivalent to the original QHawkes processes~\eqref{eq:QHawkes_review}. This means that the QHawkes dynamics has been converted into a Markovian dynamics described by the set of equations~\eqref{eq:Markov_embedding_QHawkes_set} in terms of the field variable $\{\hz(t,x)\}_x$. Correspondingly, we obtain the field ME for the QHawkes processes as 
		\begin{equation}
			\frac{\partial P_t[z]}{\partial t} = \int_0^\infty dx \frac{\delta }{\delta z(x)}\left(\frac{z(x)}{x}P_t[z]\right) + \int_{-\infty}^\infty dy\rho(y)\left\{G[z-y1]P_t[z-y1] - G[y]P_t[z]\right\}
		\end{equation} 
		with the indicator function $1(x) = 1$ for any $x \in (0,\infty)$ and the functional intensity map $G[z]$ defined by
		\begin{equation}
			G[z] := \lambda_0 +\int_0^\infty dx \tl{L}(x) z(x)+ \int_0^\infty dx \int_0^\infty dx' \tl{K}(x,x') z(x)z(x'). 
		\end{equation}
		
		\subsubsection*{Further formal generalisation}
		Obviously, this method can be readily generalised for any functional series expansion, at least formally, such as 
		\begin{align}
			\hat{\lambda}(t) &= \lambda_0 + \sum_{j=1}^{J} \int_{-\infty}^t ds_1\dots \int_{-\infty}^t ds_j K_j(t-s_1,\dots,t-s_j)\hxi^{\mrP}_{\rho(y);\hat{\lambda}(s_1)} \dots \hxi^{\mrP}_{\rho(y);\hat{\lambda}(s_j)}\notag \\
			&= \lambda_0 + \sum_{j=1}^{J} \int_0^\infty ds_1\dots \int_0^\infty ds_j \tl{K}_j(x_1,\dots,x_j)\hz(t,x_1) \dots \hz(t,x_j).
		\end{align}
		Here, we have introduced the field variable $\{\hz(t,x)\}$ obeying the Markovian SPDE
		\begin{equation}
			\frac{\partial \hz(t,x)}{\partial t} = -\frac{\hz(t,x)}{x} + \hxi^{\mrP}_{\rho(y);\hat{\lambda}(t)},
		\end{equation}
		and the Laplace decomposition
		\begin{equation}
			K_j(t-s_1,\dots,t-s_j) = \int_0^\infty dx_1 \dots \int_0^\infty dx_j ~e^{-\sum_{l=1}^j\frac{t-s_l}{x_l}}\tl{K}_j(x_1,\dots,x_j).
		\end{equation}
		The corresponding field ME can be derived in the same manner. This implies that our formulation has the potential to cover a wide variety of NLHawkes families beyond the QHawkes processes. We leave to future studies the derivation of explicit analytical solutions for general QHawkes processes and beyond, based on our formulation.

\section{Conclusion}\label{sec:conclusion}

	In this article, we have studied various analytical solutions to NLHawkes processes by generalising the field ME approach recently developed in Refs.~\cite{KzDidier2019PRL,KzDidier2019PRR}. We have derived the field ME for the general NLHawkes processes and have formulated its functional KM expansion and the corresponding diffusive approximation. We then proceeded with deriving various exact solutions of the steady-state intensity distributions for an exponential memory kernel in the absence and presence of inhibitory effects. Some of the robust asymptotic solutions have been generalised for a wide class of memory kernels, such as (i)~the non-universal power law with an arbitrary exponent for the ramp Hawkes process in the absence of inhibitory effects, (ii)~the robust Zipf law for the superexponential intensity family in the presence of symmetric inhibitory and excitatory effects, and (iii)~the ubiquitous power law for the fast-acceleration intensity (FAI) Hawkes models in the regime of zero or negative mean mark.

	The summary table~\ref{table:summary_classification_discussion} exemplifies our systematic analysis of the NLHawkes processes. However, there are two missing items in the last column of the table for general memory kernels $h(t)$. This is because our focus has been mainly on FAI Hawkes processes and the ramp Hawkes process with non-positive mean mark and the QHawkes processes with symmetric mark are not FAI Hawkes processes. It is likely that different perturbative solutions are needed to solve these two cases for general memory kernels $h(t)$, which we leave for a future work.

	While only a few analytical solution for limited cases have been derived in the past for NLHawkes processes due to their nonlinear and non-Markovian nature, we have significantly extended the set of solutions, obtaining exact and robust asymptotic expressions with the help of our formulation in terms of a field ME. This demonstrates the power of this approach in addressing non-Markovian stochastic processes. It would be interesting to generalise this framework for more general non-Markovian stochastic processes, such as non-Markovian point processes that have arbitrary intensities depending on the full past history. In addition, our results imply that the NLHawkes family can accommodate various power law relations in the intensity distribution, which could be useful for data calibration in various complex systems. 	 

\begin{acknowledgements}
	This work was supported by (i) JST, PRESTO Grant Number JPMJPR20M2, Japan, (ii) the Japan Society for the Promotion of Science KAKENHI (Grant No.~20H05526 and No.~22H04830), (iii) Intramural Research Promotion Program in the University of Tsukuba, and (iv) the National Natural Science Foundation of China under grant No. U2039202. We thank Y. Terada and J.-P. Bouchaud for fruitful discussions.
\end{acknowledgements}

\appendix

\section{Formal properties of the Dirac delta function}
\label{sec:app:dirac_delta}
	\subsection{Dirac delta function}
		The Dirac delta function is formally defined by the following relationships for real numbers $x,y$:
		\begin{equation}
			\delta(x-y) = \begin{cases}
				0 & (x\neq y) \\
				\infty & (x=y)
			\end{cases}, \>\>\>
			\int_{-\infty}^\infty f(x)\delta(x-y)dx = f(y),
		\end{equation}
		which is the continuous analogue to the Kronecker delta, defined by 
		\begin{equation}
			\delta_{ij} = \begin{cases}
				0 & (i\neq j) \\
				1 & (i=j)
			\end{cases}, \>\>\>
			\sum_i f_i\delta_{ij}dx = f_i,
		\end{equation}
		for integer $i,j$.

		There are several formal methods to construct the Dirac delta function. In this paper, we construct the Dirac delta function via a formal continuous limit from the discrete picture. Let us consider the lattice coordinate $x_i := i dx$ for an integer $i$ with the lattice interval $dx$. The Dirac delta function can be formally introduced by 
		\begin{equation}
			\delta(x_i-x_j) = \lim_{dx \downarrow 0} \frac{1}{dx}\delta_{i,j},
		\end{equation}
		which satisfies 
		\begin{equation}
			\int_{-\infty}^\infty f(x)\delta(x-y)dx := \lim_{dx\downarrow 0}\sum_{i}f(x_i)\left[\frac{1}{dx}\delta_{ij}\right]dx = f(y) \>\>\> \mbox{for $y=x_j$.}
		\end{equation}

	\subsection{Functional derivative}
		The functional derivative is an analogue to the partial differential, such that 
		\begin{equation}
			\frac{\delta }{\delta z(x)}z(y) = \delta(x-y),\label{eq:app:delta_functionalder}
		\end{equation}
		which is similar to $(\partial/\partial z_i)z_j = \delta_{ij}$. The functional derivative can be constructed via a formal continuous limit, 
		\begin{equation}
			\frac{\delta}{\delta z(x_k)}[...] = \lim_{dx\downarrow 0}\frac{1}{dx}\frac{\partial}{\partial z_k} [...].
		\end{equation}
		Indeed, this definition satisfies the relationship~\eqref{eq:app:delta_functionalder}, such that 
		\begin{equation}
			\frac{\delta }{\delta z(x)}z(y) = \lim_{dx\downarrow 0}\frac{1}{dx}\frac{\partial}{\partial z_i}z_j = \lim_{dx\downarrow 0}\frac{1}{dx}\delta_{ij} = \delta(x_i-x_j)
		\end{equation}
		for $x=x_i$ and $y=x_j$.

	\subsection{Functional Taylor expansion}
		For a finite-dimensional vector $\bm{z}:=(z_1,\dots, z_N)$, the Taylor expansion is given by 
		\begin{equation}
			f(\bm{z}) = \sum_{n=1}^\infty \frac{1}{n!}\left(\sum_{i=1}^N z_i\frac{\partial}{\partial x_i}\right)^n f(\bm{x})\bigg|_{\bm{x}=\bm{0}}. 
		\end{equation}
		As its continuous analogue, the functional Taylor expansion for a function $f[z]:= f(\{z(x)\}_{x})$ reads 
		\begin{equation}
			f[z] = \sum_{n=1}^\infty \frac{1}{n!}\left(\int_{-\infty}^\infty dx z(x)\frac{\delta}{\delta y(x')}\right)^n f[y]\bigg|_{y=0}. 
		\end{equation}

\section{Another derivation of the field master equation~\eqref{eq:functional_master_gen}}\label{sec:app:field_master}
	Here, we provide another derivation of the field ME~\eqref{eq:functional_master_gen} by direct manipulation of PDFs. Let us consider the time evolution of any functional $f[\hz]:= f(\{\hz(t,x)\}_x)$, given by
	\begin{align}
		&df(\{\hz(t,x)\}_x) := f(\{\hz(t+dt,x)\}_x) - f(\{\hz(t,x)\}_x) \notag \\
		=& 	\begin{cases}
				\displaystyle -dt\int_0^\infty dx \frac{\hz(t,x)}{x}\frac{\delta f[\hz(t,x)]}{\delta \hz(t,x)} & (\mbox{No jump during }[t,t+dt)\mbox{: probability} = 1-\hat{\lambda}(t) dt) \\
				\displaystyle f(\{\hz(t,x) + \hat{y}\tilh(x) \}_x) - f(\{\hz(t,x)\}) & (\mbox{Jump in }[t,t+dt) \mbox{ with } \hat{y} \in [y,y+dy)\mbox{: probability} = \hat{\lambda}(t)\rho(y) dtdy)
			\end{cases}
	\end{align}
	with intensity 
	\begin{equation}
		\hat{\lambda}(t) = G[\hz]:= g\left(\int_0^\infty dx \hz(t,x)\right).
	\end{equation}
	By taking the ensemble average on both sides of the equation, we obtain
	\begin{align}
		\int \mcD z f[z]\frac{\partial P_t[z]}{\partial t}dt = \int \mcD z \left[-dt\int_0^\infty dx \frac{z(x)}{x}\frac{\delta f[z]}{\delta z(x)} + dt\int dy\rho(y)G\left[  z \right] \left\{f[z+y\tilh ]-f[z]\right\} \right]P_t[z].
	\end{align}
	By integration by parts and performing a variable transformation $z+y\tilh \to z$, we obtain an identity
	\begin{equation}
		\int \mcD z f[z]\frac{\partial P_t[z]}{\partial t} = \int \mcD z f[z]\left[\int dx \frac{\delta }{\delta z}\frac{z}{x}P_t[z] + \int dy\rho(y) G\left[ z-y\tilh\right] P_t\left[ z-y\tilh\right]-G\left[ z\right] P_t[z] \right].
	\end{equation}
	Since this identity holds for any functional $f[z]$, we obtain Eq.~\eqref{eq:functional_master_gen}.

\section{Integral identities for exponential mark distributions}
	\label{sec:app:integral_identity}
	Here we provide the detailed derivation of the identities~\eqref{eq:identity_one_sided_expon} and~\eqref{eq:identity_one_sided_expon2}. 
	\subsection{For positive contribution}
		Let us consider the following quantity
		\begin{equation}
			\mathcal{I}_+(\nu):= \int_0^\infty dy \frac{e^{-y/y^*}}{y^*} f(\nu-\eta y/\tau)
		\end{equation}
		with the boundary condition $\lim_{\nu \to \pm \infty}f(\nu)=0$. Let us differentiate both hand sides as 
		\begin{equation}
			\frac{d}{d\nu}\mathcal{I}_+(\nu) = \int_0^\infty dy \frac{e^{-y/y^*}}{y^*} \frac{d}{d\nu}f(\nu-\eta y/\tau).
		\end{equation}
		The identity 
		\begin{equation}
			\frac{d}{dy} f(\nu-\eta y/\tau) = -\frac{\eta}{\tau}\frac{d}{d\nu}f(\nu-\eta y/\tau),
		\end{equation}
		leads to
		\begin{equation}
			\frac{d}{d\nu}\mathcal{I}_+(\nu) = -\frac{\tau}{\eta}\int_0^\infty dy \frac{e^{-y/y^*}}{y^*} \frac{d}{dy} f(\nu-\eta y/\tau)
			= -\frac{\tau}{\eta}\left[\frac{e^{-y/y^*}}{y^*} f(\nu-\eta y/\tau)\right]_0^\infty - \frac{\tau}{\eta y^*}\int_0^\infty dy \frac{e^{-y/y^*}}{y^*} f(\nu-\eta y/\tau),
		\end{equation}
		where we have performed an integration by part. This means that 
		\begin{equation}
			\frac{d}{d\nu}\mathcal{I}_+(\nu) =  c f(\nu) - c \mathcal{I}_+(\nu), \>\>\> c:= \frac{\tau}{\eta y^*},
		\end{equation}
		which is equivalent to 
		\begin{equation}
			\left(1+\frac{1}{c}\frac{d}{d\nu}\right)\mathcal{I}_+(\nu) = f(\nu).
		\end{equation}
		We thus have the identity~\eqref{eq:identity_one_sided_expon}. 

	\subsection{For negative contribution}
		Let us consider the following quantity
		\begin{equation}
			\mathcal{I}_-(\nu):= \int_{-\infty}^0 dy \frac{e^{y/y^*}}{y^*} f(\nu-\eta y/\tau)
		\end{equation}
		with the boundary condition $\lim_{\nu \to \pm \infty}f(\nu)=0$. Let us differentiate both hand sides as 
		\begin{align}
			\frac{d}{d\nu}\mathcal{I}_-(\nu) &= \int_{-\infty}^0 dy \frac{e^{y/y^*}}{y^*} \frac{d}{d\nu}f(\nu-\eta y/\tau) \notag \\
			&= -\frac{\tau}{\eta}\int_{-\infty}^0 dy \frac{e^{y/y^*}}{y^*} \frac{d}{dy} f(\nu-\eta y/\tau) \notag \\
			&= -\frac{\tau}{\eta}\left[\frac{e^{y/y^*}}{y^*} f(\nu-\eta y/\tau)\right]_{-\infty}^0 + \frac{\tau}{\eta y^*}\int_0^\infty dy \frac{e^{y/y^*}}{y^*} f(\nu-\eta y/\tau),
		\end{align}
		where we have performed an integration by part. This means that 
		\begin{equation}
			\frac{d}{d\nu}\mathcal{I}_+(\nu) = - c f(\nu) + c \mathcal{I}_-(\nu), \>\>\> c:= \frac{\tau}{\eta y^*},
		\end{equation}
		which is equivalent to 
		\begin{equation}
			\left(1-\frac{1}{c}\frac{d}{d\nu}\right)\mathcal{I}_+(\nu) = f(\nu).
		\end{equation}
		We thus have the identity~\eqref{eq:identity_one_sided_expon2}.

\section{Summary of special functions}\label{app:specialFunctions}
	Here we summarise special functions used in this paper. 
	\subsection{Modified Bessel functions}\label{app:specialFunctions:modiedBessel}
		The modified Bessel functions of the first and second kinds, denoted by $I_{\gamma}(x)$ and $K_{\gamma}(x)$, are defined by 
		\begin{subequations}
		\begin{align}
			I_{\gamma}(x) &:= \sum_{k=0}^\infty \frac{1}{k!\Gamma(k+\gamma+1)}\left(\frac{x}{2}\right)^{2k+\gamma}, \\
			K_{\gamma}(x) &:= \frac{\pi}{2}\frac{I_{-\gamma}(x)-I_{\gamma}(x)}{\sin(\gamma \pi)}.
		\end{align}
		\end{subequations}

	\subsection{Confluent hypergeometric function}\label{app:specialFunctions:confluentHyperGeometric}
		The confluent hypergeometric functions of the first and second kinds are defined by
		\begin{subequations}
		\begin{align}
			{}_1F_1(a,b;x) &:= \frac{\Gamma(b)}{\Gamma(b-a)\Gamma(a)}\int_0^1 dte^{xt}t^{a-1}(1-t)^{b-a-1}, \\
			{}_1U_1(a,b;x) &:= \frac{1}{\Gamma(a)}\int_0^{\infty} dte^{-xt}t^{a-1}(1+t)^{b-a-1},
		\end{align}
		\end{subequations}
		respectively. For positive $c$ and $\beta$, an asymptotic formula is available for large $x$
		\begin{equation}\label{eq:special_CH_asymp}
			{}_1U_1\left(1+\frac{c}{\beta},1+\frac{2c}{\beta}; x\right) \propto x^{-1-c/\beta}.
		\end{equation}

	\subsection{Generalised Laguerre function}\label{app:specialFunctions:gLaguerre}
		The generalised Laguerre function $L_a^b(x)$ is defined as the solution of the following differential equation: 
		\begin{equation}
			x\frac{d^2}{dx^2} L_a^b(x) + (b+1-x)\frac{d}{dx} L_a^b(x) + a L_a^b(x) = 0.
		\end{equation}
		For positive $c$ and $\beta$, an asymptotic formula is available for large $x$
		\begin{equation}\label{eq:special_LG_asymp}
			L_{-1-c/\beta}^{2c/\beta}\left(x\right) \simeq \frac{\Gamma(c/\beta)}{\Gamma(-c/\beta)\Gamma(1+c/\beta)}e^{x}x^{-c/\beta}.
		\end{equation}

	\subsection{Hypergeometric function}\label{app:specialFunctions:hyperGeometric}
		The hypergeometric function is defined as the analytic function whose expansion is given by
		\begin{equation}
			{}_2F_1(a,b,c; z) := \sum_{k=0}^\infty \frac{(a)_k(b)_k}{(c)_k}\frac{z^k}{k!}
		\end{equation}
		for $|z|<1$ with the Pochhammer symbol $(a)_k:=\Gamma(a+k)/\Gamma(a)$ and $
		(a)_0=1$. The hypergeometric function has the integral representation
		\begin{equation}
			{}_2F_1(a,b,c; z) := \frac{\Gamma(c)}{\Gamma(b)\Gamma(c-b)}\int_0^1t^{b-1}(1-t)^{c-b-1}(1-tz)^{-a}dt
		\end{equation}
		for $0<b<c$ and $|z|<1$ with real numbers $a$, $b$, $c$, and $z$. There is an identity
		\begin{equation}
			{}_2F_1(a,b,c; z) = (1-z)^{-a}{}_2F_1\left(a,c-b,c; \frac{z}{z-1}\right).
		\end{equation}

		\subsubsection{Useful identities}
			We state the following useful asymptotic formulas: For $n>2$ and large $\nu \to +\infty$, the asymptotic formula holds,
			\begin{subequations}\label{eq:app:hyperGeometricAsymptotics_poly}
			\begin{align}\label{eq:app:hyperGeometricAsymptotics_poly:a}
				\nu^2 {}_2F_1\left(1,\frac{2}{n},1+\frac{2}{n}; -\frac{k\nu^n}{\nu_0}\right) 
				=
				\underbrace{\frac{\frac{2\pi}{n}}{\sin \frac{2\pi}{n}}
				\left(\frac{\nu_0}{k}\right)^{\frac{2}{n}}}_{\mbox{const.}} + o(\nu^0).
			\end{align}
			In addition, for $n=2$ and large $\nu$, we obtain 
			\begin{align}\label{eq:app:hyperGeometricAsymptotics_poly:b}
				\nu^2 {}_2F_1\left(1,1,2; -\frac{k\nu^2}{\nu_0}\right) 
				=
				\frac{\nu_0}{k}\log \left(\frac{k\nu^2}{\nu_0}\right) + o(\nu^0),
			\end{align}
			where we have used the identity ${}_2F_1(1,1,2;x)=-x^{-1}\log(1-x)$. 
			\end{subequations}

		\subsubsection{Crossover between $n=2$ and $n>2$}
			\label{app:specialFunctions:hyperGeometric:crossover}
			The formulas~\eqref{eq:app:hyperGeometricAsymptotics_poly} are qualitatively different for $n=2$ and $n>2$ because of the analytical singularity of ${}_2F_1$ at $n=2$. Here we consider the crossover between $n=2$ and $n>2$. Since the expansion holds for $n>2$
			\begin{equation}\label{eq:app:hyperGeometricAsymptotics_poly:c}
				{}_2F_1\left(1,\frac{2}{n},1+\frac{2}{n}; -x\right) 
				= \frac{\frac{2\pi}{n}}{\sin \frac{2\pi}{n}}x^{-2/n} - \frac{\frac{2}{n}}{\frac{2}{n}-1}x^{-1} + O(x^{-2})
			\end{equation}
			for large $x$, by the substitution $x=k\nu^n/\nu_0$, we obtain the following expansion in terms of the small parameter $1/x$, valid for any $n>2$: 
			\begin{align}
				\nu^2 {}_2F_1\left(1,\frac{2}{n},1+\frac{2}{n}; -\frac{k\nu^{n}}{\nu_0}\right) 
				\simeq 	\nu^2 \frac{\frac{2\pi}{n}}{\sin \frac{2\pi}{n}}x^{-1}\left(x^{1-2/n}
								-\frac{\sin \frac{2\pi}{n}}{\pi\left(\frac{2}{n}-1\right)}\right) 
				= 	\frac{\nu_0}{k}\nu^{-2\eps/(1-\eps)} \frac{\pi(1-\eps)}{\sin \pi \eps}
						\left(x^{-\eps}-\frac{\sin \pi\eps}{\pi \eps}\right)
			\end{align}
			with $\eps:=1-2/n>0$. By taking the limit $\eps\downarrow 0$ for a large but fixed $\nu$, we can apply
			\begin{equation}
				\label{eq:app:cross_over_expansion}
				\frac{x^{-\eps}-1}{\eps} \simeq -\log x + \frac{\eps}{2} \left(\log x\right)^2 + \dots 
			\end{equation}
			to obtain 
			\begin{align}
				\lim_{\eps\downarrow 0}\left\{\nu^2 {}_2F_1\left(1,\frac{2}{n},1+\frac{2}{n}; -\frac{k\nu^{n}}{\nu_0}\right) \right\}
				\simeq -\frac{\nu_0}{k}\log \left(\frac{k\nu^2}{\nu_0}\right),
			\end{align}
			recovering formula~\eqref{eq:app:hyperGeometricAsymptotics_poly:b}. 
			
			In contrast, even if $\eps>0$ is small, the first-order truncation of expansion~\eqref{eq:app:cross_over_expansion} is not applicable for too large $\nu$'s. Indeed, the truncation is only valid for 
			\begin{equation}
				\frac{\eps}{2}\left(\log x\right)^2 \ll \left|\log x\right|
				\>\>\> \Longrightarrow \>\>\> 
				\frac{k}{\nu_0}\nu^{n} \ll e^{2/\eps}.
			\end{equation}
			We thus obtain the characteristic intensity of this crossover as $\lambda^* := \nu_0 e^{2/\eps}$.

\section{Solution of an integral equation}\label{sec:app:sol_integralSolution}
	Here we study the solution of the integral equation with the following form 
	\begin{equation}
		\int_{-\infty}^\infty dy\rho(y)\phi(\nu-y) - \phi(\nu) = 0,
		\label{eq:app:integral_eq}
	\end{equation}
	which repeatedly appear in this paper, with the assumption that $\phi(\nu)$ is nonnegative. We assume that the mark distribution is two-sided 
	\begin{equation}
		p_+ := \int_{0}^\infty \rho(y)dy > 0, \>\>\> 
		p_- := \int_{-\infty}^0 \rho(y)dy > 0
	\end{equation}
	and that the mean mark is nonpositive 
	\begin{equation}
		m:= \int_{-\infty}^\infty y\rho(y)dy \leq 0. 
	\end{equation}

	\subsection{For negative mean mark $m<0$}
		Let us first consider the case with negative mean mark $m<0$. Let us assume that a special solution of Eq.~\eqref{eq:app:integral_eq} is given by an exponential, 
		\begin{equation}
			\phi(\nu) \simeq C_0 e^{-c\nu} \>\>\> \mbox{ for large $\nu$}. 
		\end{equation}
		By substituting this solution into Eq.~\eqref{eq:app:integral_eq}, we obtain 
		\begin{align}
			\Phi(c) = 0, \>\>\> \Phi(x) := \int_{-\infty}^\infty \rho(y)(e^{xy}-1)dy.
		\end{align}
		Based on this fact, we decompose the general solution of Eq.~\eqref{eq:app:integral_eq} as the superposition of exponentials: 
		\begin{equation}
			\phi(\nu) \simeq  \sum_{i}C_i e^{-c_i \nu} \>\>\> \mbox{ for large $\nu$}, 
		\end{equation}
		where $c_i$ is the $i$-th root of $\Phi(x) = 0$. Because $\phi(\nu)$ is nonnegative, oscillatory solutions (corresponding to $c_i$ being a complex number) are excluded and thus $c_i$ must be a real number. According to Appendix~\ref{sec:app:phi_symmetric}, the solutions of $\Phi(x)=0$ are given by $x=0$ and $x=c^*>0$. Thus, we obtain 
		\begin{equation}
			\phi(x) \simeq C_0 + C_1e^{-c^* \nu} \>\>\> \mbox{ for large $\nu$}.
			\label{eq:app:sol_integralEq_negativeMean}
		\end{equation}

	\subsection{For zero mean mark $m=0$}\label{app:sec:Phi_zeroMeanMark}
		We next consider the case with zero mean mark  $m\uparrow 0$. According to Appendix~\ref{sec:app:phi_symmetric}, the roots of $\Phi(x)=0$ are given by $x=0$ and $x=c^*>0$. 
		
		In the zero mean mark limit $m\uparrow 0$, the positive root $c^*$ approaches zero as shown here. Let us assume that the mean mark is negative but very small, such that $m=-\eps$ with a small positive parameter $\eps>0$. The moment-generating function is expanded around $x=0$ as 
		\begin{equation}
			\Phi(x) = -\eps x + \frac{\la y^2\ra x^2}{2} + \dots, \>\>\> \la y^2\ra:= \int_{-\infty}^\infty y^2\rho(y)dy. 
		\end{equation}
		For small $\eps$, the positive root of $\Phi(x)=0$ is given by
		\begin{equation}
			c^* = \frac{2\eps}{\la y^2\ra} + o(\eps)> 0,
		\end{equation}
		which converges to zero for small $\eps$ limit: $\lim_{\eps\downarrow 0}c^*=0$.
		
		Assuming a small positive $\eps$, let us expand the solution~\eqref{eq:app:sol_integralEq_negativeMean} to obtain 
		\begin{equation}
			\phi(x) \simeq C'_0 + C'_1 x + O(C_1\eps^2), \>\>\> C_0' := C_0 + C_1, \>\>\> C_1' := -C_1c^* = O(C_1\eps).
		\end{equation}
		Since $C_0$ and $C_1$ are arbitrary constants, we can assume $C_1=O(\eps^{-1})$, $C_0'=O(\eps^0)$, and $C_1' = O(\eps^0)$, where the divergence of $C_1$ is absorbed by an appropriate selection of $C_0$. Under this assumption, the solution is given by
		\begin{equation}
			\phi(x) \simeq C'_0 + C'_1\nu + O(\eps^1), \>\>\> C_0' = O(\eps^0), \>\>\> C_1' = O(\eps^0).
		\end{equation}
		By taking the zero mean limit $\eps\downarrow 0$, we obtain 
		\begin{equation}
			\phi(x) \simeq C'_0 + C'_1\nu \>\>\> \mbox{ for large $\nu$}
		\end{equation}
		as the general solution.

\section{Analytical properties of the moment-generating function $\Phi(x)$}\label{sec:app:phi_symmetric}
	Here we summarise the analytical properties of the moment-generating function 
	\begin{equation}
		\Phi (x) := \int_{-\infty}^\infty dy\rho(y)(e^{xy}-1)
	\end{equation}
	in the regime where the mean mark is non positive
	\begin{equation}
		m:= \int_{-\infty}^\infty y\rho(y)dy \leq 0
	\end{equation}
	and the mark distribution is two-sided
	\begin{equation}
		p_+ := \int_{0}^\infty \rho(y)dy > 0, \>\>\> 
		p_- := \int_{-\infty}^0 \rho(y)dy > 0.
	\end{equation}
	The moment-generating function $\Phi(x)$ is a strictly convex funtion because 
	\begin{equation}
		\frac{d^2\Phi(x)}{dx^2} = \int_{-\infty}^\infty y^2\rho(y)e^{xy}dy > 0,
	\end{equation}
	implying that $\Phi(x)$ has no more than one minimum. 

	\subsection{For negative mean mark $m<0$}
		\begin{figure}
			\centering
			\includegraphics[width=160mm]{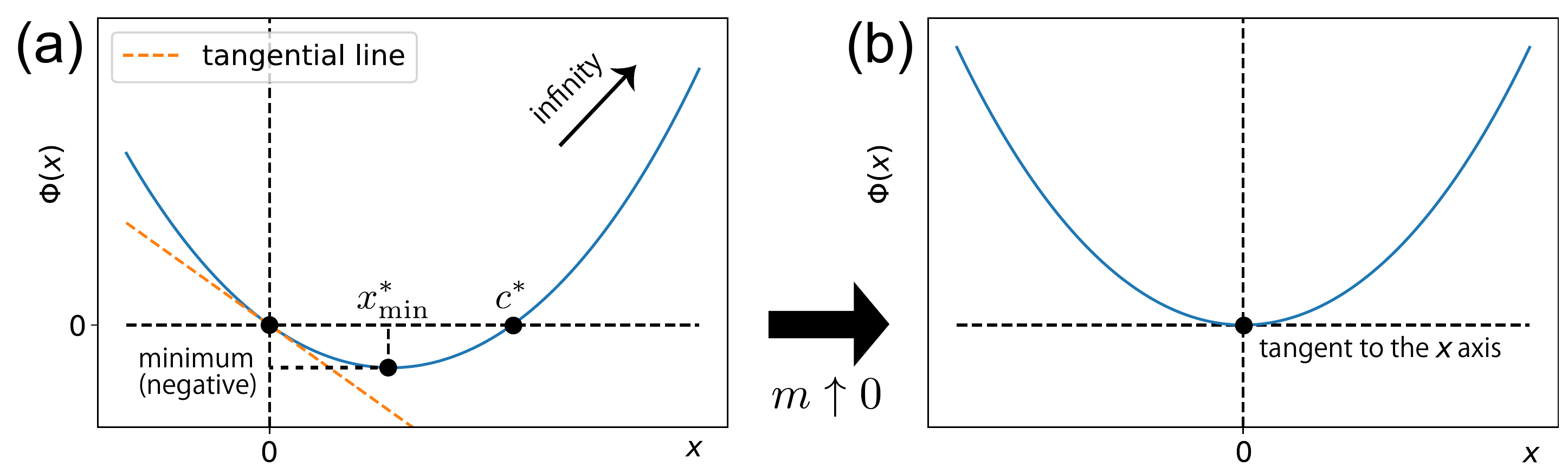}
			\caption{
				Schematic of the shape of the moment-generating function $\Phi(x)$. $\Phi(x)$ is strictly convex and takes specific values $\Phi(0)=0$, $\Phi(+\infty)=\infty$, and $d\Phi(0)/dx = m$. (a)~For negative $m$, the tangential line at $x=0$ has a negative slope $d\Phi(0)/dx = m < 0$. The minimum occurs at $x=x^*_{\min}>0$ and the roots of $\Phi(x)=0$ are given by $x=0$ and $x=c^*>0$. (b)~In the zero mean mark limit $m\uparrow 0$, $c^*$ approaches zero and thus the only root of $\Phi(x)=0$ is given by $x=0$. }
			\label{fig:CGF}
		\end{figure}
		
		Because we have assumed $p_+ =\int_0^\infty dy\rho(y)>0$, $\rho(y)\neq 0$ in some finite region of $y>0$. Therefore, the relation
		\begin{equation}
			m_+ := \int_{0}^\infty y\rho(y)dy > 0
		\end{equation}
		must hold. Based on this fact, the following three properties hold: 
		\begin{screen}
			\begin{itemize}
				\item[(i)] The tangential line of the curve $\Phi(x)$ at $x=0$ has a negative slope: 
				\begin{equation}
					\frac{d\Phi(x)}{dx}\bigg|_{x=0} = \int_{-\infty}^\infty y\rho(y)dy = m < 0.
				\end{equation}
				\item[(ii)] The moment-generating function is zero at $x=0$: 
				\begin{equation}
					\Phi(0) = 0. 
				\end{equation}
				\item[(iii)] The moment-generating function diverges to infinity for large $x$ 
				\begin{equation}
					\lim_	{x\to\infty}\Phi(x)=\infty
				\end{equation}
				because 
				\begin{align}
					\Phi(x) &= \int_0^\infty dy\rho(y)(e^{xy}-1) + \int_{-\infty}^0 dy\rho(y)(e^{xy}-1) \notag\\
					&\geq \int_0^\infty dy\rho(y)xy + \int_{-\infty}^0 dy\rho(y)(0-1) \notag \\
					&= m_+ x - p_- \to \infty, 
				\end{align}
				where we have used $e^{xy} \geq xy + 1$ for $x\geq 0$ and $e^{xy}>0$ for all $x$. These properties imply that the minimum of $\Phi(x)$ exists at some $x^*_{\min}>0$ as depicted in Fig.~\ref{fig:CGF}a and that the root of $\Phi(x)=0$ are given by $x=0$ and $x=c^*>0$. 
			\end{itemize}
		\end{screen}

	\subsection{For zero mean mark $m=0$}
		Let us consider the case where the mean mark is zero $m=0$, which is realised typically for symmetric mark distributions $\rho(y)=\rho(-y)$. Under this condition, the minimum occurs at $x=0$ because 
		\begin{equation}
			\frac{d\Phi(x)}{dx}\bigg|_{x=0} = 0.
		\end{equation}
		This implies that the solution of $\Phi(x)=0$ is $x=0$, which is a double root (see Fig.~\ref{fig:CGF}b). The appearance of the double root can be understood as the zero mean mark limit of the negative mean mark regime. While $c^*$ is positive for $m<0$, it approaches zero in the zero mean mark limit: $\lim_{m\uparrow 0}c^* = 0$ (see the discussion in Appendix.~\ref{app:sec:Phi_zeroMeanMark}).

		In the special case of symmetric mark distributions $\rho(y)=\rho(-y)$, the moment-generating function can be transformed into 
		\begin{equation}
			\Phi(x) = \int_{-\infty}^\infty \rho(y)(\cosh xy - 1)dy
		\end{equation}
		Also, considering the geometrical shape (see Fig.~\ref{fig:CGF}b), the equation
		\begin{equation}
			\Phi(x) = ax
		\end{equation}
		has a single posive root for positive $a$.

	\subsection{Special cases}
		For reference, we summarise several results for specific mark distributions. 
		\subsubsection{Gaussian mark distribution}
			Let us consider the Gaussian mark distribution
			\begin{equation}
				\rho(y) = \frac{1}{\sqrt{2\pi \sigma^2}}e^{-(y-m)^2/(2\sigma^2)}
			\end{equation}
			with mean mark $m$ and variance $\sigma^2$. The corresponding moment-generating function is given by 
			\begin{equation}
				\Phi(x) = e^{mx+\sigma^2 x^2/2} -1,
			\end{equation}
			which leads to the explicit formula for the root of $\Phi(c^*)=0$ as 
			\begin{equation}
				c^* = -\frac{2m}{\sigma^2}.
			\end{equation}
			
		\subsubsection{Two-sided exponential asymmetric mark distribution}
			We next consider the case of the two-sided exponential asymmetric mark distribution
			\begin{equation}
				\rho(y) = \begin{cases}
					\displaystyle \frac{p_+}{y^*_+} e^{-y/y^*_+} & (y\geq 0) \\
					\displaystyle \frac{p_-}{y^*_-} e^{y/y^*_-} & (y < 0)
				\end{cases},
			\end{equation}
			where $p_+ + p_- = 1$, $y_+^* >0$, and $y_-^* >0$. The mean mark is given by
			\begin{equation}
				m:= y^*_+ p_+ - y^*_- p_- < 0.
			\end{equation}
			We obtain 
			\begin{equation}
				\Phi(x) = \frac{p_+ y^*_+ x}{1-y^*_+ x} - \frac{p_- y^*_- x}{1+y^*_- x},
			\end{equation}
			which leads to 
			\begin{equation}
				c^* = \frac{p_-y_-^* - p_+y_+^*}{y^*_+ y^*_-} = -\frac{m}{y^*_+ y^*_-} > 0
				\label{eq:app:root_c*_two-sided_exponential}
			\end{equation}
			as the unique positive root of $\Phi(c^*)=0$.

\section{Proofs of mathematical properties of $\bm{H}$ \eqref{wrnhmnnh3}}	
\label{sec:app:proof_of_H}
	Here, we summarize the proofs of the main mathematical properties of $\bm{H}$ defined by Eq.~\eqref{wrnhmnnh3}. 
	
	\subsection{Proof that eigenvalues are real}\label{app:sec:proof_eigenvalues_H_real}
		We show that all eigenvalues of $\bm{H}$ are real numbers as follows. First, $\bm{H}$ can be symmetrized as $\bar{\bm{H}}$, defined by
		\begin{align}
			\bar{\bm{H}} 
			:= \bm{A}\bm{H}\bm{A}^{-1} = 
							\begin{pmatrix}
								\frac{1}{\tau_1}-\tilh_1,& -\sqrt{\tilh_1\tilh_2},& \dots& -\sqrt{\tilh_1\tilh_K} \\
								-\sqrt{\tilh_2\tilh_1},& \frac{1}{\tau_2}-\tilh_2,& \dots& -\sqrt{\tilh_2\tilh_K} \\
								\vdots& \vdots& \ddots& \vdots \\
								-\sqrt{\tilh_K\tilh_1},& -\sqrt{\tilh_K\tilh_2},& \dots& \frac{1}{\tau_K}-\tilh_K
							\end{pmatrix}, \>\>\>\>\>
			\bm{A} :=		\begin{pmatrix}
								\sqrt{\tilh_1},& 0,& \dots& 0 \\
								0,& \sqrt{\tilh_2},& \dots& 0 \\
								\vdots& \vdots& \ddots& \vdots \\
								0,& 0,& \dots& \sqrt{\tilh_K}
							\end{pmatrix}. \>\>\>
		\end{align}
		Indeed, by representing all the matrices by their elements $\bar{\bm{H}}:=(\bar{H}_{ij})$, $\bm{H}:=(H_{ij})$, and $\bm{A}:= A_{ij}$, we obtain
		\begin{equation}
			\bar{H}_{ij} = \sum_{k,l}A_{ik}H_{kl}A^{-1}_{lj} = 
			\sum_{k,l}\sqrt{\tilh_i}\delta_{ik}\left(\frac{\delta_{kl}}{\tau_k}-\tilh_l\right)\sqrt{\frac{1}{\tilh_j}}\delta_{lj} = \frac{\delta_{ij}}{\sqrt{\tau_i\tau_j}} - \sqrt{\tilh_i\tilh_j}.
		\end{equation}
		Since $\bar{\bm{H}}$ is a symmetric matrix, all their eigenvalues are real. 
		We obtain
		\begin{equation}
			\bm{H}\bm{e}_i = \lambda_i \bm{e}_i \>\>\> \Longleftrightarrow \>\>\> \bar{\bm{H}}\left(\bm{A}\bm{e}_i\right) = \lambda_i \left(\bm{A}\bm{e}_i\right).
		\end{equation}
		This relationship implies that any the eigenvalues of $\bm{H}$ are the same as that of $\bar{\bm{H}}$. Thus, all the eigenvalues of $\bm{H}$ are real likewise. 
		
	\subsection{Determinant}\label{app:sec:proof_determinant_H}
		The determinant $\det \bm{H}$ is derived as follows. Let us recall the invariance of determinants
		\begin{align}
			\det \bm{H} &=
					 	\det \begin{pmatrix}
							\bm{a}_1 \\
							\bm{a}_2 \\
							\vdots \\
							\bm{a}_j \\
							\vdots \\
							\bm{a}_K
						\end{pmatrix}
						=
					 	\det \begin{pmatrix}
							\bm{a}_1 \\
							\bm{a}_2 \\
							\vdots \\
							\bm{a}_j + c \bm{a}_k \\
							\vdots \\
							\bm{a}_K
						\end{pmatrix}
		\end{align}
		for any constant $c$. This implies
		\begin{align}
			\det \bm{H} &=
					 	\det \begin{pmatrix}
							\bm{a}_1 \\
							\bm{a}_2 \\
							\bm{a}_3 \\
							\vdots \\
							\bm{a}_K
						\end{pmatrix}
						=
					 	\det \begin{pmatrix}
							\bm{a}_1 \\
							\bm{a}_2-\bm{a}_1 \\
							\bm{a}_3 \\
							\vdots \\
							\bm{a}_K
						\end{pmatrix}
						=
					 	\det \begin{pmatrix}
							\bm{a}_1 \\
							\bm{a}_2-\bm{a}_1 \\
							\bm{a}_3-\bm{a}_1 \\
							\vdots \\
							\bm{a}_K
						\end{pmatrix}
						=
						\dots
						=
					 	\det \begin{pmatrix}
							\bm{a}_1 \\
							\bm{a}_2-\bm{a}_1 \\
							\bm{a}_3-\bm{a}_1 \\
							\vdots \\
							\bm{a}_K-\bm{a}_1
						\end{pmatrix}
						:= 
					 	\det \begin{pmatrix}
							\bm{a}_1' \\
							\bm{a}_2' \\
							\bm{a}_3' \\
							\vdots \\
							\bm{a}_K'
						\end{pmatrix}
		\end{align}
		and 
		\begin{align}
						\det \bm{H} =
					 	\det \begin{pmatrix}
							\bm{a}_1' \\
							\bm{a}_2' \\
							\vdots \\
							\bm{a}_K'
						\end{pmatrix}
						=
					 	\det \begin{pmatrix}
							\bm{a}_1'+\tau_2\tilh_2\bm{a}_2' \\
							\bm{a}_2' \\
							\vdots \\
							\bm{a}_K'
						\end{pmatrix}
						=
					 	\det \begin{pmatrix}
							\bm{a}_1'+\tau_2\tilh_2\bm{a}_2'+\tau_3\tilh_3\bm{a}_3' \\
							\bm{a}_2' \\
							\vdots \\
							\bm{a}_K'
						\end{pmatrix}
						=
						\dots 
						=
						\det \begin{pmatrix}
							\bm{a}_1'+\sum_{k=2}^K\tau_k\tilh_k\bm{a}'_k \\
							\bm{a}_2' \\
							\vdots \\
							\bm{a}_K'
						\end{pmatrix}
		\end{align}
		with constants $\{\tau_{k}\tilh_{k}\}_k$. Using these relations, the determinant of $\bm{H}$ is given by
		\begin{align}
			\det \bm{H} &=
					 	\det \begin{pmatrix}
										-\tilh_1 + 1/\tau_1,& -\tilh_2,& \dots,& -\tilh_K \\
										-\tilh_1,& -\tilh_2 + 1/\tau_2,& \dots,& -\tilh_K \\
										\vdots& \vdots& \ddots& \vdots \\
										-\tilh_1,& -\tilh_2,& \dots,& -\tilh_K  + 1/\tau_K
						\end{pmatrix}
						\begin{matrix}
							\leftarrow \bm{a}_1 \\
							\leftarrow \bm{a}_2 \\
							\vdots \\
							\leftarrow \bm{a}_K \\
						\end{matrix}\notag\\
						&=
						\det \begin{pmatrix}
							(1-\tau_1\tilh_1)/\tau_1,& -\tilh_2,& \dots& -\tilh_K \\
							-1/\tau_1,& 1/\tau_2,& \dots& 0\\
							\vdots& \vdots& \ddots& \vdots\\
							-1/\tau_1,& 0,& \dots& 1/\tau_K
						\end{pmatrix}
						\begin{matrix}
							\leftarrow \bm{a}_1' &=\bm{a}_1 \\
							\leftarrow \bm{a}_2' &=\bm{a}_2&-&\bm{a}_1 \\
							\vdots \\
							\leftarrow \bm{a}_K' &=\bm{a}_K&-&\bm{a}_1 \\
						\end{matrix}\notag\\
						&=
						\det \begin{pmatrix}
							(1-\sum_{k=1}^K\tau_k\tilh_k)/\tau_1,& 0,& \dots& 0 \\
							-1/\tau_1,& 1/\tau_2,& \dots& 0 \\
							\vdots& \vdots& \ddots& \vdots\\
							-1/\tau_1,& 0,& \dots& 1/\tau_K
						\end{pmatrix}
						\begin{matrix}
							\leftarrow \bm{a}_1'' &= \bm{a}_1' &+& \sum_{k=2}^K\tau_k\tilh_k\bm{a}_k' \\
							\leftarrow \bm{a}_2'' &= \bm{a}_2' \\
							\vdots \\
							\leftarrow \bm{a}_K'' &= \bm{a}_K' \\
						\end{matrix}\notag\\
						&= \frac{1-\sum_{k=1}^K \tau_k\tilh_k}{\tau_1\dots \tau_K}. 
		\end{align}
		Notably, $\det \bm{H}=0$ at criticality $\eta=1$. This singularity is consistent with the singularity of the inverse matrix $\bm{H}^{-1}$, as discussed in Appendix.~\ref{app:sec:inverse_matrix_H}. 
	
	\subsection{Inverse matrix}\label{app:sec:inverse_matrix_H}
		The inverse matrix of $\bm{H}$ is derived from the method of row reduction:
		\begin{align}
			&\left(
				\begin{array}{cccc|cccc}
					-\tilh_1 + 1/\tau_1,& -\tilh_2,& \dots& -\tilh_K& 1,& 0,&\dots,& 0 \\
					-\tilh_1,& -\tilh_2 + 1/\tau_2,& \dots& -\tilh_K& 0,& 1,&\dots,& 0 \\
					\vdots& \vdots& \ddots & \vdots& \vdots& \vdots & \ddots& \vdots \\
					-n_1/\tau_1,& -n_2/\tau_2,& \dots,& -n_K/\tau_K  + 1/\tau_K& 0,& 0,&\dots,& 1
				\end{array}
			\right)
						\begin{matrix}
							\leftarrow \bm{b}_1 \\
							\leftarrow \bm{b}_2 \\
							\vdots \\
							\leftarrow \bm{b}_K \\
						\end{matrix}\notag\\
			\to
			&\left(
				\begin{array}{cccc|cccc}
					(1-\tau_1\tilh_1)/\tau_1,& -\tilh_2,& \dots& -\tilh_K& 1,& 0,&\dots& 0 \\
					-1/\tau_1,& 1/\tau_2,& \dots& 0& -1,& 1,&\dots& 0 \\
					\vdots& \vdots& \ddots& \vdots& \vdots& \vdots& \ddots& \vdots\\
					-1/\tau_1,& 0,& \dots& 1/\tau_K& -1,& 0,&\dots& 1
				\end{array}
			\right)
						\begin{matrix}
							\leftarrow \bm{b}_1'&=\bm{b}_1& \\
							\leftarrow \bm{b}_2'&=\bm{b}_2&-&\bm{b}_1 \\
							\vdots \\
							\leftarrow \bm{b}_K'&=\bm{b}_K&-&\bm{b}_1 \\
						\end{matrix}\notag\\
			\to
			&\left(
				\begin{array}{cccc|cccc}
					(1-\eta)/\tau_1,& 0,& \dots& 0& 1-\sum_{k=2}^K\tau_k \tilh_k,& \tau_2 \tilh_2,&\dots& \tau_K \tilh_K \\
					-1/\tau_1,& 1/\tau_2,& \dots& 0& -1,& 1,&\dots& 0 \\
					\vdots& \vdots& \ddots& \vdots& \vdots& \vdots& \ddots& \vdots\\
					-1/\tau_1,& 0,& \dots& 1/\tau_K& -1,& 0,&\dots& 1
				\end{array}
			\right)
						\begin{matrix}
							\leftarrow \bm{b}_1'' &= \bm{b}_1' &+& \sum_{k=2}^K\tau_k \tilh_k\bm{b}'_k \\
							\leftarrow \bm{b}_2'' &= \bm{b}_2' \\
							\vdots \\
							\leftarrow \bm{b}_K'' &= \bm{b}_K' \\
						\end{matrix}\notag\\
			\to
			&\left(
				\begin{array}{cccc|cccc}
					1,& 0,& \dots& 0& \tau_1+\tau_1n_1/(1-\eta),& \tau_1n_2/(1-\eta),&\dots& \tau_1n_K/(1-\eta) \\
					-\tau_2/\tau_1,& 1,& \dots& 0& -\tau_2,& \tau_2,&\dots& 0 \\
					\vdots& \vdots& \ddots& \vdots& \vdots& \vdots& \ddots& \vdots\\
					-\tau_K/\tau_1,& 0,& \dots& 1& -\tau_K,& 0,&\dots& \tau_K
				\end{array}
			\right)
						\begin{matrix}
							\leftarrow \bm{b}_1''' &= \frac{\tau_1}{(1-\eta)}\bm{b}_1'' \\
							\leftarrow \bm{b}_2''' &= \tau_2\bm{b}_2'' \\
							\vdots \\
							\leftarrow \bm{b}_K''' &= \tau_K\bm{b}_K'' \\
						\end{matrix}\notag\\
			\to
			&\left(
				\begin{array}{cccc|cccc}
					1,& 0,& \dots& 0& \tau_1+\tau_1^2\tilh_1/(1-\eta),& \tau_1\tau_2\tilh_2/(1-\eta),&\dots& \tau_1\tau_K\tilh_K/(1-\eta) \\
					0,& 1,& \dots& 0& \tau_2\tau_1\tilh_1/(1-\eta),& \tau_2 + \tau_2^2\tilh_2/(1-\eta),&\dots& \tau_2\tau_K\tilh_K/(1-\eta) \\
					\vdots& \vdots& \ddots& \vdots& \vdots& \vdots& \ddots& \vdots\\
					0,& 0,& \dots& 1& \tau_K\tau_1\tilh_1/(1-\eta),& \tau_K\tau_2\tilh_2/(1-\eta),&\dots& \tau_K + \tau_K^2\tilh_K/(1-\eta)
				\end{array}
			\right)
						\begin{matrix}
							\leftarrow \bm{b}_1'''' &= \bm{b}_1''' \\
							\leftarrow \bm{b}_2'''' &= \bm{b}_2''' &+& \frac{\tau_2}{\tau_1}\bm{b}_1''' \\
							\vdots \\
							\leftarrow \bm{b}_K'''' &= \bm{b}_K''' &+& \frac{\tau_2}{\tau_1}\bm{b}_1'''\\
						\end{matrix}
		\end{align}
		with the branching ratio $\eta$ defined by 
		\begin{equation}
			\eta:= \sum_{k=1}^K \tau_k \tilh_k.
		\end{equation}
		This relation implies 
		\begin{align}
			\bm{H}^{-1} = 
			\begin{pmatrix}
				\tau_1+\tau_1^2\tilh_1/(1-\eta),& \tau_1\tau_2\tilh_2/(1-\eta),&\dots& \tau_1\tau_K\tilh_K/(1-\eta) \\
				\tau_2\tau_1\tilh_1/(1-\eta),& \tau_2 + \tau_2^2\tilh_2/(1-\eta),&\dots& \tau_2\tau_K\tilh_K/(1-\eta) \\
				\vdots& \vdots& \ddots& \vdots\\
				\tau_K\tau_1\tilh_1/(1-\eta),& \tau_K\tau_2\tilh_2/(1-\eta),&\dots& \tau_K + \tau_K^2\tilh_K/(1-\eta)
			\end{pmatrix}
		\end{align}
		or equivalently
		\begin{equation}
			H^{-1}_{ij}=\tau_i \delta_{ij} + \frac{\tau_i\tau_j \tilh_j}{1-\eta}
		\end{equation}
		in the representation by matrix elements. 
		The above calculation can be directly confirmed as follows: 
		\begin{equation}
			\bm{H}\bm{H}^{-1}=\bm{I} \>\>\> \Longleftrightarrow \>\>\> 
			\sum_{j=1}^K H_{ij}H^{-1}_{jk}=\sum_{j=1}^K \left(-\tilh_j+\frac{1}{\tau_j}\delta_{ij}\right)\left(\tau_j \delta_{jk} + \frac{\tau_j \tau_k \tilh_k}{1-\eta}\right) = \delta_{ik}. 
		\end{equation}
		The inverse matrix has a singularity at $\eta=1$, corresponding to the criticality of the ramp Hawkes process.

	\subsection{Eigenvectors of $\bm{H}$}
		Since $\bm{H}$ is directly associated with the real symmetric matric $\bm{\tilde{H}}$, $\bm{H}$ can be diagnalised, such that
		\begin{equation}
			\bm{P}:= (\bm{e}_1,\dots,\bm{e}_K), \>\>\> \bm{P}^{-1}\bm{H}\bm{P} = 
			\begin{pmatrix}
				\lambda_1,& 0, & \dots& 0 \\
				0,& \lambda_2,& \dots& 0 \\
				\vdots& \vdots& \ddots& \vdots \\
				0,& 0,& \dots& \lambda_K
			\end{pmatrix}
		\end{equation}
		with eigenvectors $\{\bm{e}_k\}_{k=1,\dots,K}$ and corresponding eigenvalues $\{\lambda_k\}_{k=1,\dots,K}$. 

		At criticality $\eta=1$, the smallest eigenvalues is zero, such that $\lambda_1=0$. In addition, the zero eigenvector $\bm{e}_1$ is explicitly given by 
		\begin{equation}
			\bm{e}_1 = \begin{pmatrix}
						\tau_1 \\
						\tau_2 \\
						\vdots \\
						\tau_K
					\end{pmatrix}
		\end{equation}
		Indeed, we can directly confirm the following relationship:
		\begin{equation}
			\bm{H}\bm{e}_1 = \begin{pmatrix}
								\frac{1}{\tau_1}-\tilh_1,& -\tilh_2,& \dots& -\tilh_K \\
								-\tilh_1,& \frac{1}{\tau_2}-\tilh_2,& \dots& -\tilh_K \\
								\vdots& \vdots& \ddots& \vdots \\
								-\tilh_1,& -\tilh_2,& \dots& \frac{1}{\tau_K}-\tilh_K
							\end{pmatrix}
							\begin{pmatrix}
								\tau_1 \\
								\tau_2 \\
								\vdots \\
								\tau_K
							\end{pmatrix}
							=\begin{pmatrix}
								1-\eta \\
								1-\eta \\
								\vdots \\
								1-\eta
							\end{pmatrix}
							= \bm{0}    ~~~~~{\rm for}~\eta:=\sum_{k=1}^K \tau_k\tilh_k=1.
		\end{equation}

		We next consider the representation based on the eigenvectors: 
		\begin{equation}
			\bm{X} = \begin{pmatrix}
				X_1 \\
				X_2 \\
				\vdots \\
				X_K
			\end{pmatrix} := \bm{P}^{-1}\bm{s}, \>\>\>\>~~~~~
			\bm{P}^{-1} =
			\begin{pmatrix}
				\bm{g}_1 \\
				\bm{g}_2 \\
				\vdots \\
				\bm{g}_K
			\end{pmatrix}.
		\end{equation}
		On the basis of this representation, we obtain 
		\begin{equation}
			\frac{dX_1}{dl} = 0 + O(\bm{X}^2), \>\>\> \frac{dX_k}{dl} = -\lambda_k X_k + O(\bm{X}^2) \>\>\> \mbox{ for }k\geq 2. 
		\end{equation}
		This implies that the leading-order contribution comes from the $X_1$ direction, such that 
		$|X_1| \gg |X_k|$ with $k\geq 2$. We can approximate 
		\begin{equation}
			\bm{X} = 
			\begin{pmatrix}
				X_1 \\
				0 \\
				\vdots \\
				0
			\end{pmatrix}
			+ O\left(\bm{X}^2\right) 
			\>\>\> \Longrightarrow \>\>\> 
			\bm{s} = \bm{P}\bm{X} \simeq X_1 \bm{e}_1 + O(\bm{X}^2).
		\end{equation}

		By direct substitution, we can confirm that $\bm{g}_1$ is given by 
		\begin{equation}
			\bm{g}_1 = \left(\frac{\tau_1 \tilh_1}{\la\tau\ra}, \>\dots\> , \frac{\tau_K \tilh_K}{\la\tau\ra}\right), \>\>\> 
			\la\tau \ra := \sum_{k=1}^K \tau_k^2\tilh_k. 
			\label{eq:app:matrix_H_def_g1}
		\end{equation}
		Indeed, this implies that $X_1$ is given by 
		\begin{equation}
			X_1 = \bm{g}_1 \cdot \bm{s} = \frac{1}{\la\tau \ra}\sum_{k=1}^K \tau_k \tilh_k s_k,
		\end{equation}
		which leads to 
		\begin{equation}
			\frac{dX_1}{dl} = \frac{1}{\la\tau \ra}\sum_{k=1}^K \tau_k \tilh_k \frac{ds_k}{dl} 
			= 0 + O(\bm{X}^2). 
			\label{eq:app:martix_H_leading_X1}
		\end{equation}
		Thus, we find that the first-order contribution is absent in Eq.~\eqref{eq:app:martix_H_leading_X1}, confirming the correctness of the representation of Eq.~\eqref{eq:app:matrix_H_def_g1}. In addition, this representation~\eqref{eq:app:matrix_H_def_g1} is consistent with the following identity: 
		\begin{equation}
			\bm{P}^{-1}\bm{P} = 
			\begin{pmatrix}
				\bm{g}_1 \\
				\bm{g}_2 \\
				\vdots \\
				\bm{g}_K
			\end{pmatrix}
			\begin{pmatrix}
				\bm{e}_1, \bm{e}_2, \dots, \bm{e}_K
			\end{pmatrix}
			=
			\begin{pmatrix}
				\tau_1 \tilh_1/\la \tau\ra, & \tau_2 \tilh_2/\la \tau\ra,& \dots& \tau_K \tilh_K/\la \tau\ra \\
				\bigcirc,& \bigcirc,& \dots& \bigcirc \\
				\vdots& \vdots& \ddots& \vdots \\
				\bigcirc,& \bigcirc,& \dots& \bigcirc
			\end{pmatrix}
			\begin{pmatrix}
				\tau_1,& \bigcirc,& \dots& \bigcirc \\
				\tau_2,& \bigcirc,& \dots& \bigcirc \\
				\vdots & \vdots& \ddots& \vdots \\
				\tau_2,& \bigcirc,& \dots& \bigcirc
			\end{pmatrix}
			=
			\begin{pmatrix}
				1,& 0,& \dots& 0 \\
				0,& 1,& \dots& 0 \\
				\vdots& \vdots& \ddots& \vdots \\
				0,& 0,& \dots& 1
			\end{pmatrix},
		\end{equation}
		where $\bigcirc$ represents some unspecified value\footnote{
				If we set $\bm{g}_1 = \left(\frac{\tau_1 \tilh_1}{c}, \>\dots\> , \frac{\tau_K \tilh_K}{c}\right)$ with some constant $c\neq \la\tau\ra$, the identity $\bm{P}^{-1}\bm{P}=\bm{E}$ does not hold with the unit vector $\bm{E}$, while the relation $dX_1/dl = 0 +O(\bm{s}^2)$ still holds. Therefore, $c$ must be $\la\tau \ra$ for the self consistency.} . Thus, we find that Eq.~\eqref{eq:app:matrix_H_def_g1} is the correct and consistent representation.

\section{Summary of asymptotic forms of the Laplace transform}\label{sec:app:Laplace}
	Here we summarise the asymptotic forms of the Laplace transform, in particular for power law distributions. Let us first recall the Tauberian theorem for the Laplace transform of asymptotic power law functions~\cite{KlafterB}:
	\begin{screen}
		Let us consider a function $f(x)$ satisfying the asymptotic form
		\begin{equation}
			f(x) \simeq x^{\rho-1}L(x) \>\>\> \mbox{for large }x
		\end{equation}
		with $0<\rho<\infty$ and slowly varing function $L(x)$. By definition, a slowly varying function satisfies $\lim_{x \to \infty}(L(Cx)/L(x)) = 1$ for any positive constant $C$. 
		The Laplace transform of $f(x)$ has the asymptotic form
		\begin{equation}
			\tl{f}(s):=\mathcal{L}_1[f(x);s] \simeq \Gamma(\rho)s^{-\rho}L(1/s) \>\>\> \mbox{for small }s.
		\end{equation}
	\end{screen}
	Using this theorem, let us consider the Laplace transform of power law functions $f(x)\simeq Ax^{-1-a}$ for various $a$ and positive constant $A$. 

	\subsection{Negative case: $a<0$}\label{sec:app:Laplace_0}
		For $a<0$, the Tauberian theorem can  be readily applied to obtain
		\begin{equation}
			\tl{f}(s) \simeq A's^{a}, \>\>\> A':= A\Gamma(-a) > 0 \>\>\> \mbox{for small }s.
		\end{equation}

	\subsection{Positive case: $0<a<1$}\label{sec:app:Laplace_1}
		Let us consider the following relation
		\begin{equation}
			\tl{f}(s) = \int_0^\infty f(x)e^{-sx}dx = \left[-F^{(0)}(x)e^{-sx}\right]_0^\infty - s\int_0^\infty F^{(0)}(x)e^{-sx}dx = F^{(0)}(0) - s\mathcal{L}_1\left[F^{(0)}(x);s\right], \>\>\> F^{(0)}(x):=\int_{x}^{\infty} f(x')dx'. 
		\end{equation}
		Here we notice that the asymptotic tail of $F^{(0)}(x)$ satisfies the condition of the Tauberian theorem, such that
		\begin{equation}
			F^{(0)}(x) \simeq \int_x^\infty \frac{Adx}{x^{1+a}} = \frac{A}{a}x^{-a} \>\>\> \mbox{for large }x.
		\end{equation}
		By applying the Tauberian theorem, we obtain
		\begin{equation}
			\tl{f}(s) \simeq F^{(0)}(0) - A's^{a} + o(s^a), \>\>\> A':= \frac{A}{a}\Gamma(1-a) > 0.
		\end{equation}
		When $f(x)$ is a PDF, $F^{(0)}(0)=\int_0^\infty dxf(x)=1$ and we obtain
		\begin{equation}
			\tl{f}(s) \simeq 1 - A's^a + o(s^a). 
		\end{equation}
		
	\subsection{General positive case: non-integer $0<a$}\label{sec:app:Laplace_2}
		Let us introduce the integer $m := \lfloor a \rfloor = \max\{k \in \bm{Z}\>|\> k\leq a\}$ with the set of integers $\bm{Z}$, satisfying $m\leq a \leq m+1$, and the iterated integral of $f(x)$: 
		\begin{equation}
			F^{(l)}(x) := \int_x^\infty dx_{l+1} \int_{x_{l+1}}^\infty dx_{l}\dots \int_{x_2}^\infty dx_1 f(x).
		\end{equation}
		We find that we can apply the Tauberian theorem to $F^{(m)}(x)$, because
		\begin{equation}
			F^{(m)}(x) \simeq \frac{A(-1)^{m+1}}{(-a)(1-a)\dots (m-a)}x^{m-a} =A(-1)^{m+1}\frac{\Gamma(-a)}{\Gamma(m-a+1)}x^{m-a}
		\end{equation}
		with $-1<m-a<0$. Due to the identity
		\begin{equation}
			\mcL_1\left[F^{(l)}(x); s\right] = F^{(l+1)}(0) - s\mcL_1\left[F^{(l+1)}(x); s\right],
		\end{equation}
		we obtain an asymptotic relation for small $s$, 
		\begin{align}
			\mcL_1\left[f(x); s\right] &= \sum_{k=0}^{m} (-s)^{k}F^{(k)}(0) +(-s)^{m+1}\mcL_1\left[F^{(m)}(x);s\right] \notag\\
			& \simeq \sum_{k=0}^{m} (-s)^{k}F^{(k)}(0) - A(-1)^{m}\Gamma(-a)s^{a} + o(s^a).
		\end{align}

		Since the iterated integral satisfies the identity
		\begin{equation}
			F^{(l)}(x) = \frac{1}{l!}\int_{x}^\infty (x'-x)^{l}f(x')dx',
		\end{equation}
		one finds that $F^{l}(0)$ is proportional to the $l$-th order moment $M_l$ when $f(x)$ is a PDF:
		\begin{equation}
			M_l := \int_{0}^\infty x^{l}f(x)dx = l! F^{(l)}(0).
		\end{equation}
		In other words,
		\begin{equation}
			\mcL_1\left[f(x); s\right] \simeq \sum_{k=0}^{m} \frac{(-1)^{k}M_k}{k!}s^k  + A's^{a} + o(s^a), \>\>\> A': = A\Gamma(-a).\label{eq:app:Laplace_gen}
		\end{equation}
		Since $A$ is positive for $f(x)$ to be a PDF (i.e., $f(x)\geq 0$), $A'$ must be negative (positive) for even (odd) $m$.

\section{Numerical implementation}\label{sec:app:numerical_implementation}

	This appendix describes our numerical method for the simulation of the NLHawkes processes. Our starting point is the Markovian SDE~\eqref{eq:der_K-expon_memory_Markov_embedding} for the discrete sum of exponentials. We introduce a discretised time series $\{t_i\}_i$ and time steps $\Delta t_i:= t_{i+1}-t_i$, satisfying $0 = t_{0} < t_1 < \dots < t_{N_{\mathrm{T}}}=T_{\mathrm{tot}}$. Given that the intensity of the state-dependent Poisson process is given by $\hat{\lambda}(t)= G(\hbmz(t))$, the discrete version of the SDE~\eqref{eq:der_K-expon_memory_Markov_embedding} is given by
	\begin{equation}
		\hz_k(t_i+\Delta t_i) = \hz_k(t_i)e^{-\frac{\Delta t_i}{\tau_k}} + 
			\begin{cases}
				0 & (\mbox{Probability: }1-G(\hbmz(t_k))\Delta t_k) \\
				\tilh_k\hat{y}_i  & (\mbox{Probability: }G(\hbmz(t_k))\Delta t_k)
			\end{cases}
	\end{equation}
	with the IID random number sequence $\{\hat{y}_i\}_i$ obeying the mark distribution $\rho(y)$. Given that $G(\hbmz(t_k))\Delta t_k$ must be sufficiently small such that $G(\hbmz(t_k))\Delta t_k\ll 1$ for a proper probability interpretation, we employ an adaptive scheme for the time discretisation
	\begin{equation}
		\Delta t_i = \min \left\{ \Delta t^{(1)}_{\max}, \frac{\Delta t^{(2)}_{\max}}{G(\hbmz(t_i))} \right\}, 
	\end{equation}
	because the intensity $G(\hbmz(t_i))$ sometimes takes extremely large values near criticality. For this setup, we obtained an empirical intensity distribution by assuming ergodicity as
	\begin{equation}
		P_{\mrss}(\lambda) = \lim_{t\to \infty}\la \delta(\lambda - \hat{\lambda}(t))\ra = \lim_{T_{\mathrm{tot}}\to \infty} \frac{1}{T_{\mathrm{tot}}}\int_0^{T_{\mathrm{tot}}}\delta(\lambda- G(\hbmz(t)))dt 
		\simeq \frac{1}{T_{\mathrm{tot}}}\sum_{i=0}^{N_{\mathrm{T}}-1} \delta(\lambda- G(\hbmz(t_i)))\Delta t_i. 
	\end{equation}
	For its practical implementation, we have applied a parallel computing technique for better convergence. We have obtained the empirical intensity distribution as
	\begin{equation}
		P_{\mrss}(\lambda) = \left< \lim_{T_{\mathrm{tot}}\to \infty} \frac{1}{T_{\mathrm{tot}}}\int_0^{T_{\mathrm{tot}}}\delta(\lambda- G(\hbmz(t)))dt \right>
		\simeq \frac{1}{N_{\mathrm{PC}}}\sum_{j=1}^{N_{\mathrm{PC}}}
		\left[
			\frac{1}{T_{\mathrm{tot}}}\sum_{i=0}^{N_{\mathrm{T}}^{(j)}-1} \delta(\lambda- G(\hbmz^{(j)}(t_i^{(j)})))\Delta t_i^{(j)}
		\right],
	\end{equation}
	where $\hbmz^{(j)}(t)$ is the trajectory obtained in the $j$-th parallel thread and $N_{\mathrm{PC}}$ is the number of total parallel threads. 
	
	\subsubsection{Ramp intensity map without inhibitory effect (Fig.~\ref{fig:numerical_ramp})}
		We describe the setup for Fig.~\ref{fig:numerical_ramp}, where the intensity function is given by the ramp function and the mark takes a single value as
		\begin{equation}
			G(\hbmz):= \max \left\{\sum_{k=1}^K \hz_k - \nu_1, \nu_0 \right\}, \>\>\>
			\rho(y) = \delta(y-1)
		\end{equation}
		with any positive number $\nu_0$ and any real number $\nu_1$. 

		\paragraph*{Figure~\ref{fig:numerical_ramp}a.}
			The parameters are given by $K=2$, $(\tau_1, \tau_2)=(1, 2)$, $(\tilh_1,\tilh_2)=(0.7,0.14995)$, $\eta=0.9999$, $\nu_0=0.01$, $\nu_1\simeq 0.385$, $T_{\mathrm{tot}}=5\times 10^6$, $\Delta t^{(1)}_{\max}=0.1$, and $\Delta t^{(2)}_{\max}=0.01$ with the initial condition $(\hz_1(0), \hz_2(0)) = (1,1)$. Since $\alpha_2=1$, we obtain the power law exponent $a\simeq 1.0$ from Eq.~\eqref{eq:power-law_gen_memory_K_one-sidedMark_ramp}. 
			The total number of parallel threads is given by $N_{\mathrm{PC}}=4$.

		\paragraph*{Figure~\ref{fig:numerical_ramp}b.}
			The parameters are given by $K=3$, $(\tau_1, \tau_2, \tau_3)=(1, 2, 3)$, $(\tilh_1,\tilh_2,\tilh_3)=(0.5, 0.15, 0.1999/3)$, $\eta=0.9999$, $\nu_0=0.01$, $\nu_1\simeq 0.147$, $T_{\mathrm{tot}}=5\times 10^6$, $\Delta t^{(1)}_{\max}=0.1$, and $\Delta t^{(2)}_{\max}=0.01$ 
			with the initial condition $(\hz_1(0), \hz_2(0),\hz_3(0)) = (1,1,1)$. Since $\alpha_2=1$, we obtain the power law exponent $a\simeq 0.5$ from Eq.~\eqref{eq:power-law_gen_memory_K_one-sidedMark_ramp}. The total number of parallel threads is given by $N_{\mathrm{PC}}=4$.

		\paragraph*{Figure~\ref{fig:numerical_ramp}c.}
			The parameters are given by $K=3$, $(\tau_1, \tau_2, \tau_3)=(1, 2, 3)$, $(\tilh_1,\tilh_2,\tilh_3)=(0.5, 0.15, 0.1999/3)$, $\eta=0.9999$, $\nu_0=0.01$, $\nu_1=0$, $T_{\mathrm{tot}}=5\times 10^5$, $\Delta t^{(1)}_{\max}=0.1$, and $\Delta t^{(2)}_{\max}=0.01$ with the initial condition $(\hz_1(0), \hz_2(0),\hz_3(0)) = (1,1,1)$. Since $\alpha_2=1$, we obtain the power law exponent $a=0$ from Eq.~\eqref{eq:power-law_gen_memory_K_one-sidedMark_ramp}. The total number of parallel threads is given by $N_{\mathrm{PC}}=4$.
		
	\subsubsection{MSA intensity map with inhibitory effect (Fig.~\ref{fig_numerical_MSA})}
		We describe the setup for Fig.~\ref{fig_numerical_MSA}, where the intensity map and the mark distribution are given by the exponential function with finite cutoff and the normal distribution, respectively, such that
		\begin{equation}
			G(\hbmz) = \min \left\{ \lambda_0 \exp\left[ \beta \sum_{k=1}^K\hz_k\right], \lambda_{\max}\right\}, \>\>\>
			\rho(y) = \frac{1}{\sqrt{2\pi \sigma^2}}e^{-\frac{y^2}{2\sigma^2}}.
		\end{equation}

		\paragraph*{Figure~\ref{fig_numerical_MSA}a.}		
		The parameters are given by $K=3$, $(\tau_1, \tau_2, \tau_3)=(1, 0.5, 2)$, $(\tilh_1,\tilh_2,\tilh_3)=(0.5,0.6,0.1)$, $\lambda_0=0.01$, $\beta=6$, $\lambda_{\max}=10^6$, $\sigma=0.3$, $T_{\mathrm{tot}}=5\times 10^6$, $\Delta t^{(1)}_{\max}=0.1$, and $\Delta t^{(2)}_{\max}=0.01$ with the initial condition $(\hz_1(0), \hz_2(0),\hz_3(0)) = (0, 0, 0)$. The total number of parallel threads is given by $N_{\mathrm{PC}}=8$.

		\paragraph*{Figure~\ref{fig_numerical_MSA}b.}		
		The parameters are given by $K=3$, $(\tau_1, \tau_2, \tau_3)=(1, 0.5, 2)$, $(\tilh_1,\tilh_2,\tilh_3)=(0.5,0.6,0.1)$, $\lambda_0=0.001$, $\beta=10$, $\lambda_{\max}=10^6$, $\sigma=0.3$, $T_{\mathrm{tot}}=5\times 10^5$, $\Delta t^{(1)}_{\max}=0.1$, and $\Delta t^{(2)}_{\max}=0.01$ with the initial condition $(\hz_1(0), \hz_2(0),\hz_3(0)) = (0, 0, 0)$.

	\subsubsection{Brownian motion with $\hnu$-dependent diffusion constant (Fig.~\ref{fig_numerical_BrownianMSA})}
		For Fig.~\ref{fig_numerical_BrownianMSA}, we describe the numerical method for the Brownian motion with $\hnu$-dependent diffusion constant governed by the SDE~\eqref{eq:discuss_Brownian_positionDependence}. The numerical simulation is based on the following discrete version
		\begin{equation}
			\hnu(t_i+\Delta t_i) = 
			\begin{cases}
				\hnu(t_i) + \sqrt{\tl{g}(\hnu(t_i)) \Delta t_i} \hxi_i^{\mrG} & (\hnu(t_i)\geq 0) \\
				0 & (\hnu(t_i) < 0)
			\end{cases}
		\end{equation}
		with independent Gaussian random number $\hxi_i^{\mrG}$. For Fig.~\ref{fig_numerical_BrownianMSA}, we employ the following model:  
		\begin{equation}
			\tl{g}(\hnu) = \lambda_0 e^{\beta \hnu}, \>\>\> \Delta t_i = \min \left\{   \Delta t^{(1)}_{\max}, \frac{\Delta t^{(2)}_{\max}}{\tl{g}(\hnu(t_i))}  \right\}
		\end{equation}
		with $\lambda_0=10^{-4}$, $\beta=3$, $\Delta t^{(1)}_{\max}=0.1$, and $\Delta t^{(2)}_{\max}=0.01$. The total time of the simulation was $T=5\times 10^5$ and the total number of parallel threads was $N_{\mathrm{PC}}=4$.

\end{document}